\documentclass[useAMS,usenatbib,fleqn]{mn2e}
\usepackage{graphicx}
\usepackage{amssymb}
\usepackage{amsmath}
\usepackage{placeins}

\title[Non-Gaussianities in the WMAP data]
  {Scale dependent non-Gaussianities in the CMB data identified with Minkowski Functionals and Scaling Indices}
\author[Modest et al.]
       {H. I.~Modest$^1$\thanks{E-mail: hmodest@mpe.mpg.de}, C. ~R\"ath$^1$, A. J. ~Banday$^{2,3}$, G. ~Rossmanith$^1$, R. ~S\"utterlin$^1$, \newauthor S.~Basak$^{4,5}$, J.~Delabrouille$^4$,   K. M. ~G\'{o}rski$^{6,7}$ and  G. E. ~Morfill$^1$\\
        $^1$ Max-Planck Institut f\"ur extraterrestrische Physik, Giessenbachstr. 1, 85748 Garching, Germany\\
        $^2$ CNRS, IRAP, 9 Av. colonel Roche, BP 44346, F-31028 Toulouse cedex 4, France\\
        $^3$ Universit\'e de Toulouse, UPS-OMP, IRAP,  F-31028 Toulouse cedex 4, France \\
        $^4$ CNRS, Laboratoire APC, 10 Rue Alice Domon et L\'{e}onie Duquet, 75205 Paris cedex 13, France\\
        $^5$ Laboratoire AIM, UMR CEA-CNRS-Paris 7, Irfu, SAp/SEDI, Service d'Astrophysique, CEA Saclay, F-91191 GIF-SUR-YVETTE\\ CEDEX, France\\
        $^6$ Jet Propulsion Laboratory, California Institute of Technology, Pasadena, CA 91109, USA \\
        $^7$ Warsaw University Observatory, Aleje Ujazdowskie 4, 00 - 478 Warszawa, Poland}
\date{Accepted for publication in MNRAS 2012 September 20.}
     
\pagerange{\pageref{firstpage}--\pageref{lastpage}} \pubyear{2012}

\def\LaTeX{L\kern-.36em\raise.3ex\hbox{a}\kern-.15em
    T\kern-.1667em\lower.7ex\hbox{E}\kern-.125emX}

\begin{document}

\label{firstpage}

\maketitle

\begin{abstract}
We present further investigations of the \emph{Wilkinson Microwave Anisotropy Probe} (WMAP) data by means of the Minkowski functionals and the scaling index method. 
In order to test  for non-Gaussianities (NGs) with respect to scale-dependencies we use so-called surrogate maps, in which possible phase correlations of the Fourier phases of the original WMAP data and the simulations, respectively, are destroyed by applying a shuffling scheme to the maps. A statistical comparison of the original maps with the surrogate maps then allows to test for the existence of higher order correlations (HOCs) in the original maps, also and especially on well-defined Fourier modes.

We calculate the $\sigma$-normalised deviation between the Minkowski functionals of original data and $500$ surrogates for different hemispheres in the sky and find ecliptic hemispherical asymmetries between the northern and southern ecliptic sky. Using Minkowski functionals as an image analysis technique sensitive to HOCs we find deviations from Gaussianity in the WMAP data with an empirical probability $p>99.8\%$ when considering the low-$\ell$ range with $\Delta \ell = [2,20]$. The analysis technique of the scaling indices leads to the same results for this $\ell$ interval with a slightly lower deviation but still at $p>99.8\%$. Although the underlying foreground reduction methods of the maps differ from each other, we find similar results for the WMAP seven-year ILC map and the WMAP seven-year (needlet-based) NILC map for deviations from Gaussianity in the low-$\ell$ range. Our results point once more to a cosmological nature of the signal. For a higher $\ell$ range with $\Delta \ell = [120,300]$ the results differ between the two image analysis techniques and between the two maps which makes an intrinsic nature of the signal on this $\ell$ range less likely. When we decrease the size of the analysed sky regions for the low-$\ell$ study, we do not find signatures of NG in the northern ecliptic sky. In the south we find individual spots which show deviations from Gaussianity.

In addition, we investigate non-Gaussian CMB simulations that depend on the $f_{\mathrm{NL}}$-parameter of the local type. These simulations with $f_{\mathrm{NL}}^{\mathrm{local}} = [0,\pm 100, \pm 1000]$ cannot account for the detected signatures on the low-$\ell$ range. 
\end{abstract}

\section{Introduction}
The primary anisotropies of the Cosmic Microwave Background (CMB) are caused by the primordial inhomogeneities of the Universe. They are assumed to be generated during a period of cosmic inflation and became the seeds for the structure of the density distribution in the Universe we observe today. The CMB was emitted from the surface of last scattering about 380,000 years after the Big Bang and the analysis of its $\Delta T/T \simeq 10^{-5}$ temperature fluctuations help unraveling the mysteries of inflation. 

The WMAP high resolution maps of the CMB allow detailed tests on the nature of the primordial density perturbations. The analysis of the CMB power spectrum is allowing for a high-precision determination of the fundamental cosmological parameters. However, any cosmological information that is encoded in the phases and correlations among them is not contained in the power spectrum and has to be extracted from measurements of higher-order correlations (HOCs).

%INFLATIONARY MODELS:
Evidence for the existence or non-existence of non-Gaussianity (NG) is necessary to choose the correct class of scenarios for the very early Universe. 
The inflationary scenario \citep{1981PhRvD..23..347G, 1982PhLB..108..389L, 1982PhRvL..48.1220A} was proposed about 30 years ago to solve the flatness, monopole and horizon problems of standard Big Bang cosmology. It endorses homogeneity and isotropy of the Universe, provides a mechanism for the generation of primordial, approximately scale-invariant and approximately Gaussian density perturbations and its predictions are consistent with the observed power spectrum \citep{2011ApJS..192...18K}. 

The Gaussian distribution of scalar (density) and tensor (metric) perturbations is a generic prediction of single-field slow-roll inflationary models. Yet, it has been shown that single field as well as two-/multi-field models \citep{1997PhRvD..56..535L,2002PhRvD..65j3505B, 2002PhRvD..66j3506B,2006JCAP...05..019V} generate a small amount of non-Gaussianity, below our current experimental limits though \citep{2003JHEP...05..013M, 2003NuPhB.667..119A}. Models where the primordial density perturbations are generated by a curvaton field may predict a high level of NG \citep{2001PhLB..522..215M,2002NuPhB.626..395E,2003PhRvD..67b3503L}. There are many more alternative inflationary scenarios which could generate NG at an observable level, e.g. Dirac-Born-Infeld inflation \citep{2004PhRvD..70j3505S,2004PhRvD..70l3505A} or ghost inflation \citep{2004JCAP...04..001A}. See also review articles on NGs from inflationary models written by \citet{2004PhR...402..103B}  and \citet{2010AdAst2010E..72C}. The determination of non-Gaussianity would make it possible to distinguish between these different inflationary models. However, inflation still remains as a paradigm and a determination of NGs would also constrain alternatives, e.g. ekpyrotic and cyclic models  \citep{2008PhRvL.100q1302B,2008PhRvD..78b3506L}, and help unraveling the nature of the primordial density perturbations in general. 

Various methods of statistical analysis of the WMAP data claim the detection of non-Gaussianity and different anomalies like hemispherical asymmetries, lack of power at large angular scales, alignment of multipoles, detection of the Cold Spot, etc. \citep{2004MNRAS.349..313P,2004ApJ...605...14E,2005ApJ...622...58E,2007ApJ...660L..81E,2004MNRAS.354..641H,2009ApJ...704.1448H,2004PhRvD..69f3516D,2004ApJ...609...22V,2009MNRAS.399.1921R}. In these studies, the level of NG is assessed by comparing WMAP data with simulated model-dependent CMB maps based on cosmological models and/or specific assumptions about the nature of non-Gaussianities as parametrised with e.g. the scale-independent scalar parameters $f_{\mathrm{NL}}$ and $g_{\mathrm{NL}}$. Fewer studies are testing the random phase hypothesis for Gaussian random fields analysing the distribution of the Fourier phases of the data \citep{2003ApJ...590L..65C,2004MNRAS.350..989C,2005PhRvD..72f3512N,2007ApJ...664....8C}. These model-independent tests also show signatures of anomalies and non-Gaussianities. 

In this paper, we apply the well established method of surrogate data sets to analyse Fourier phase correlations in the WMAP data. The method of the surrogates has first been established in \citet{Theiler199277} in order to detect weak nonlinearities in time series. For given possibly nonlinear data, so-called surrogate data sets are generated, which mimic the linear properties of the original time series. A comparison of original data set and surrogates through statistical measures sensitive to higher order correlations (HOCs) can reveal a significant deviation from linearity in the data. Extensions to this formalism to three-dimensional galaxy distributions \citep{2002MNRAS.337..413R} and two-dimensional simulated flat CMB maps \citep{2003MNRAS.344..115R} have been proposed and discussed. In a prior report,  we introduced a two-step surrogatisation scheme for full sky CMB observations which allows for a scale-dependent analysis of the data \citep{2009PhRvL.102m1301R}. This method yields significant signatures for both NG and ecliptic hemispherical asymmetries \citep{2009PhRvL.102m1301R,2011MNRAS.415.2205R}, especially on largest scales ($\ell\leq20$). In these papers, the HOCs are measured with the weighted scaling index method (SIM) introduced in \citet{2002MNRAS.337..413R}, \citet{2003MNRAS.344..115R} and \citet{2007MNRAS.380..466R}. The image data is represented as a point distribution comprising the spatial and scalar information of each pixel. The scaling indices then offer the possibility to estimate the local scaling properties of such a point distribution and depend on a scale parameter $r$.

We concentrate on the set of statistics known as Minkowski Functionals (MFs) as a comparative measure sensitive to HOCs. A general theorem of integral geometry states that three Minkowski Functionals quantify the integrated geometry and topology and therefore morphology of a two-dimensional density field. For a pixelised temperature map $\Delta T/T(n)$, we consider the excursion sets of the map, defined as the set of all map pixels with value of $\Delta T/T(n)$ greater than some threshold level $\nu$. The Minkowski functionals of these excursion sets completely describe the morphological properties of the underlying temperature map $\Delta T/T(n)$.
These measures embody the information from all orders of the correlation functions. They are additive measures which allows us to calculate them efficiently by summing up their local contributions. The calculation of n-point correlation functions is much more expensive computationally \citep{2005ApJ...622...58E,2009LNP...665..457S}. In comparison to the SIM the Minkowski functionals are scale-independent, i.e. they do not depend on any scale parameter. 

We apply our methods to WMAP seven-year ILC and NILC maps. Furthermore, we simulate and analyse non-Gaussian CMB maps that depend on the scale-independent $f_{\mathrm{NL}}^{\mathrm{local}}$ parameter and serve as first toy models for further tests on inflationary models. 

The paper is organised as follows. In Section 2 we briefly describe the observational and simulated data we use in our study. The method of generating surrogate maps is reviewed in Section 3. A discussion and comparison of the used test statistics, Minkowski functionals and scaling indices, can be found in Section 4. In Section 5 we present  our results and we draw our conclusions in Section 6.  

\section{Data Sets and Simulations}
For our studies we use the WMAP seven-year foreground-cleaned internal linear combination map (in the following: ILC7) 
generated and provided by \citet{2011ApJS..192...15G}. This map can be directly downloaded from the LAMBDA-website\footnote{http://lambda.gsfc.nasa.gov \label{lambdaNote}} with a HEALPix\footnote{http://healpix.jpl.nasa.gov} \citep{2005ApJ...622..759G} resolution parameter of $N_{\mathrm{side}}=512$, corresponding to $3,145,728$ sky pixels. The ILC7 map has one degree FWHM. For comparison we also include the seven-year needlet-based ILC map (in the following: NILC7) produced by \citet{2011MNRAS.tmp.1768B} pursuing a different approach for foreground removal. Because the needlet implementation of the ILC allows for optimising weights both as a function of sky direction and as a function of scale, the NILC7 is significantly less contaminated by foregrounds than other existing maps obtained from WMAP, in particular at low Galactic latitude (see \citet{2009A&A...493..835D} for a description of the needlet ILC method). It features a better total resolution, corresponding to the WMAP W-band resolution, and has an original HEALPix resolution of $N_{\mathrm{side}}=1024$. We  downgrade the map to $N_{\mathrm{side}} = 512$ in the employed HEALPix software. The ILC7 as well as the NILC7 map are weighted linear combinations of the five frequency channels $K, Ka, Q, V$ and $W$ that recover the CMB signal even in heavily foreground affected regions. The weights are calculated by requiring minimum variance in a given region of the sky under the constraint that the sum of the weights is unity.  Full sky maps ensure the required orthogonality of the set of basis functions $Y_{\ell m}$ when generating surrogates.

For our comparison of the ILC7 and NILC7 maps on larger scales, $120 \leq \ell \leq 300$, we choose to decrease the resolution of the original NILC7 map down to one degree FWHM before downgrading to a HEALPix resolution of $N_{\mathrm{side}} = 512$. In these higher $\ell$-ranges the beam effects have stronger influences. For our studies of the low-$\ell$ ranges with $0 \leq \ell \leq 20$ this influence is negligible.

The primordial non-Gaussianity that might arise during cosmic inflation has often been parametrised by the $f_{\mathrm{NL}}$ parameter in the following simple form of the curvature perturbation $\Phi$ \citep{PhysRevD.22.1882} with quadratic correction 
\begin{equation}
\Phi(x)  = \Phi_L(x)+ f_{\mathrm{NL}}\Phi_{\mathrm{NL}}(x)
\end{equation}
where $\Phi_L$ are Gaussian linear perturbations and $\Phi_{\mathrm{NL}}(x)$ is defined as 
\begin{equation}
\Phi_{\mathrm{NL}}(x)=\Phi^2_L(x)-\langle \Phi^2_L(x)\rangle
\end{equation}
and $f_{\mathrm{NL}}$ characterises the dimensionless amplitude of quadratic correction. This form of $f_{\mathrm{NL}}$ describes non-Gaussianity of the local type. In order to test for possible impacts of local type NG on the data, we compare the results obtained from the original data to models based on a scale-independent $f_{\mathrm{NL}}^{\mathrm{local}}$ parameter. Therefore, we compute temperature maps as constrained realisations of non-linear fields with varying $f_{\mathrm{NL}}^{\mathrm{local}}$.

The spherical harmonic coefficients $a_{\ell m}$ of the CMB temperature ($i=T$) and polarisation ($i=E$) anisotropies in harmonic space are related to the primordial fluctuations $\Phi_{\ell m}(x)$ via
\begin{equation}
a_{\ell m}^i = \int \mathrm{d} x \, x^2 \, \Phi_{\ell m}(x) \, \alpha_\ell^i(x)
\end{equation}
as a function of comoving distance $x$, where $\alpha_l^i(x)$ is the real space transfer function. A set of linear and non-linear spherical harmonic coefficients are a valid realisation of temperature and polarisation fluctuations for a given cosmological model, in this case $\Lambda$CDM model. The total $a_{\ell m}$ coefficients and a desired level of $f_{\mathrm{NL}}^{\mathrm{local}}$ NG are then calculated as 
\begin{equation}
a_{\ell m} = a_{\ell m}^{\mathrm{linear}} + f_{\mathrm{NL}} * a_{\ell m}^{\mathrm{non-linear}},
\end{equation}
where the linear and non-linear $a_{\ell m}$ coefficients are provided by \citet{2009ApJS..184..264E}. 

To explore the plausibility of $f_{\mathrm{NL}}$ as an explanation for the observed large-scale NGs, we use these $a_{\ell m}$ coefficients to simulate five co-added VW-band maps with $f_{\mathrm{NL}}^{\mathrm{local}} = [0,\pm 100, \pm 1000]$. As shown in \citet{2011MNRAS.415.2205R} the results for the method of surrogates of the simulated VW-band maps are similar to those of ILC-like simulations on large scales. The specific beam properties of the WMAP satellite are taken into account using the respective window functions for each differencing assembly $(V1-V2, W1-W4)$, being available again on the LAMBDA-website$^1$. For every assembly, we add Gaussian noise to these maps with a particular variance for every pixel of the sphere. This variance depends on the number of observations $N_i(\theta,\phi)$ in the respective direction and the noise dispersion per observation per different assembly, $\sigma_{0,i}$. After this procedure, we accumulate the V- and W-bands to a co-added VW-band via a noise-weighted sum \citep{2003ApJS..148....1B}:
\begin{equation} \label{GewSum}
T(\theta, \phi) = \frac{\sum_{i \in \mathcal{A}} T_i(\theta,\phi)/ \sigma^2_{0,i} } {\sum_{i \in \mathcal{A}} 1/ \sigma^2_{0,i}}
\end{equation}
In this equation, $\mathcal{A}$ characterises the set of required assemblies, for the co-added VW-map $\mathcal{A} = \lbrace V1,V2,W1,W2,W3,W4 \rbrace$. The parameters $\theta = [0,\pi]$ and $\phi=[0,2\pi]$ correspond to the co-latitude and the longitude on the sphere, while the seven-year noise per observation of the different assemblies is denoted by $\sigma_0$, given by  \citet{2011ApJS..192...14J}.
Again, the maps are decreased to a HEALPix resolution of $N_{\mathrm{side}} =  512$. Finally, we remove the residual monopole and dipole by means of the appropriate HEALPix routine.

\section{Method of Surrogates}
In order to constrain possible phase-correlations of the Fourier phases of the cosmic microwave background, we generate so-called surrogate maps as proposed in \citet{2003MNRAS.344..115R}. We destroy possible phase correlations in the data by applying a shuffling scheme to the phases. In order to test for the scale-dependence of non-Gaussianity this is done with a two-step procedure. 

The full sky CMB temperature anisotropy map of the celestial sphere with $\Delta T/T(\textbf{n}) \equiv \Delta T/T(\theta,\phi)$ can be expanded in orthonormal spherical harmonics $Y_{\ell m}$ as
\begin{equation}
\centering
\Delta T/T(\theta,\phi) = \sum\limits_{l=0}^{\infty} \sum\limits_{m=-l}^{l}a_{\ell m} Y_{\ell m}(\theta,\phi)
\end{equation}
with the complex spherical harmonic coefficients
\begin{equation}
\centering
a_{\ell m} = \int \mathrm{d} \textbf{n}\ T(\textbf{n}) Y^*_{\ell m}(\textbf{n})
\end{equation}
where $\textbf{n}$ is the unit direction vector and $a_{\ell m} = | a_{\ell m}| e^{i\phi_{\ell m}}$.
The linear properties of the underlying density field are contained in the absolute values $|a_{\ell m}|$, whereas all HOCs - if existent - are encoded in the phases $\phi_{\ell m}$ and the correlations among them. If the spherical harmonic coefficients are independent Gaussian random variables their probability density is
\begin{equation}
P(a_{\ell m}) \mathrm{d} a_{\ell m} = \frac{1}{\sqrt{2\pi C_\ell}}e^{-\frac{a_{\ell m}^2}{2C_\ell}} \mathrm{d} a_{\ell m}  
\end{equation}
(e.g. \cite{2010arXiv1008.1704R}) and the orthonormality relation holds:
\begin{equation}
\centering
\langle a_{\ell m} a^*_{\ell'm'}\rangle = \delta_{\ell \ell'} \delta_{m m'}C_{\ell},
\end{equation}
where $C_\ell$ is the angular power spectrum and $\delta$ is the Kronecker delta.
In this case, the amplitude $|a_{\ell m}|$ would be Rayleigh-distributed and the phase $\phi_{\ell m} = \arctan \left(\mathrm{Im}(a_{\ell m})/\mathrm{Re}(a_{\ell m})\right)$ would be independent and identically distributed (i.i.d.) and follow a uniform distribution in the interval $[-\pi,\pi]$.
The temperature values themselves would be normally distributed as well. The measured angular power spectrum $C_l^{obs}$ depends on the $a_{\ell m}$ coefficients by 
\begin{equation}
\centering
C_\ell^{\mathrm{obs}} = \frac{1}{2\ell+1} \sum\limits_{m=-1}^{\ell} |a_{\ell m}|^2,
\end{equation}
where $\langle C_\ell^{\mathrm{obs}}\rangle = C_\ell$.
In case the CMB variables $a_{\ell m}$ are independent and Gaussian distributed and their phases are therefore i.i.d. (independent and identically distributed) and consequentially uncorrelated, it is adequate (\citet{2009astro2010S.158K} and references therein) to only investigate the linear properties of the data described by the power spectrum $C_\ell$. In this paper, we focus on the non-linear information and test for possible \emph{phase correlations} of $\phi_{\ell m}$ in the data, which we define as non-Gaussianities of the CMB, to check if the above requirements are fulfilled.  

To test the hypothesis of independent Fourier phases we try to exclude further features of the data as e.g. artefacts due to experimental constraints. Therefore, we make sure that the data meets the following two requirements before generating the surrogate maps. The temperature distribution is Gaussian and the set of phases is uniformly distributed in the interval $[-\pi,\pi]$. To fulfil these conditions we perform two preprocessing steps. First, the almost Gaussian distributed temperature values of the original map in real space are replaced by an exact Gaussian distribution in a rank-ordered way, i.e. the lowest value of the original distribution is replaced with the lowest value of the Gaussian distribution etc. Second, in order to guarantee that the phases $\phi_{\ell m}$ are identically distributed the Fourier phases are remapped on to a set of uniformly distributed ones. No significant dependence on the specific Gaussian or uniform realisation, respectively, was found in these preprocessing steps. 

The scale dependent surrogate maps that are later analysed are obtained as follows. In our scale-dependent analysis we focus on the two $\ell$-ranges $\Delta \ell_1 = [2,20]$ and $\Delta \ell_2 = [120,300]$. The interval $\Delta \ell_1$ covers the largest spatial scales of the CMB, while the first peak of the power spectrum lies in the interval $\Delta \ell_2$. Our previous results have shown that findings of an almost scale-independent analysis with a shuffling range $\Delta \ell = [2,1024]$ are only the superposition of signals from the analysis with $\Delta \ell_1$ and $\Delta \ell_2$ shuffling ranges \citep{2011MNRAS.415.2205R}. Therefore, we assume that the results of these two intervals are of special interest in the analysis of phase correlations. We first generate a first order surrogate map, in which any correlation of phases $\phi_{\ell m}$ with $\ell$ outside the certain $\Delta \ell$-range of interest is destroyed: The phases $\phi_{\ell m}$ with $2 \leq \ell \leq 1024$ and $\ell \notin \Delta \ell = [\ell_{min},\ell_{max}]$, $0 < m\leq \ell$, are randomised through a shuffling procedure. In a second step, $N$ ($N = 500$ throughout this study) realisations of second order surrogate maps are generated from the first order surrogate map: The remaining phases $\phi_{\ell m}$ with $\ell \in \Delta \ell = [\ell_{min},\ell_{max}]$ are shuffled, while the already randomised phases in the first order surrogate for the scales not under consideration are preserved. Note that the Gaussian properties of the maps, which are given by $|a_{\ell m}|$, are exactly preserved for all surrogate maps. 

The first order surrogate preserves phase correlations in the $\Delta \ell$ range, if they were present in the original maps. In the second order surrogate map, where all phases are randomly distributed, all original correlations will be completely destroyed. The statistical comparison of the two classes of surrogates will thus reveal possible higher order correlations in the original maps on defined Fourier modes and uncover signatures for deviations from Gaussianity. 

One might argue that the above discussed phase shuffling is not the intuitive approach, since it is also possible to construct surrogates (of first as well as of second order) by just replacing the respective Fourier phases with randomly generated ones. In order to account for this aspect, we additionally investigate surrogates (for selected data sets) that  were constructed by replacing the phases with a completely new set of uniform distributed values in the interval $[-\pi,\pi]$. Note that for this replacement approach, one has to generate a new set of phases for the relative $\ell$-interval. It might be preferable to only rely on the information which is given in the underlying data set, instead of adding something new.  

Before analysing the maps with different image analysis techniques we decrease the HEALPix resolution of the maps from $N_{\mathrm{side}} = 512$ to $N_{\mathrm{side}} = 256$ for the data analysis and $N_{\mathrm{side}} = 64$ for the simulated maps. 

\section{Test Statistics}
\subsection{Minkowski Functionals}
A full morphological specification of an image requires geometrical as well as topological descriptors to characterise not only the shape and content but also the connectivity of spatial patterns. Hadwiger's theorem \citep{hadwiger:1957:vio} in the field of  integral-geometry states that any complete morphological descriptor of a set $Q$ in d-dimensional Euclidian space $\mathbb{E}^d$ is a linear combination of only $d+1$ functionals which meet some simple requirements; the so-called Minkowski Functionals $M_j$ with j ranging from $0$ to $d$. Functional $M_0$ and $M_1$ in two-dimensional space can be interpreted as the familiar geometric quantities surface area and perimeter, respectively. Functional $M_2$ is the topological Euler characteristic. In spherical, two-dimensional CMB sky map space $\mathbb{S}^2$ of radius $R$ and a constant curvature $K=R^{-2}$ they are formally defined as
\begin{alignat*}{3}
&M_0&& = \int_Q \mathrm{d} a\\
&M_1&& = \frac{1}{4}\int_{\partial Q} \mathrm{d} \ell\\
&M_2&& = \frac{1}{2\pi} \int_{\partial Q} \mathrm{d} \ell \ k_g,
\end{alignat*}
where $\mathrm{d} a$ and $\mathrm{d} l$ denote the surface element of $\mathbb{S}^2$ and the line element along the smooth boundary $\partial Q$, respectively, as explained in \citet{1998MNRAS.297..355S}. The factor $k_g$ is the geodesic curvature. The Minkowski functionals have been introduced into cosmology as descriptors for the morphological properties of large-scale structure by \citet{1994A&A...288..697M} and of CMB sky maps by \citet{1998NewA....3...75W} and \citet{1998MNRAS.297..355S}. 

In order to study the morphology of the temperature anisotropies $\Delta T/T (n)$, which can be considered as a smooth scalar field on $\mathbb{S}^2$,
we calculate the three Minkowski Functionals (MFs) of excursion sets $Q_{\nu}$ in the pixelised spherical two-dimensional maps of original data, surrogates and simulations.  $Q_{\nu}$ is defined as the set of all map pixels with value of $\Delta T/T$ greater than or equal to some temperature threshold $\nu$ by $Q_{\nu} = \lbrace n \in \mathbb{S}^2 | \Delta T/T (n) \ge \nu \rbrace$. 

The maps we are analysing are pixelised according to the HEALPix pixelisation scheme. HEALPix produces a partition of a spherical surface into exactly equal area quadrilateral pixels of varying shape which simplifies the calculation of the Minkowski functionals. The pixel size depends on the HEALPix resolution parameter of the grid equal to $N_{\mathrm{side}}=1,2,4,8,...$ corresponding to a total number of pixels of $N_{\mathrm{pix}}=12 \times N^2_{\mathrm{side}} = 12,48,192,768,...$ In this work we use a resolution parameter of $N_{\mathrm{side}} = 256$ and $N_{\mathrm{side}} = 64$, respectively. 

The temperature maps we want to analyse are divided into an active and a non-active part by running over 200 threshold steps $\nu_i$ with $-4\sigma_T \le \nu_i \le +4 \sigma_T$. At the first threshold step $\nu_0$ nearly every pixel is included in the active part besides few outliers with $T< -4\sigma_T$. The last step $\nu_{199}$ excludes most of the pixels.

We adapted an algorithm of \citet{michielsen01} to compute the Minkowski functionals of the pixelised maps. Conceptually, each active pixel is decomposed into 4 vertices, 4 edges and the interior of the pixel. We count the total number of active squares $n_s$ and edges $n_e$ and vertices $n_{v}$ between active and non-active pixels and compute the area $M_0$, the integral mean curvature or perimeter $M_1$ and the Euler characteristic $M_2$ from
\begin{alignat*}{3}
&M_0&&=n_s\\
&M_1&&=-4n_s+2n_e\\
&M_2&&=n_s-n_e+n_v.
\end{alignat*}
A technical difficulty with this procedure is to avoid counting an edge or vertex more than once. As suggested by \citet{michielsen01} we build up the original image by adding active pixels to an initially empty temporary image one by one. Depending on whether the surrounding pixels have already been activated in the temporary image or not, we then add up edges and vertices to their total numbers. The number of arithmetic operations required to compute $M_0$, $M_1$, and $M_2$ scales linearly with the number of active pixels and the total number of pixels of the image. 

\subsection{Weighted Scaling Indices}
For comparison we assess possible scale-dependent NGs in the CMB with the scaling indices method (SIM) as done in our series of earlier papers \citep{2002MNRAS.337..413R,2003MNRAS.344..115R, 2007MNRAS.380..466R,2009PhRvL.102m1301R,2009MNRAS.399.1921R,2011MNRAS.415.2205R, 2011AdAst2011E..11R}. The basic idea for this test statistic comes from the calculation of the dimensions of attractors in non-linear time series analysis \citep{Grassberger1983189}. The scaling indices have been extended to the field of image processing for texture discrimination. If the image data is represented as a point distribution in a $d$-dimensional embedding space the scaling indices represent one way to estimate the local morphological properties of this point set. Point-like, ring-like and sheet-like structures can be discriminated from each other and from a random background. 

In order to apply the SIM on the spherical CMB data, we have to transform the temperature anisotropies $\Delta T/T (n_i)$ 
with its pixels at positions $n_i=(\theta_i, \phi_i)$, $i=1,...,N_{pix}$ on the sphere 
to a three-dimensional point distribution of $N_{pix}$ points $P=\lbrace \textbf{p}_i\rbrace$ in an artificial embedding space, for which then the local scaling properties are assessed with the SIM.  

For each point the local weighted cumulative point distribution $\rho$ is calculated as 
\begin{equation}
\rho(\textbf{p}_i,r) = \sum^{N_{pix}}_{j=1}s_r[d(\textbf{p}_i, \textbf{p}_j)]
\end{equation}
where $s_r(\bullet)$ denotes a shaping function depending on a scale parameter $r$ and a distance measure $d(\bullet)$. 
In principle any differentiable shaping function and any measure can be used for calculating the scaling indices. We use the Euclidian norm as distance measure and set of Gaussian shaping functions, which leads to
\begin{equation}
\rho(\textbf{p}_i,r) = \sum^{N_{pix}}_{j=1} e^{-\frac{d_{ij}}{r}^2}, d_{ij}= \| \textbf{p}_i - \textbf{p}_j\|.
\label{eqn:rho}    
\end{equation}
The weighted scaling indices $\alpha(\textbf{p}_i,r)$ are then obtained by calculating the logarithmic derivative of $\rho(\textbf{p}_i,r)$ with respect to $r$:
\begin{equation}
\alpha(\textbf{p}_i,r) = \frac{\partial \log \rho(\textbf{p}_i,r)}{\partial \log(r)}.
\end{equation}
With the definition in Equation \ref{eqn:rho} the weighted scaling indices are expressed by 
\begin{equation}
\alpha(\textbf{p}_i,r) = \frac{\sum_{j=1}^{N_{pix}} q\left(\frac{d_{ij}}{r}\right)^q e^{-\left( \frac{d_{ij}}{r}\right)^q}}{\sum^{N_{pix}}_{j=1}e^{-\left( \frac{d_{ij}}{r}\right)^q}}.
\end{equation}

\subsection{Statistical Interpretation}

The two image analysis techniques, Minkowski functionals and scaling indices, are applied to scale-dependent full sky surrogate maps and 768 overlapping hemispherical maps, with two different solid angles: a solid angle of $2\pi$ (apex angle $\pi$) and $\sim\!0.6 \pi$ (apex angle $\pi/2$). These hemispherical maps are rotated around the full sky to study possible scale-dependent phase-correlations in the Fourier space of the maps with a certain spatial localisation of the phenomena. 
In this paper we focus on the analysis of the Minkowski functionals and compare our results to the scaling index method. 

\begin{figure*}
\includegraphics[width=2.6cm, keepaspectratio=true,angle={90}]{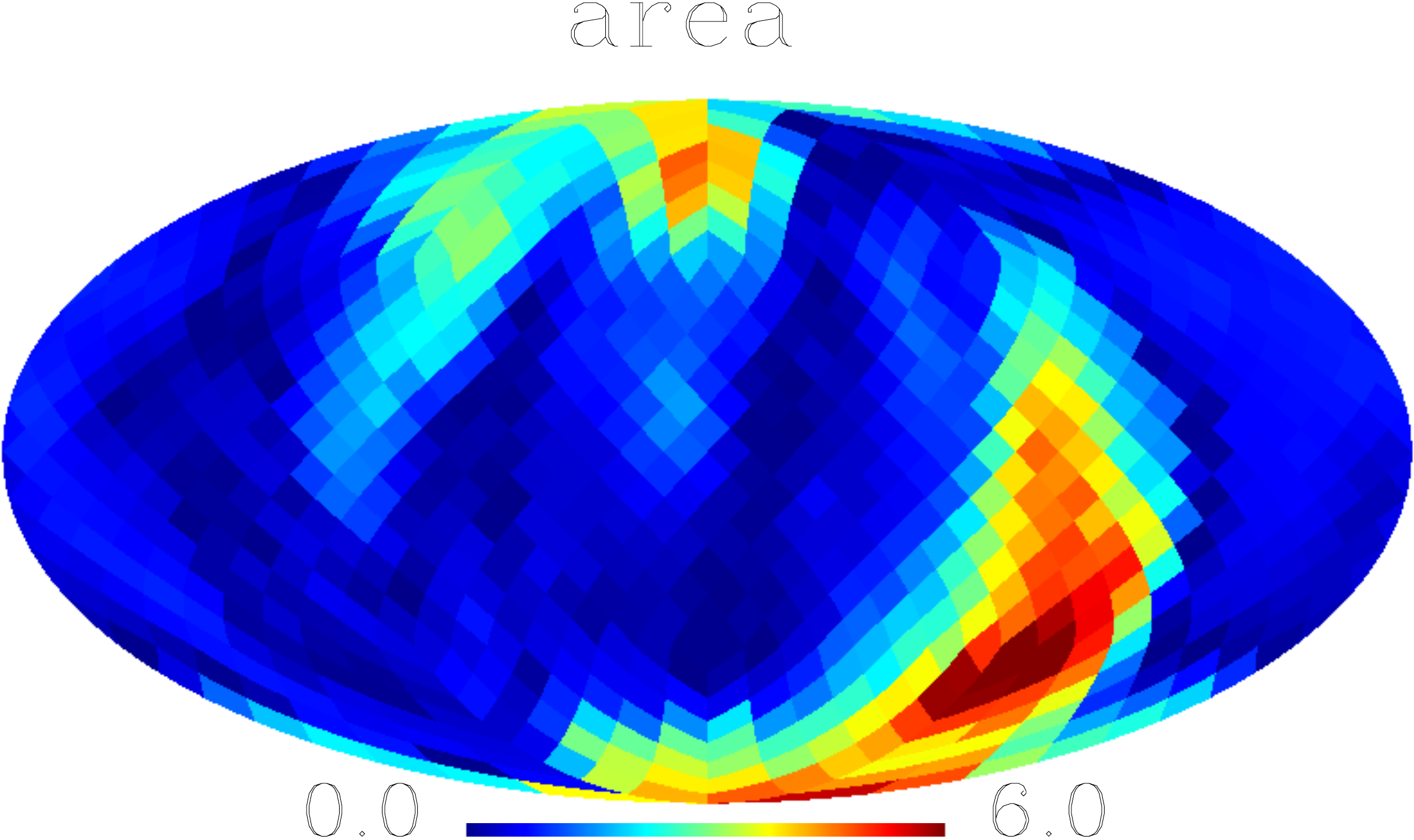}
\includegraphics[width=2.6cm, keepaspectratio=true,angle={90}]{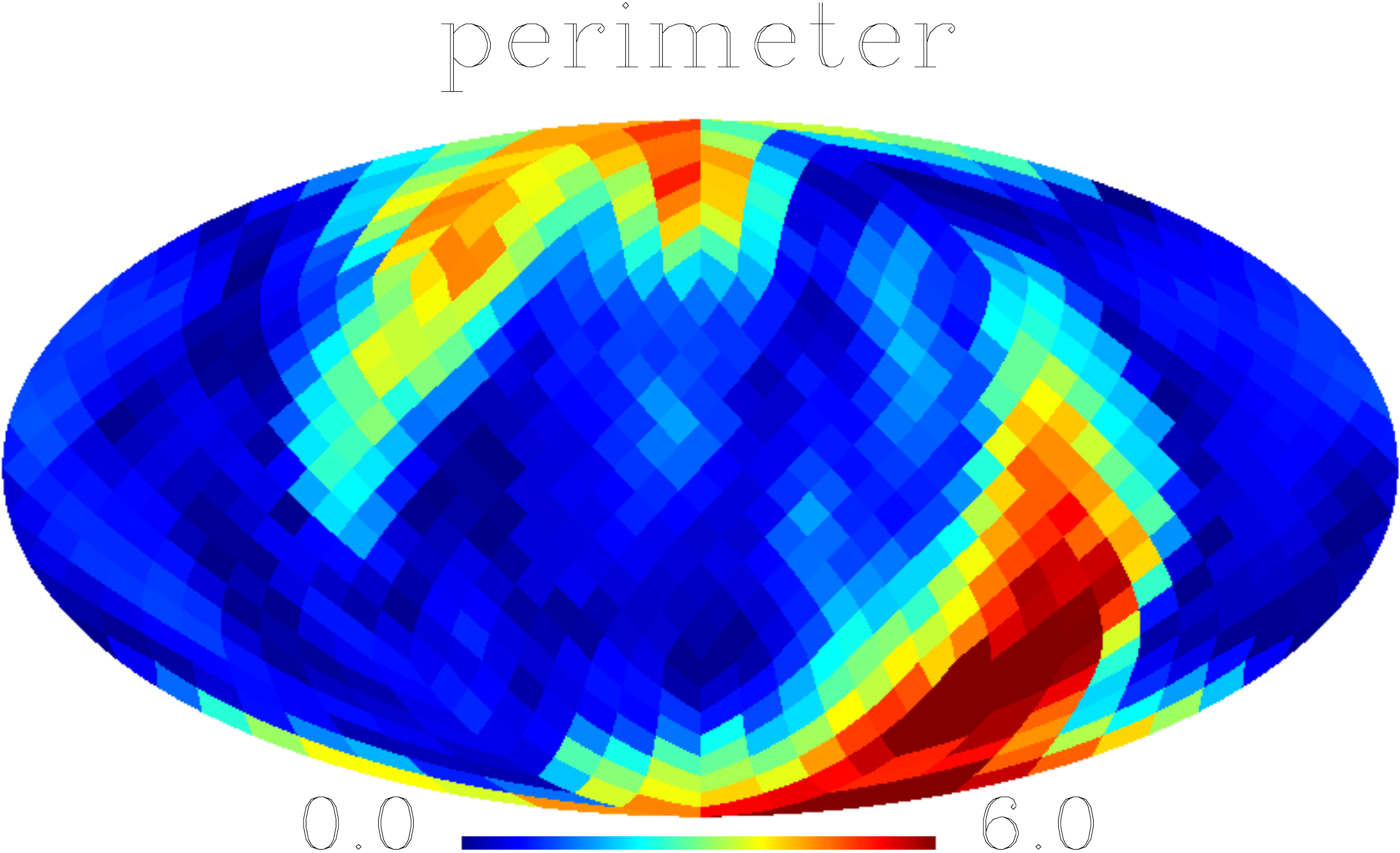}
\includegraphics[width=2.6cm, keepaspectratio=true,angle={90}]{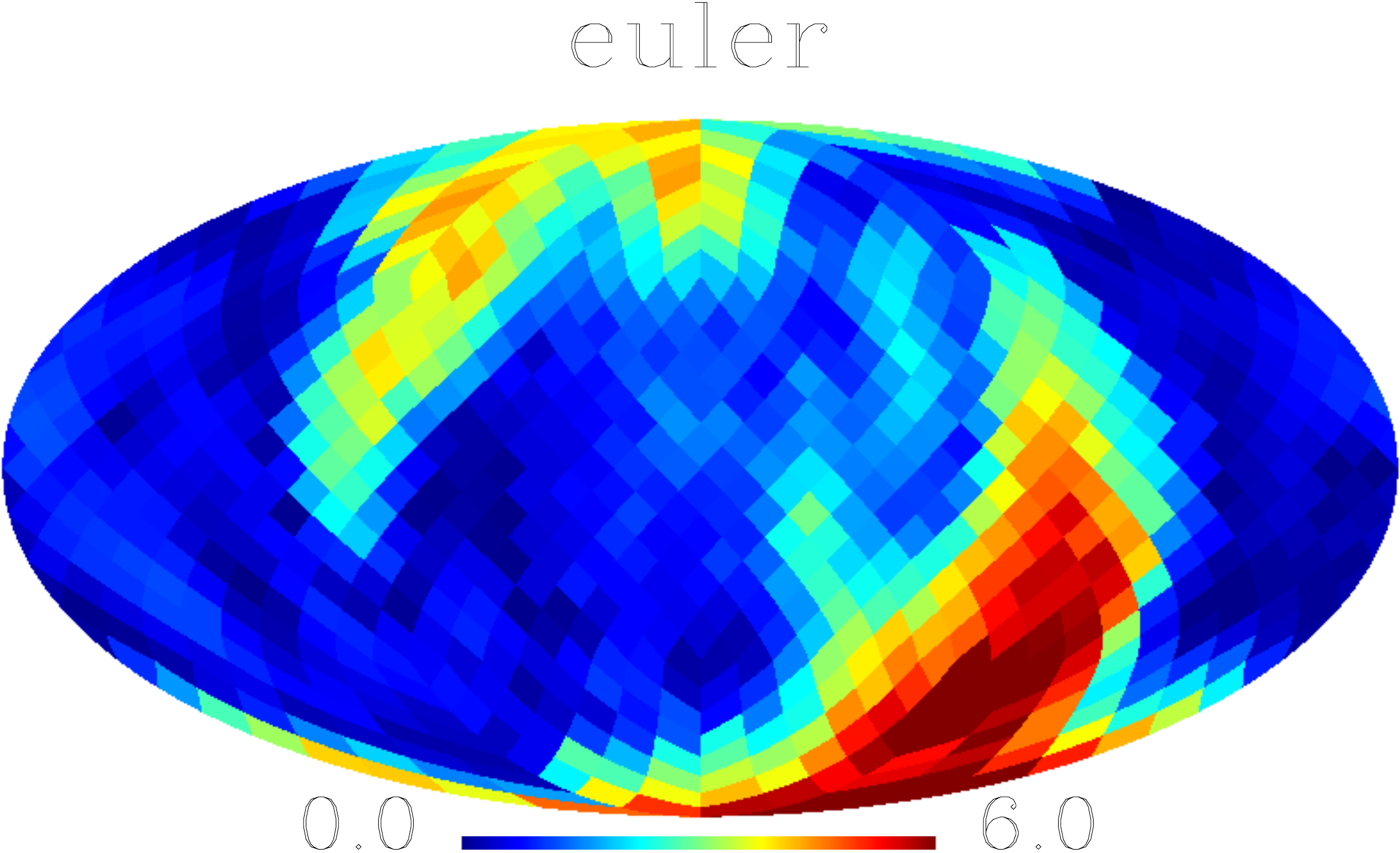}
\includegraphics[width=2.6cm, keepaspectratio=true,angle={90}]{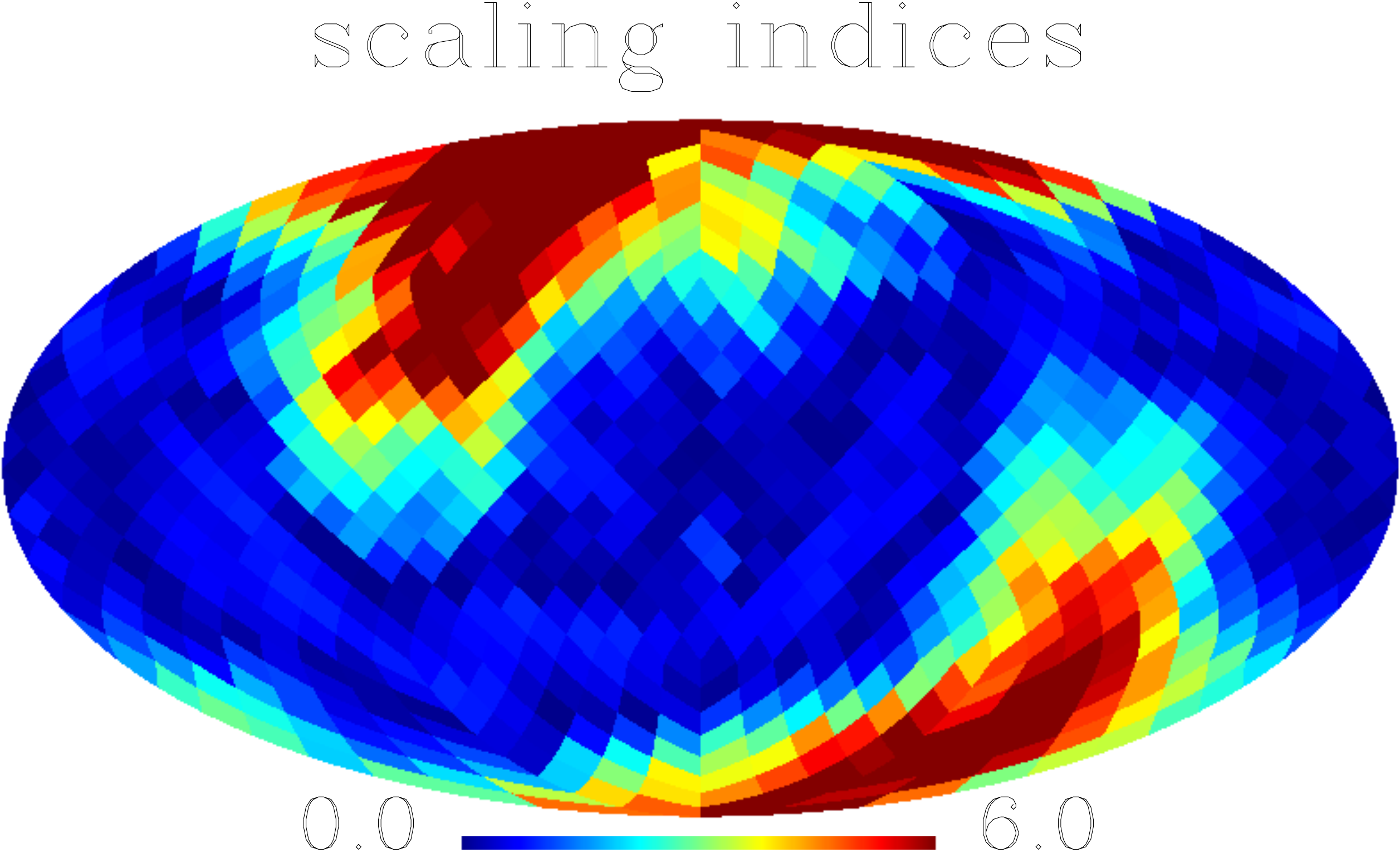}

\includegraphics[width=2.6cm, keepaspectratio=true,angle={90}]{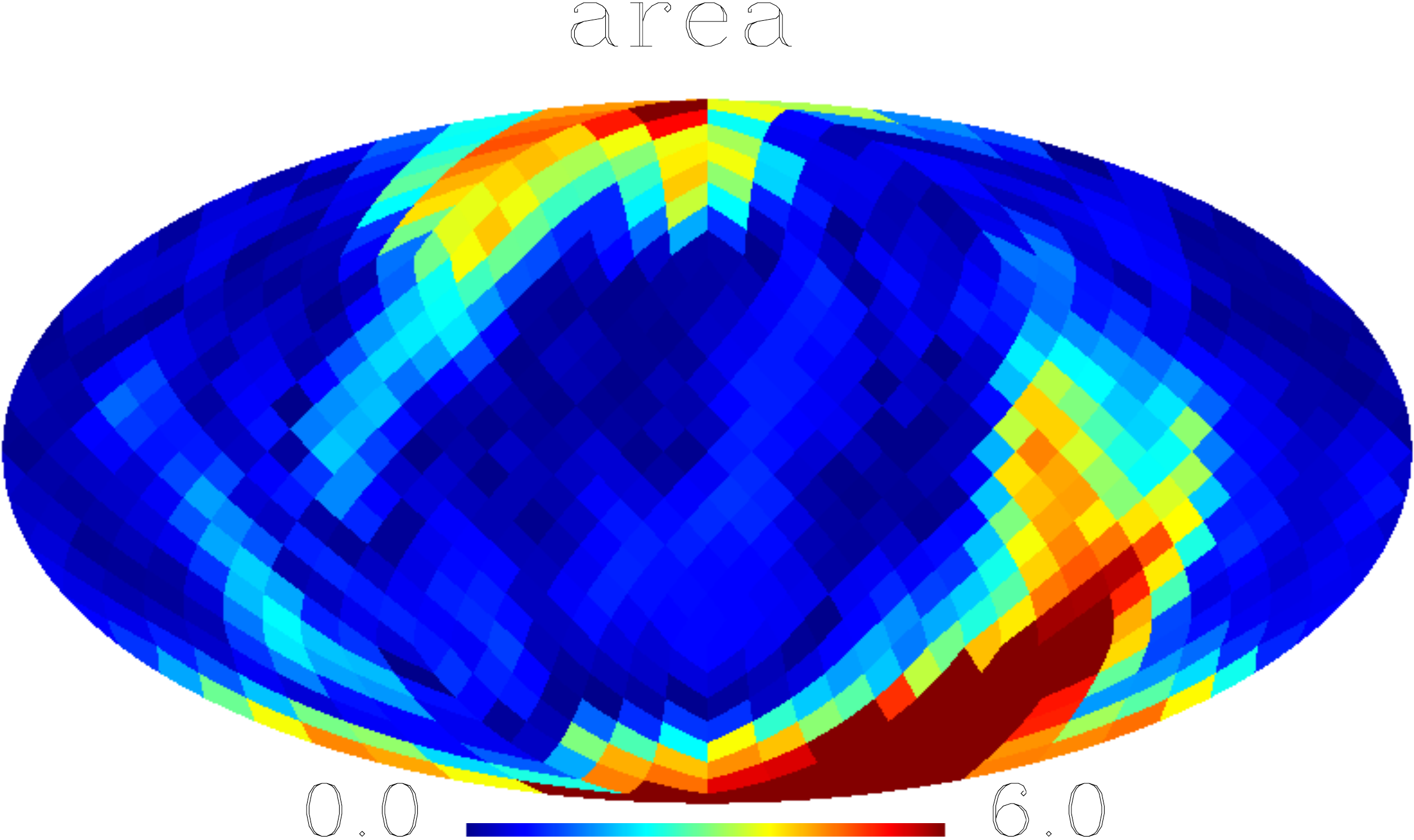}
\includegraphics[width=2.6cm, keepaspectratio=true,angle={90}]{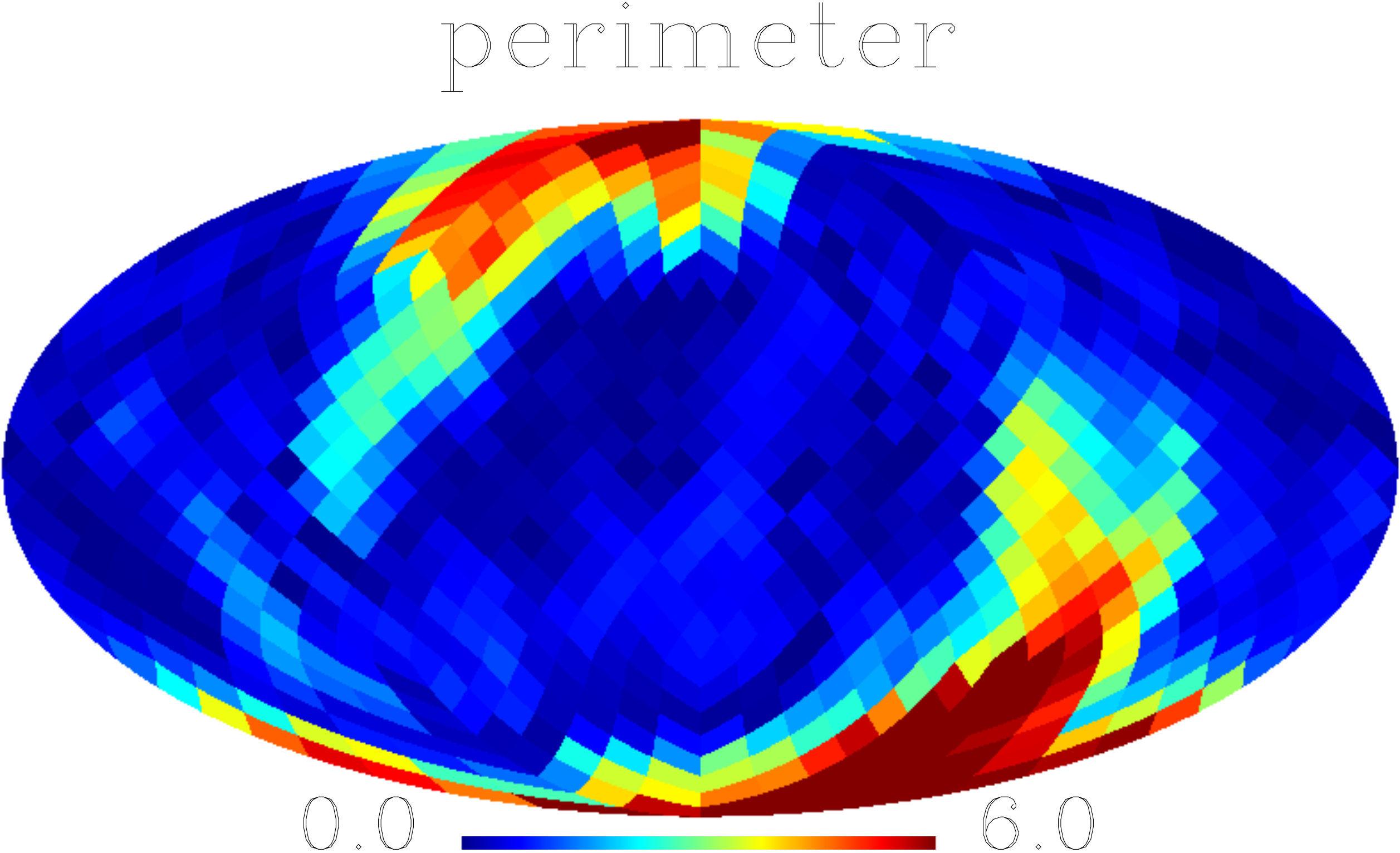}
\includegraphics[width=2.6cm, keepaspectratio=true,angle={90}]{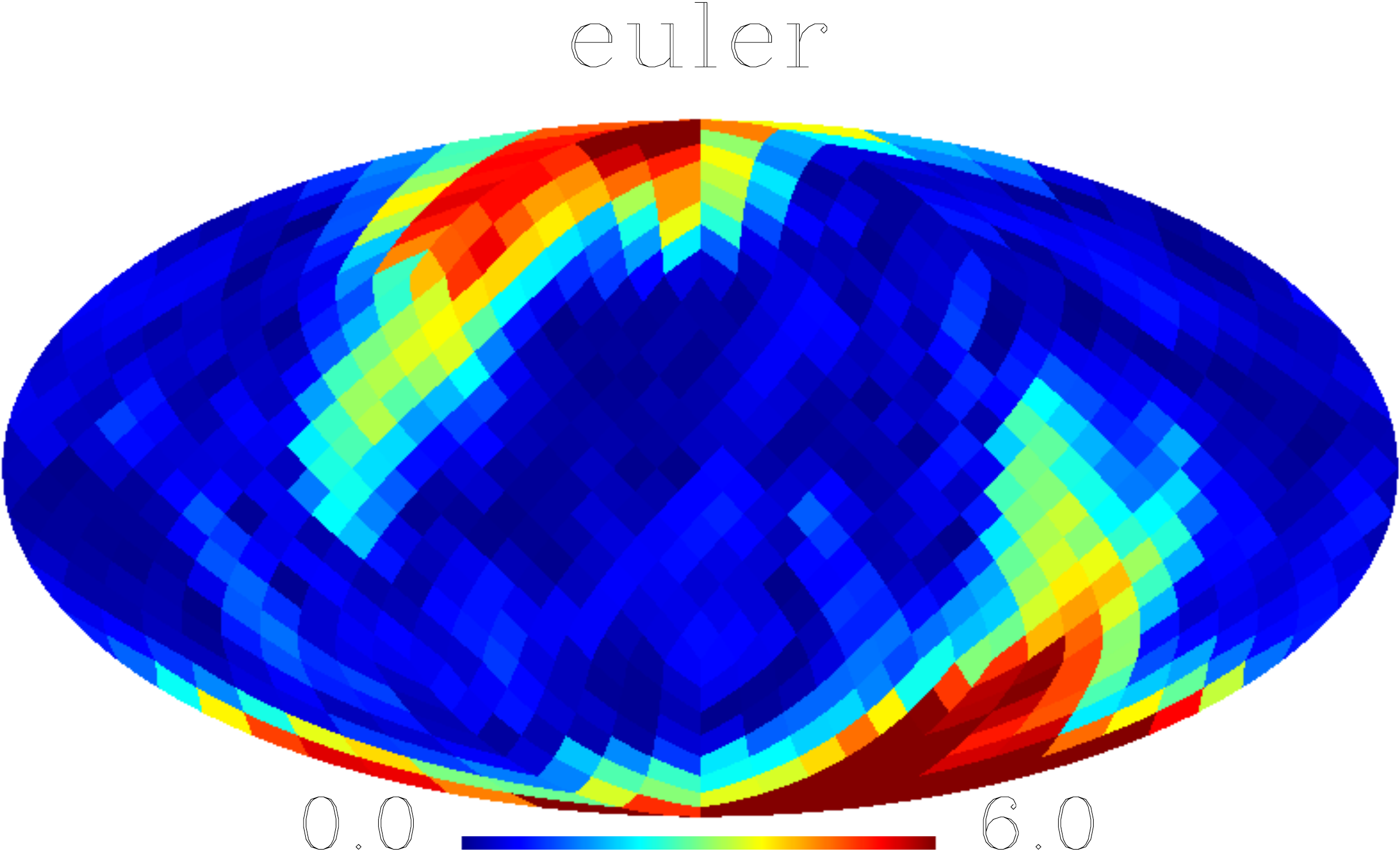}
\includegraphics[width=2.6cm, keepaspectratio=true,angle={90}]{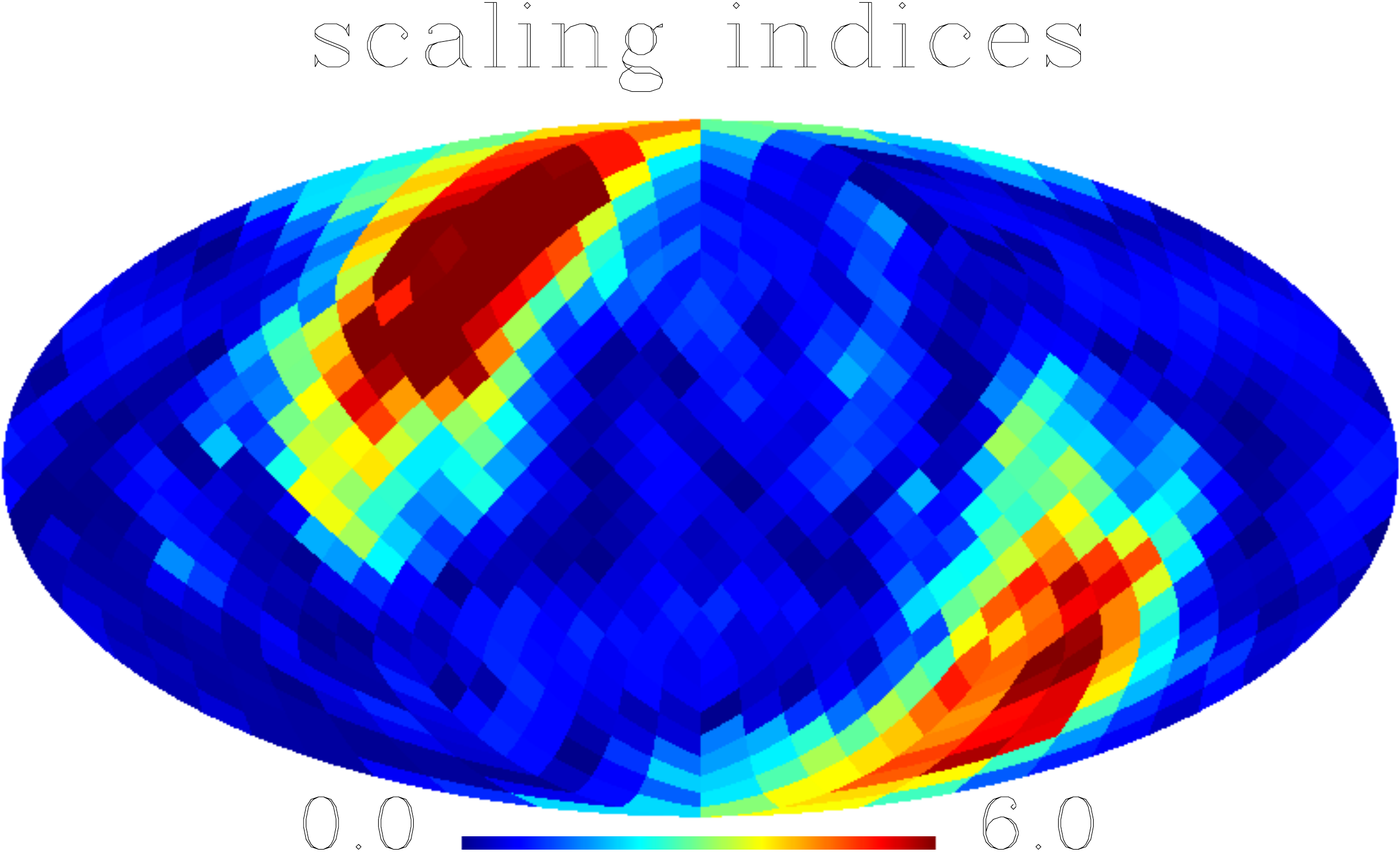}

\includegraphics[width=2.6cm, keepaspectratio=true,angle={90}]{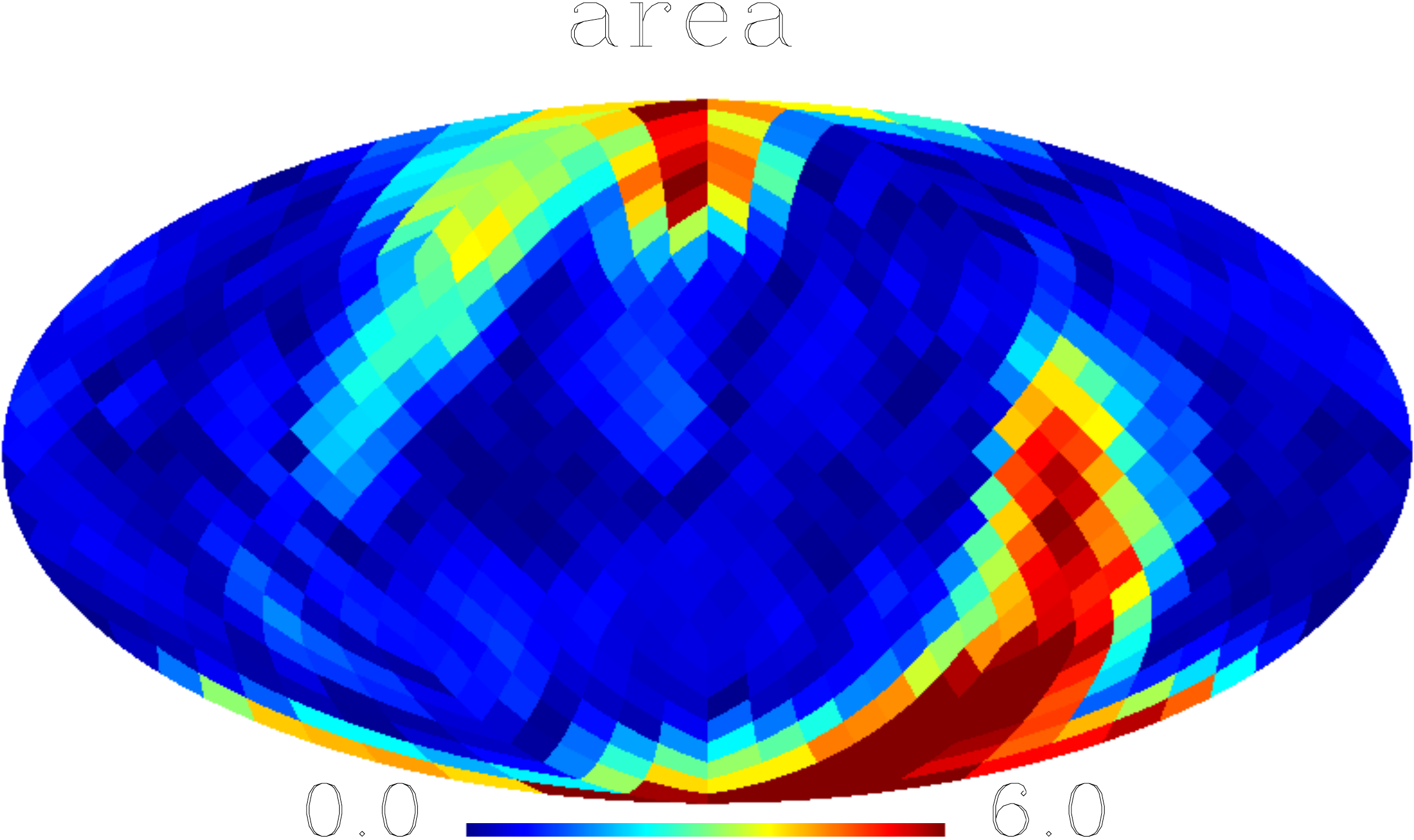}
\includegraphics[width=2.6cm, keepaspectratio=true,angle={90}]{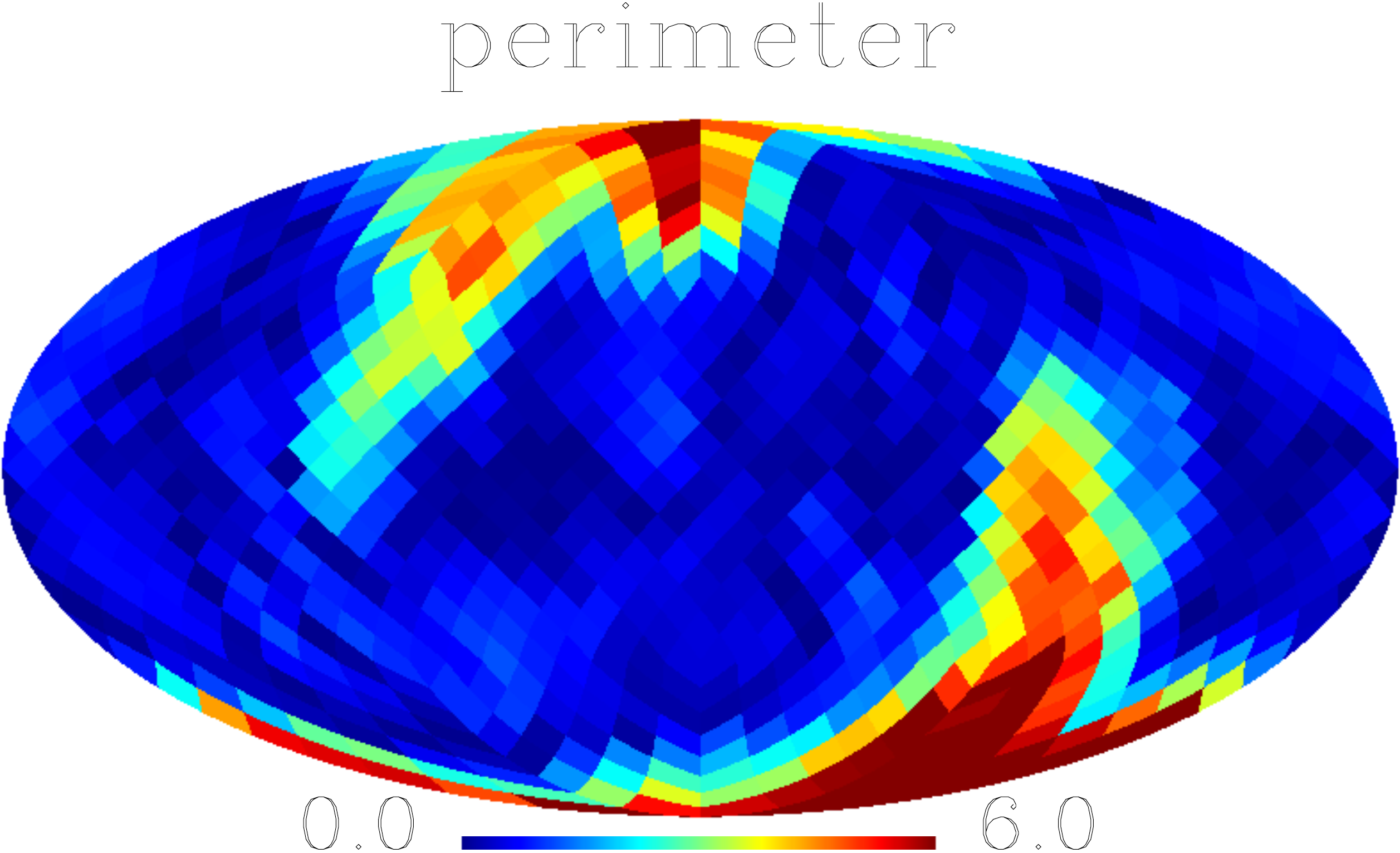}
\includegraphics[width=2.6cm, keepaspectratio=true,angle={90}]{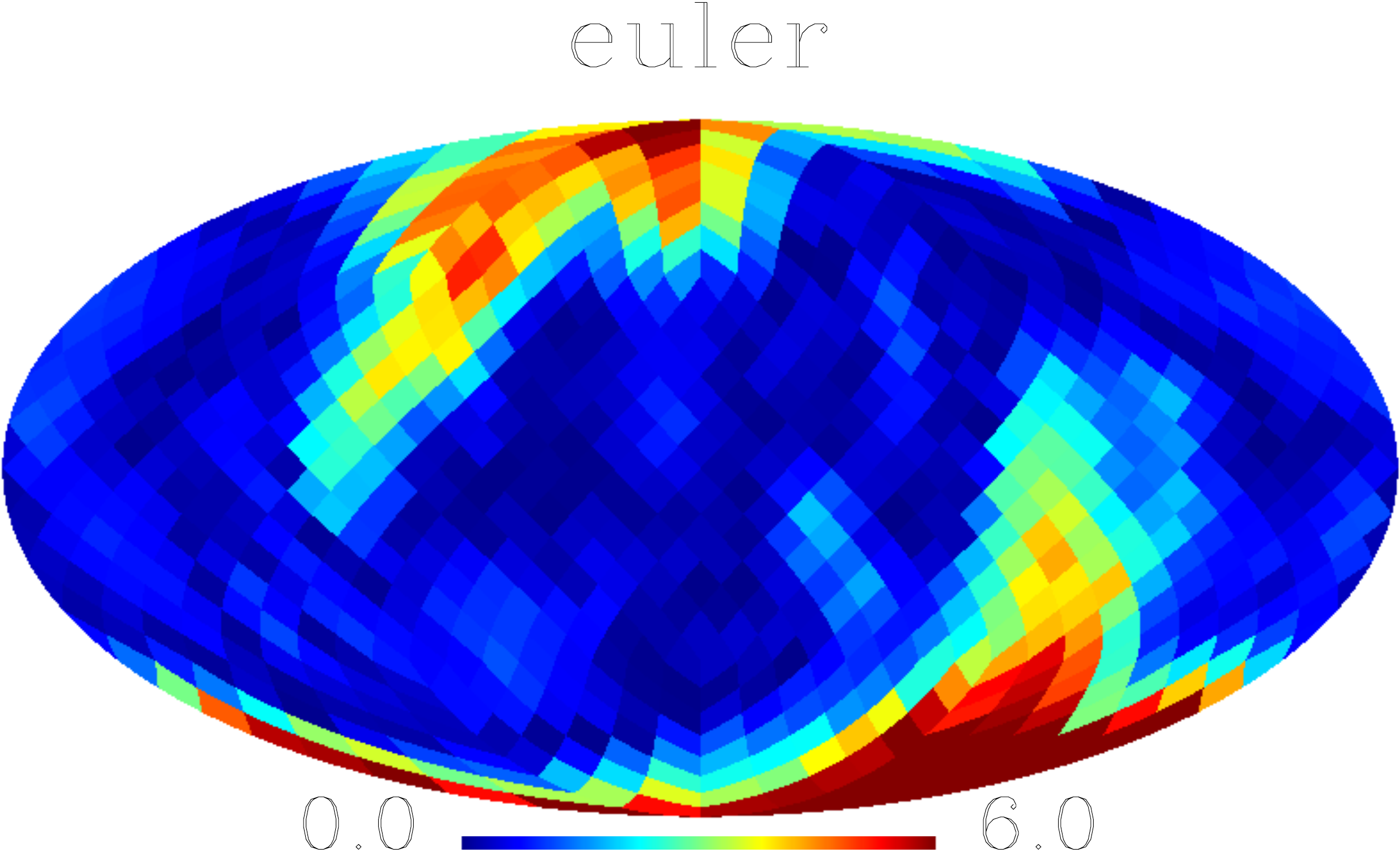}
\includegraphics[width=2.6cm, keepaspectratio=true,angle={90}]{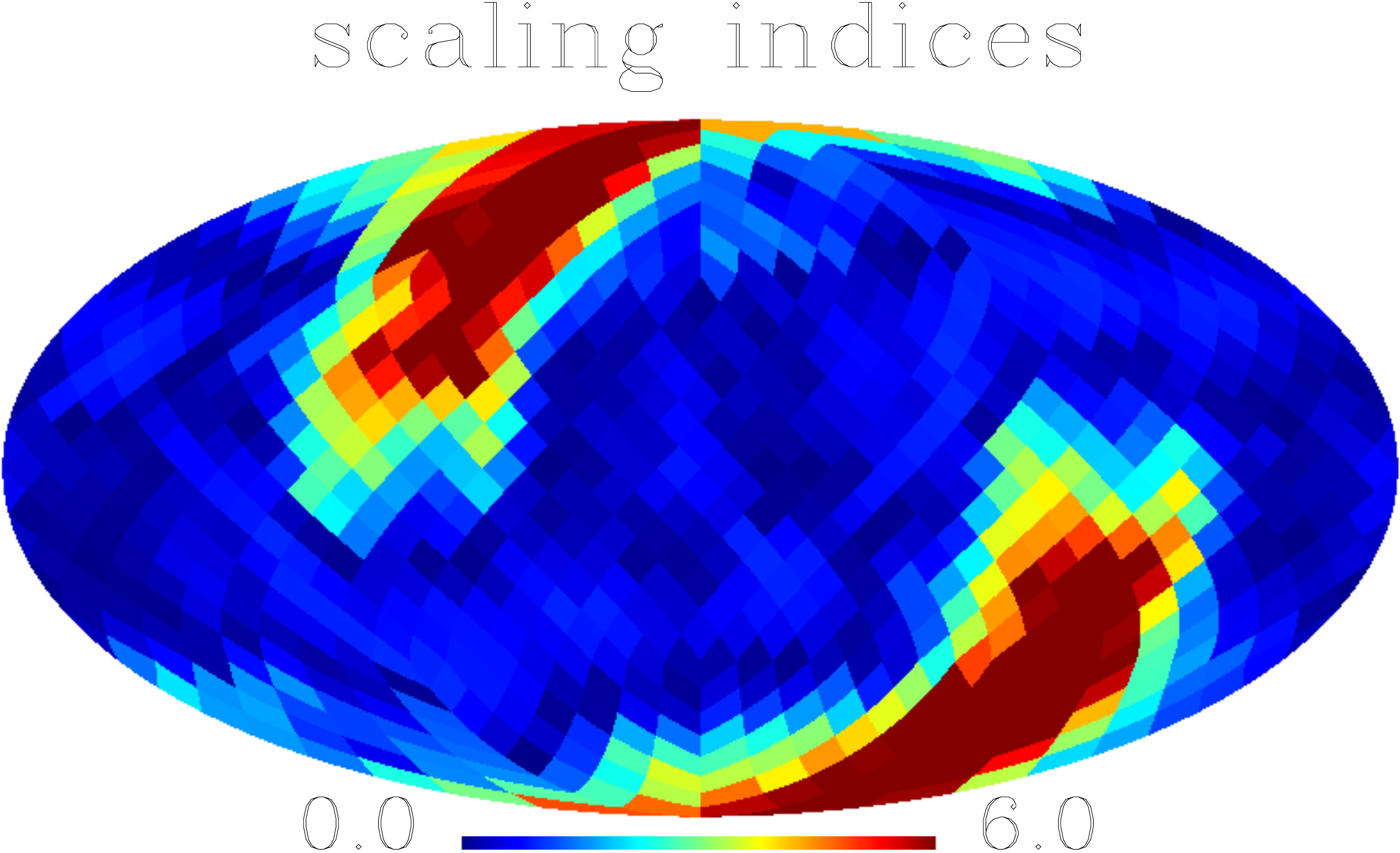}

\caption{Deviations $S(\chi^2)$ of Minkowski Functionals $M_0$, $M_1$ and $M_2$ of the rotated hemispheres for the ILC7 (upper row) and NILC7 map (middle row) in Galactic ccordiantes. In the lower row we show the results of the phase replacement method for NILC7. The plots to the very right show the corresponding results $S(\chi^2)$ for the respective maps gained by the scaling index method. The $\ell$-range for the method of the surrogates is $\Delta \ell = [2,20]$.}
\label{fig:signiMapILCand NILC020}
\end{figure*}

In order to quantify the degree of agreement between the surrogates of different orders with respect to higher order correlations found with the three Minkowski functionals $M_0, M_1$ and $M_2$ we calculate the mean of each Minkowski functional $M_{\star,surro2}$ for $N=500$ realisations of the second order surrogate per hemisphere $h$ and threshold bin $\nu$
\begin{equation*}
A:= \: \langle M_{\star,surro2}(\nu,h)\rangle  = \frac{1}{N} \sum_{m=1}^{N} M_{\star,surro2}(m,\nu,h)
\end{equation*}
 and the standard deviation 
 \begin{equation*}
 \sigma_{M_{\star,surro2}}(\nu,h) = \left( \frac{1}{N-1} \sum_{m=1}^{N} \left( M_{\star,surro2}(m,\nu,h) -  A\:  \right)^2 \right)^{1/2}
 \end{equation*}
 for $m = 1,..., N$. We combine mean and standard deviation in a diagonal $\chi^2$ statistic per hemisphere $h$ for the surrogates of first order
 \begin{equation*}
 \chi^2_{M_{\star, surro1}}(h) =  \sum_{j=0}^{\nu} \left[ \frac{M_{\star,surro1}(\nu,h) - A\: }{\sigma_{M_{\star,surro2}}(\nu,h)}\right]^2
 \end{equation*}
 and for the second order surrogate maps
 \begin{equation*}
 \chi^2_{M_{\star, surro2}}(h,m) =  \sum_{j=0}^{\nu} \left[\frac{M_{\star,surro2}(m,\nu,h)- A\:}{\sigma_{M_{\star,surro2}}(\nu,h)}\right]^2.
 \end{equation*}
Finally, the degree of agreement between the two types of surrogates is quantified by the $\sigma$-normalised deviation $S$
 \begin{equation}
 S(\chi^2_{M_{\star}}(h)) = \frac{\chi^2_{M_{\star, surro1}}(h) - \langle \chi^2_{M_{\star, surro2}}(h) \rangle}{\sigma_{\chi^2_{M_{\star, surro2}}}(h)}
 \label{eqn:S(chi2)}
 \end{equation}
 for each of the 768 hemispheres with $\langle \chi^2_{M_{\star, surro2}}(h) \rangle$ and $\sigma_{\chi^2_{M_{\star, surro2}}}(h)$ denoting the mean and the standard deviation of $\chi^2_{M_{\star, surro2}}(h)$ for the $N=500$ second order surrogates. Thus, we obtain the $\sigma$ normalised hemispherical deviations $S(\chi^2_{M_0}(h))$, $S(\chi^2_{M_1}(h))$ and $S(\chi^2_{M_2}(h))$ for the Area, Perimeter and Euler characteristics. 

As for the Minkowski functionals, we calculate the mean $\langle \alpha(r) \rangle$ and standard deviation $\sigma_{\alpha(r)}$ of the scaling indices $\alpha(\textbf{p}_i,r)$ for the set of 768 hemispherical maps. The scaling indices are calculated with a scale parameter $r=0.25$, which corresponds to $r_{10}$ as defined in our earlier works, e.g in \citet{2009MNRAS.399.1921R}. The differences of the two classes of surrogates are again quantified by the $\sigma$-normalised deviation $S$:  
\begin{equation}
S(Y) = \frac{Y_{surro1} - \langle Y_{surro2} \rangle}{\sigma_{Y_{surro2}}}
\label{eqn:S(Y)}
\end{equation}
where $Y$ represents a diagonal $\chi^2$ statistic
\begin{equation}
\chi^2_{\langle \alpha(r)\rangle, \sigma_{\alpha(r)}}= \sum_{j=1}^2 \left[ \frac{B_j - \langle B_j \rangle}{\sigma_{B_j}} \right]^2,
\end{equation}
as a combination of the mean and the standard deviation where $B_1(r) = \langle \alpha(r)\rangle$, $B_2(r) = \sigma_{\alpha(r)}$. See \citet{2007MNRAS.380..466R} for a detailed description regarding the scaling indices statistics.

The results we obtain with the hemispherical study of the sky are visualised in so-called $S$-maps. Each pixel centre of a full sky map with a HEALPix resolution of $N_{\mathrm{side}}=8$ marks one of the 768 hemispheres' poles. After calculating the deviation $S$ between one first order surrogate and 500 realisations of second order surrogates for each individual hemisphere, we plot the respective value in a sky map at that pixel position where the $z$-axis of the rotated hemisphere pierces the sky. This is done for both, the Minkowski functionals and the scaling indices. For one single $S$-map with 768 values stemming from the comparison of 500 surrogates of second order with one first order surrogate we need to calculate 384,768 hemispherical maps. 

\section{Results and Discussion}
\subsection{ILC and NILC Maps}
Figure \ref{fig:signiMapILCand NILC020} shows the $S$-values per hemisphere for a deviation from Gaussianity in ILC7 and NILC7 data on largest scales with $\Delta \ell = [2,20]$, found with the method of surrogates by using a shuffling approach and a phase replacement procedure, respectively, and analysed by the Minkowski functionals as well as the scaling index method. The $\chi^2$-statistics of the two image analysis techniques yield consistent results on largest scales and show significant signatures for ecliptic hemispherical asymmetries and non-Gaussianity in CMB sky maps. The signal for the Minkowski functionals is maximal in the southern ecliptic sky whereas for the SIM we find the maximum in the northern ecliptic sky. In Table \ref{tab:S_ILC7} and \ref{tab:S_NILC7} we summarised the deviations $S$ and the empirical probabilities $p$ of the three Minkowski functionals and the SIM. The results are shown for the full sky and a pair of hemispheres which consists of the hemisphere with maximum $S(\chi^2)$ and its opposing hemisphere on the other side of the sky.

The two CMB maps, ILC7 and NILC7, are different enough in their implementation that the presence of residual foregrounds and noise, if important in the present analysis, is not expected to result in similar NG detections. The needlet ILC is, in principle, less contaminated by foregrounds and noise, as shown on 5-year data by \citet{2009A&A...493..835D}. However, the pattern in the $S$-maps of these two maps are almost identical to each other. The $\sigma$-normalised deviations $S$ are higher for the NILC map and range up to 9.97 for the perimeter with an empirical probability above $99.8\%$.  The stronger signatures for NGs in the NILC map can further be attributed to the fact that this map did not undergo the one-degree smoothing of the ILC7 map. 
 
Expectedly, the replacement of the original phase distribution by a distribution of random values in NILC7 during the surrogates generating process yields the same pattern in the $S$-maps as for the shuffling procedure. However, the shuffling approach avoids any dependence on additional data and is therefore preferable for generating surrogates. 

We show that the scale-dependence of the scaling index method is not a limitation to that measure since the scale-independent $\chi^2$ statistic of the Minkowski functionals still yields the same results as the scale-dependent $\chi^2$ statistic of the scaling indices, that depends on scale parameter $r$. 

All three Minkowski functionals area, perimeter and Euler as well as the SIM detect phase correlations and therefore non-Gaussianities in the data with almost identical spatial signatures. However, the single values in the $S$-maps do not demonstrate local NGs but must be interpreted as an overall signal per hemisphere. Note that in this hemispherical study large overlapping sky patches have been analysed. $S$-values in the upper sky also account for signal from the lower sky and the other way around.

\begin{table}
\begin{tabular}{lcccc}
\hline \hline
 & Full Sky &  hemisphere & hemisphere \\ 
 &  & $S_{max}$ &  Opposite $S$\\ 
\hline
$\chi^2$ & $(S|\%)$ & $(S|\%)$ & $(S|\%)$ \\ \\
Area & 0.62 $|$ 86.4 & 6.72 $|$ 99.6 & 3.05 $|$ 98.8 \\
Perimeter & 0.93 $|$ 88.6 & 7.33 $|$ \textgreater 99.8 & 4.52 $|$ 99.4 \\
Euler & 1.44 $|$ 92.2 & 7.24 $|$ \textgreater 99.8 & 3.62 $|$ 99.0 \\ \\
SIM &  0.41 $|$ 57.0 & 8.9 $|$ \textgreater 99.8 & 6.1 $|$ 99.8\\
\hline \hline
\end{tabular}
\caption{Deviations $S$ calculated for the ILC7 map at a shuffling range of $\Delta \ell = [2,20]$ from Equation \ref{eqn:S(chi2)} and the empirical probabilities $p$ of the scale-independent diagonal $\chi^2$-statistics for full sky functionals and two selected single hemispheres for the three Minkowski Functionals and the scaling index method. This table corresponds to the first row of Figure \ref{fig:signiMapILCand NILC020}.}
\label{tab:S_ILC7}
\end{table}

\begin{table}
\begin{tabular}{lcccc}
\hline \hline
 & Full Sky &  hemisphere & hemisphere \\ 
 &  & $S_{max}$ &  Opposite $S$\\ 
\hline
$\chi^2$ & $(S|\%)$ & $(S|\%)$ & $(S|\%)$ \\ \\
Area & 1.03 $|$ 88.2 & 9.51 $|$ \textgreater 99.8 & 5.98 $|$ 99.8 \\
Perimeter & 0.89 $|$ 86.4  & 9.97 $|$ \textgreater 99.8 & 7.31 $|$ 99.8\\ 
Euler & 0.77 $|$ 84.4 & 9.50 $|$ \textgreater 99.8 & 7.22 $|$ \textgreater 99.8  \\
\\
SIM & 0.29 $|$ 51.4 & 7.53 $|$ \textgreater 99.8 & 6.23 $|$  \textgreater 99.8\\
\hline \hline
\end{tabular}
\caption{The same as Table \ref{tab:S_ILC7}, but for the NILC7 surrogate maps. This table corresponds to the second row of Figure \ref{fig:signiMapILCand NILC020}.}
\label{tab:S_NILC7}
\end{table}

\begin{figure*}
\includegraphics[width=5cm,keepaspectratio=true]{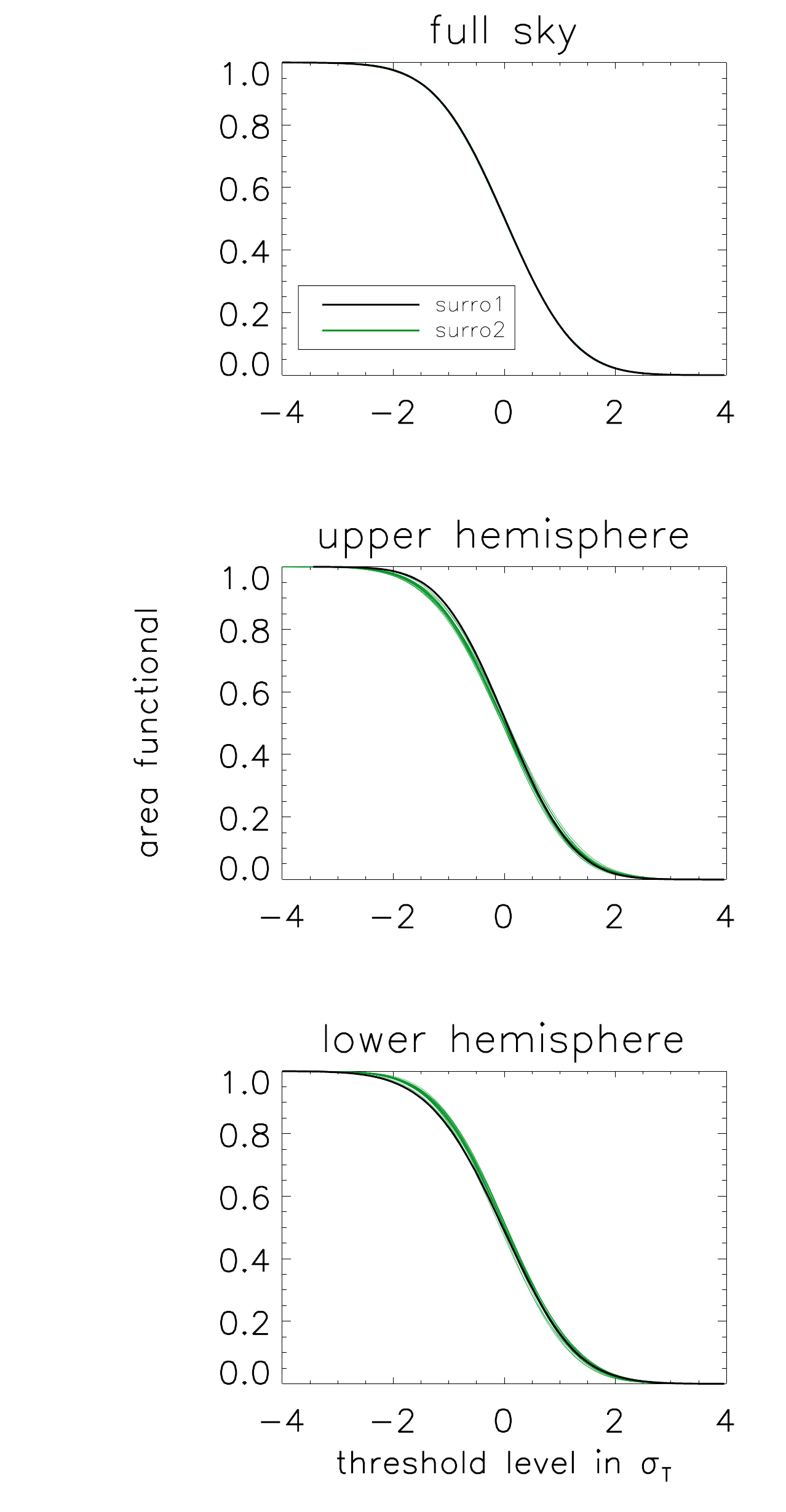}
\includegraphics[width=5cm,keepaspectratio=true]{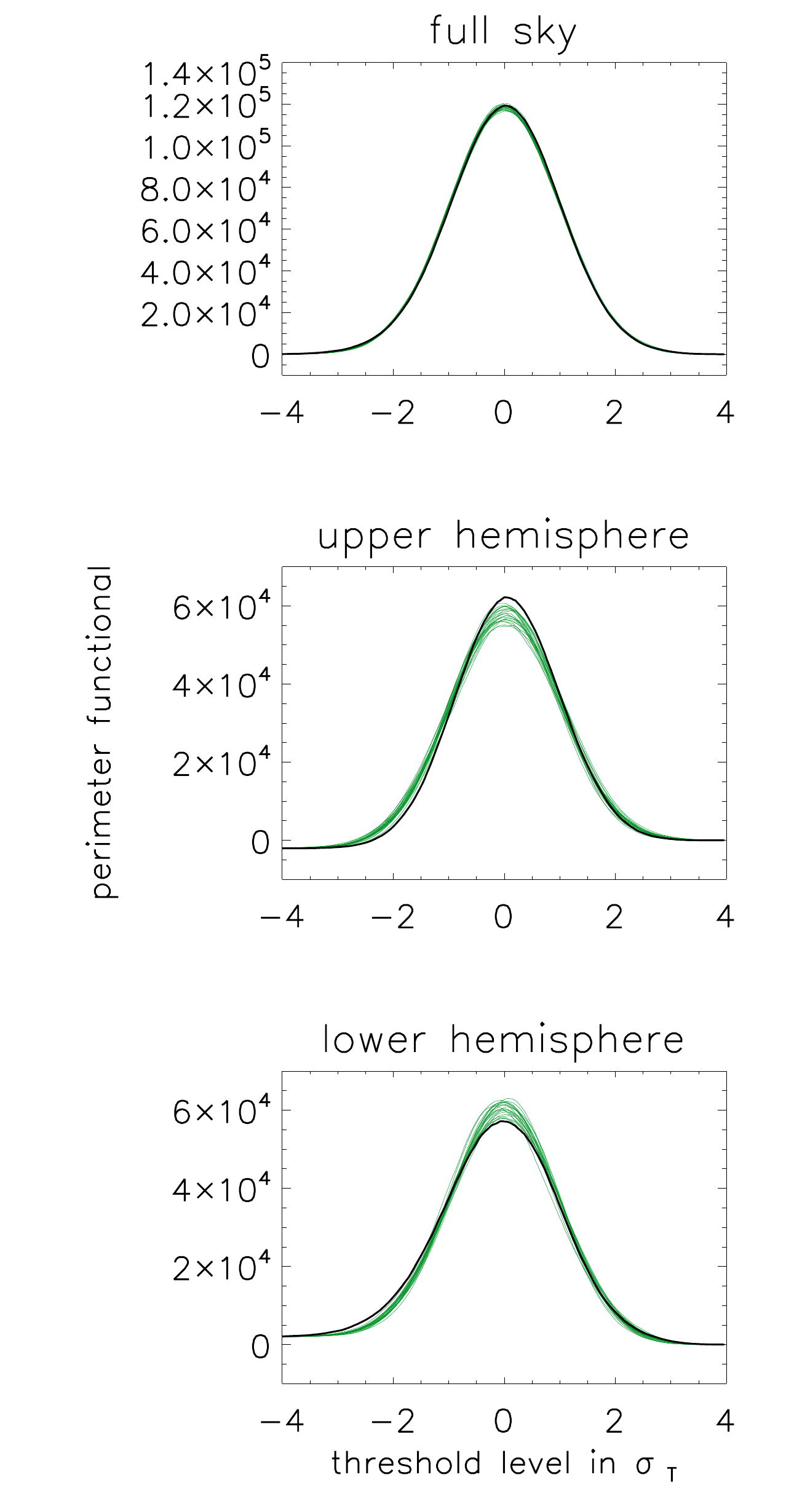}
\includegraphics[width=5cm,keepaspectratio=true]{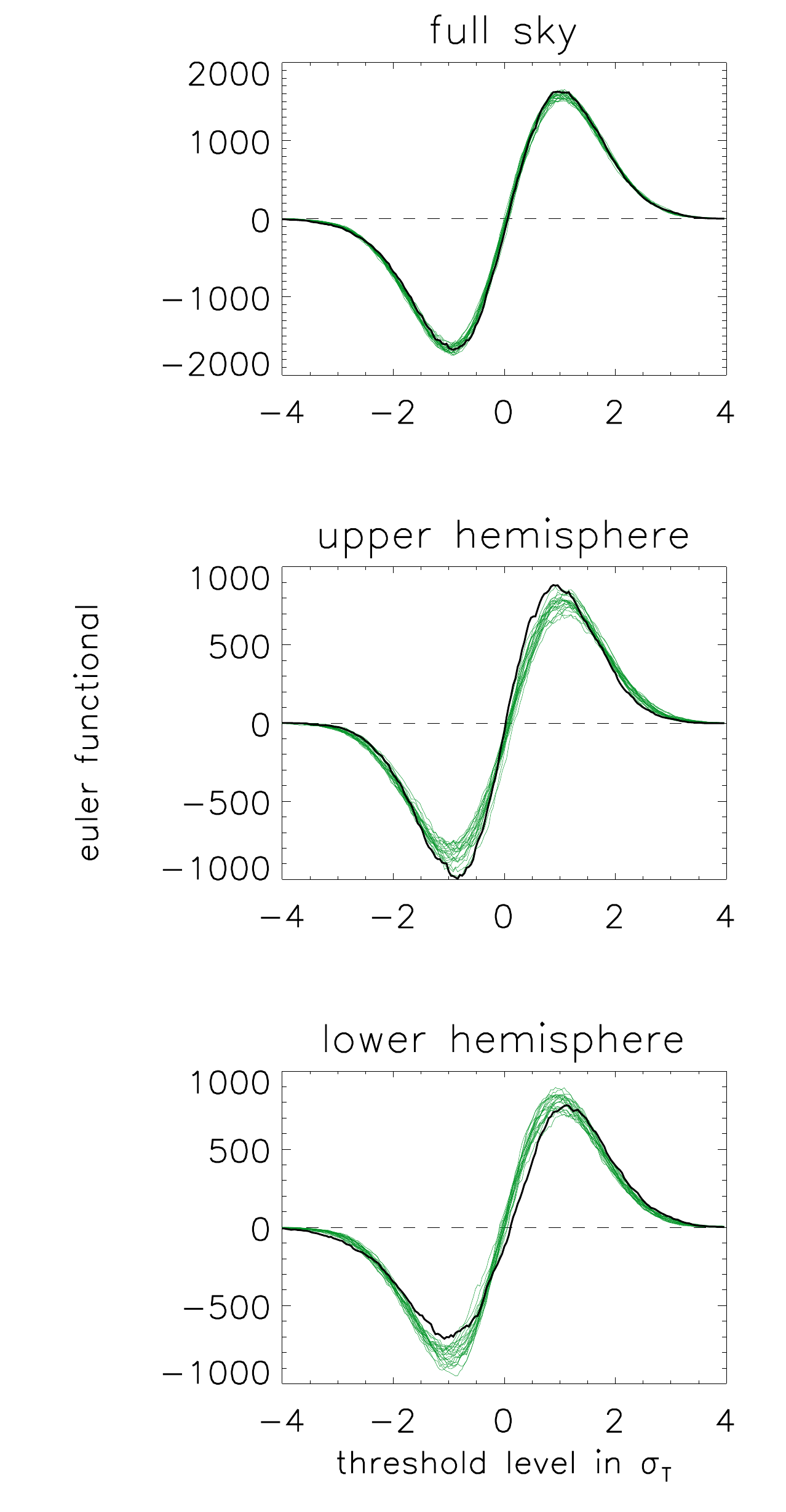}
\caption{Minkowski Functionals $M_0$ (left), $M_1$ (middle) and $M_2$ (right) of the first (black) and 20 second order surrogate maps (green), calculated for $\Delta \ell = [2,20]$. The surrogate maps were derived from the ILC7 map. The upper row shows the Minkowski functionals for the full sky, the middle row for an ecliptic northern sky hemisphere and the lower row for an ecliptic southern sky hemisphere as selected from the $S$-maps. The coordinates are given in Table \ref{tab:S_ILC7}.}
\label{fig:minkoFunctionsILC7020}
\end{figure*}

\begin{figure*}
\includegraphics[width=5cm,keepaspectratio=true]{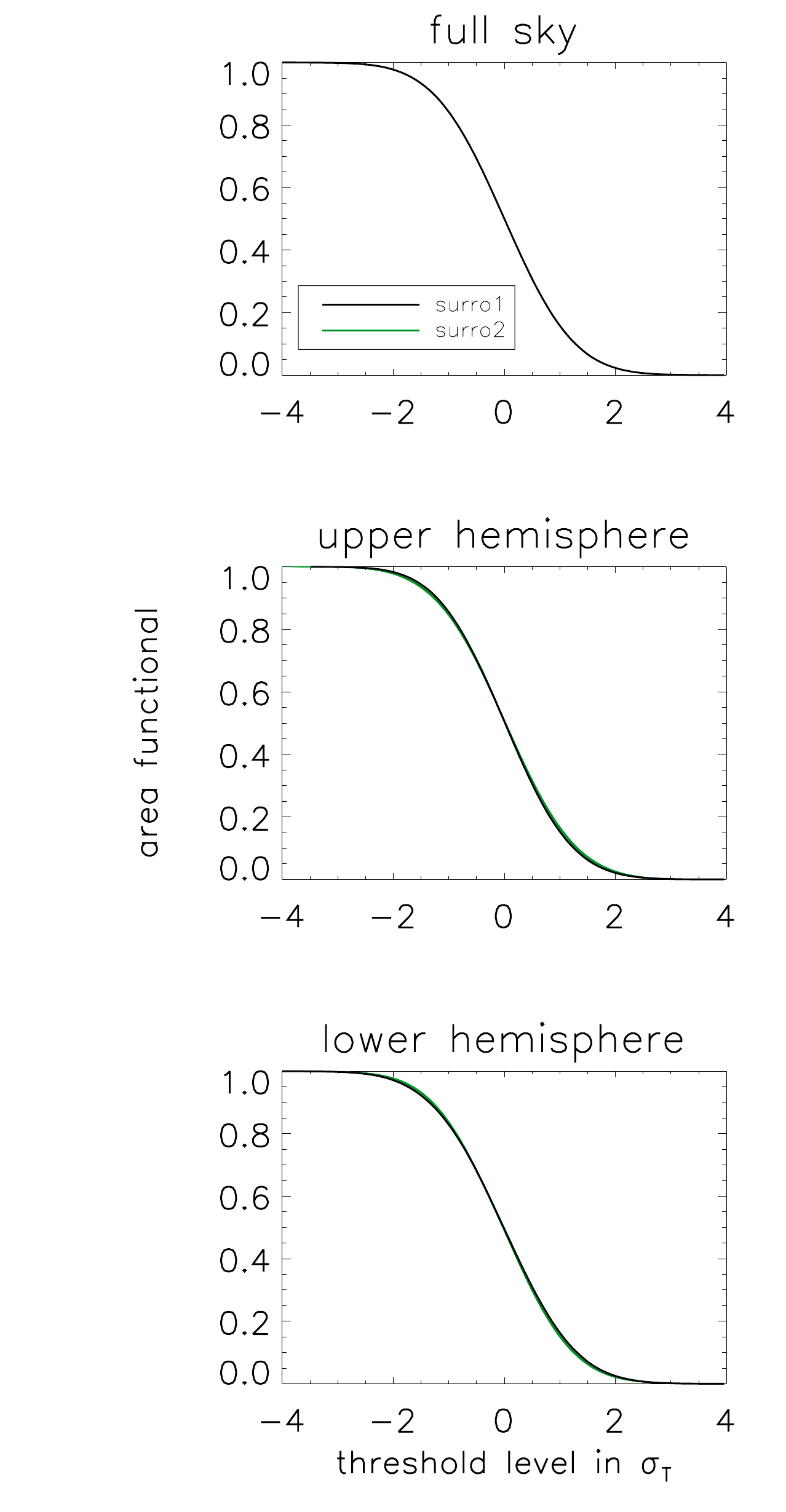}
\includegraphics[width=5cm,keepaspectratio=true]{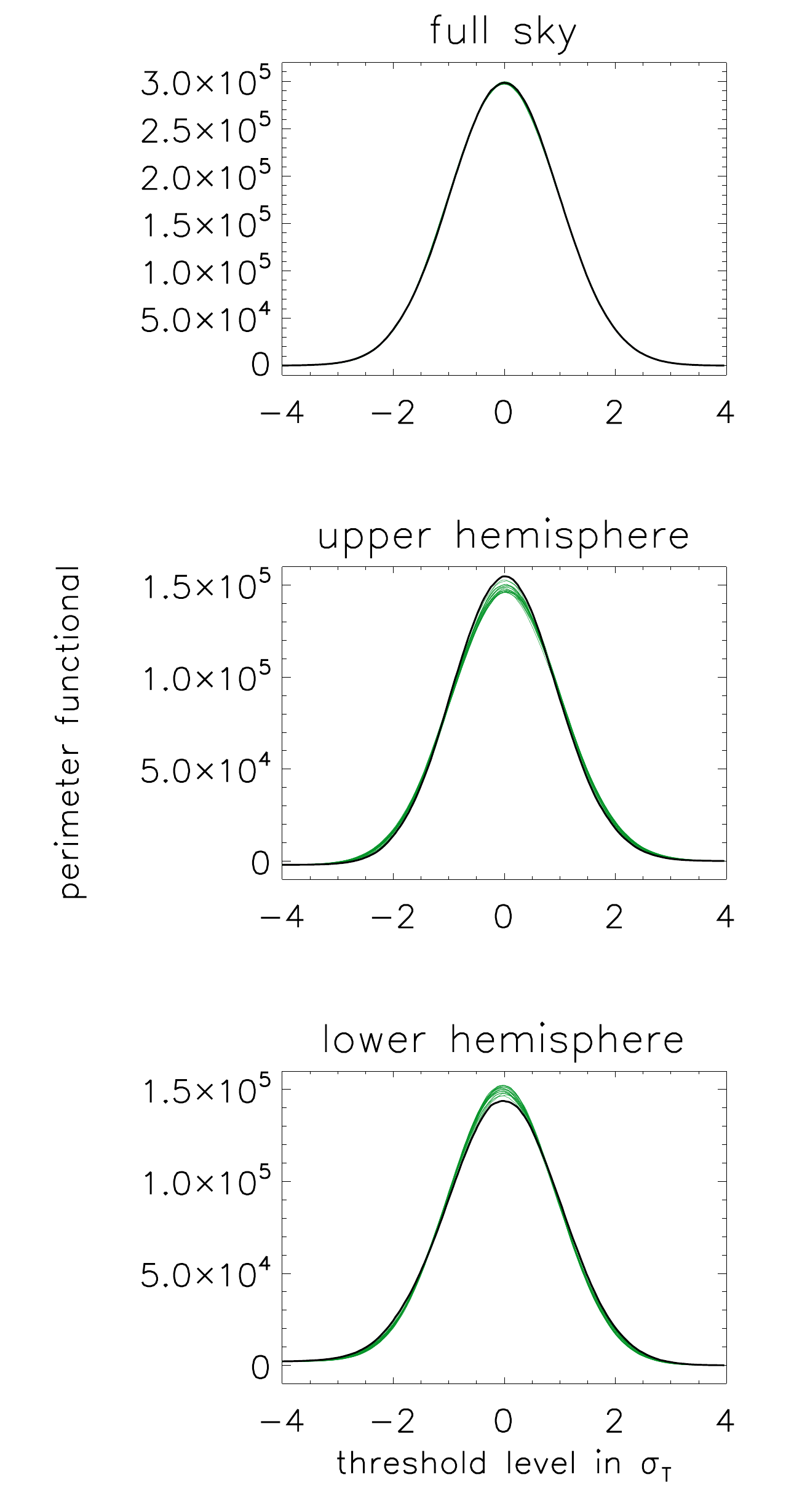}
\includegraphics[width=5cm,keepaspectratio=true]{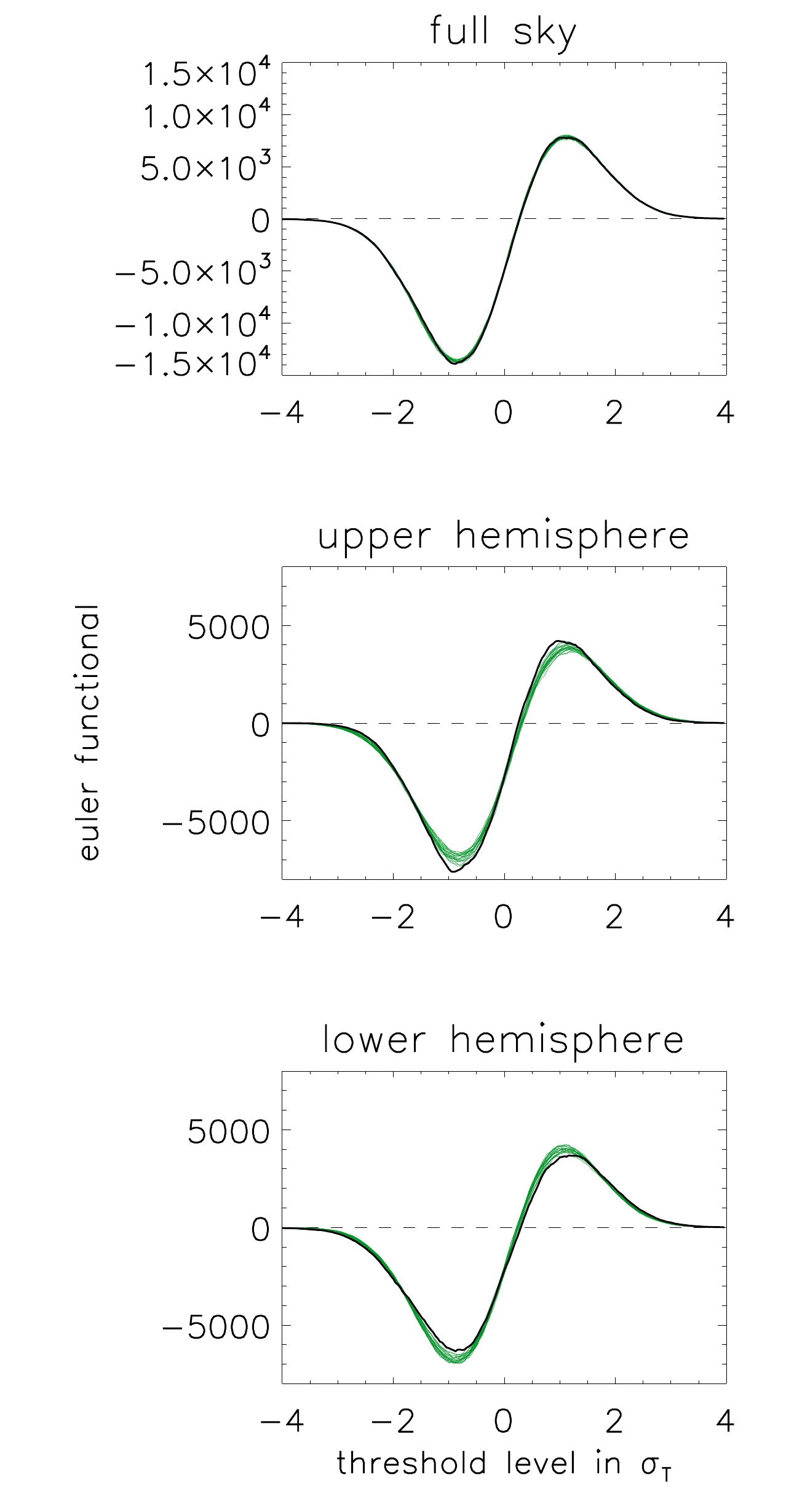}
\caption{Same as Figure \ref{fig:minkoFunctionsILC7020}, but for the NILC7 surrogate maps.}
\label{fig:minkoFunctionsNILC7020}
\end{figure*}

In Figure \ref{fig:minkoFunctionsILC7020} and \ref{fig:minkoFunctionsNILC7020}, the three Minkowski functionals area, perimeter and Euler characteristic of the ILC7 and NILC7 map, respectively, are plotted as a function of threshold values $\nu_i$. We compare one surrogate of first order with 20 realisations of  second order surrogates generated for the $\ell$-range of interest, here $\Delta \ell = [2,20]$. The full sky Minkowski functionals do not show differences between the two classes of surrogates. Single hemispheres though show clear deviations between first and second order surrogate maps. For the plots we choose again the pair of pixels as defined above: the hemisphere with the highest $S(\chi^2)$-value and its opposing hemisphere. In general, we refer to hemispheres mainly located in the northern Galactic sky as upper hemispheres, southern sky hemispheres are called lower hemispheres. 

The first order surrogate functionals of the upper or lower hemisphere differ from the respective second order surrogates for most of the threshold levels $\nu$. These deviations between the two classes of surrogates lead to the assumption that phase correlations of Fourier phases $\phi_{\ell m}$ with $2 \leq l \leq 20$, manifested as certain morphological structures in the temperature distribution in real space, are existent in the first order surrogates, in which only phases outside $\Delta \ell = [2,20]$ are randomised, but are destroyed in the second order surrogate maps.

The first order surrogates of area, perimeter and Euler functionals show contrary behavior comparing the upper and lower sky. Where the surrogates of first order of the chosen upper sky hemisphere lie above the second order surrogates it is the lower sky first order surrogates' Minkowski functionals that lie below. The algebraic sign of the deviations between the two classes of surrogates depends on the analysed sky region. The results of NILC7 in Figure \ref{fig:minkoFunctionsNILC7020} show in principle the same deviations of the two surrogates and between northern and southern sky. The absolute amplitude of the Minkowski functional $M_2$ (Euler) for the NILC7 map though is larger for negative thresholds compared to positive thresholds. In the case of the ILC7 map the two amplitudes are nearly equal. 

\begin{figure}
\centering
\includegraphics[width=8cm, keepaspectratio=true]{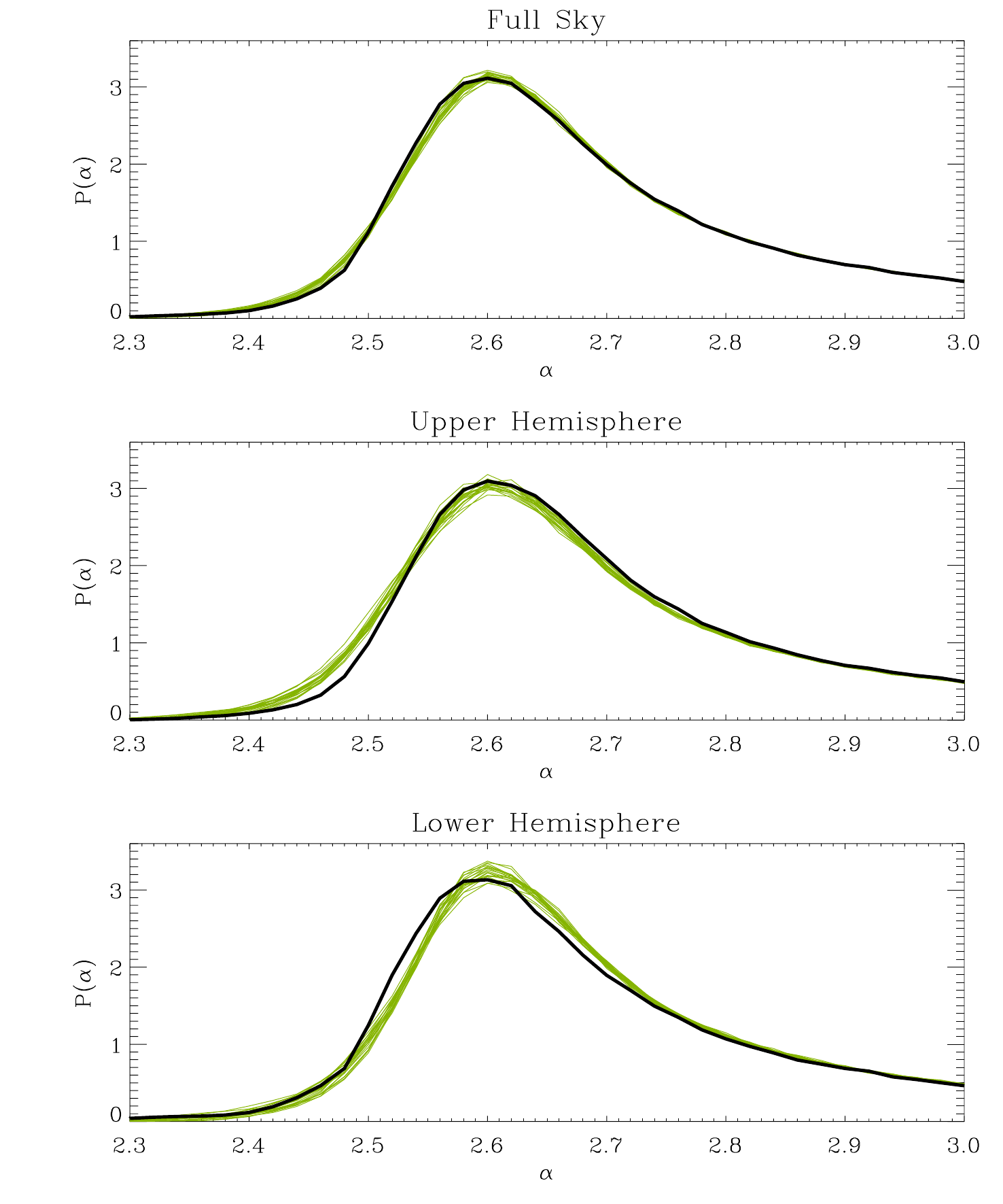}
\caption{Probability density $P({\alpha})$ of the first and second order surrogates for the scaling indices calculated for surrogate maps with the $\ell$ interval $\Delta \ell = [2,20]$. The black lines denote the first order surrogates derived from the NILC7 map. The green lines are the densities for the corresponding 20 realisations of second order surrogates.}
\label{fig:probDensityNILC7}
\end{figure}

Figure \ref{fig:probDensityNILC7} shows the probability densities $P(\alpha)$ of the scaling index method of one first and 20 second order surrogate maps for the $\ell$-interval $\Delta \ell = [2,20]$, again for the full sky analysis as well as for two opposing hemispheres for comparison. The density distributions of the second order surrogates with fully randomised phases are shifted towards lower (higher) $\alpha$ values for the upper (lower) hemisphere. This shift must be interpreted as a global trend indicating that the first order surrogate map has less (more) structure than the respective set of second order surrogates. As these deviations of different algebraic sign cancel each other for the full sky, we do no longer see significant differences in the probability densities in the full sky analysis. The SIM applied to the NILC7 map shows again that the the morphology of the temperature distribution depends on the analysed sky region and differs in algebraic sign between northern and southern Galactic sky,  as was the case with the Minkowski functionals. The deviations between first and second order surrogates demonstrate the existence of HOCs in the WMAP Fourier phases. In our earlier works we found similar results for the ILC7 and NILC5 (five-year needlet-based ILC) map (see \cite{2011MNRAS.415.2205R}).

\begin{figure}
\centering
\includegraphics[width=2.5cm, keepaspectratio=true,angle={90}]{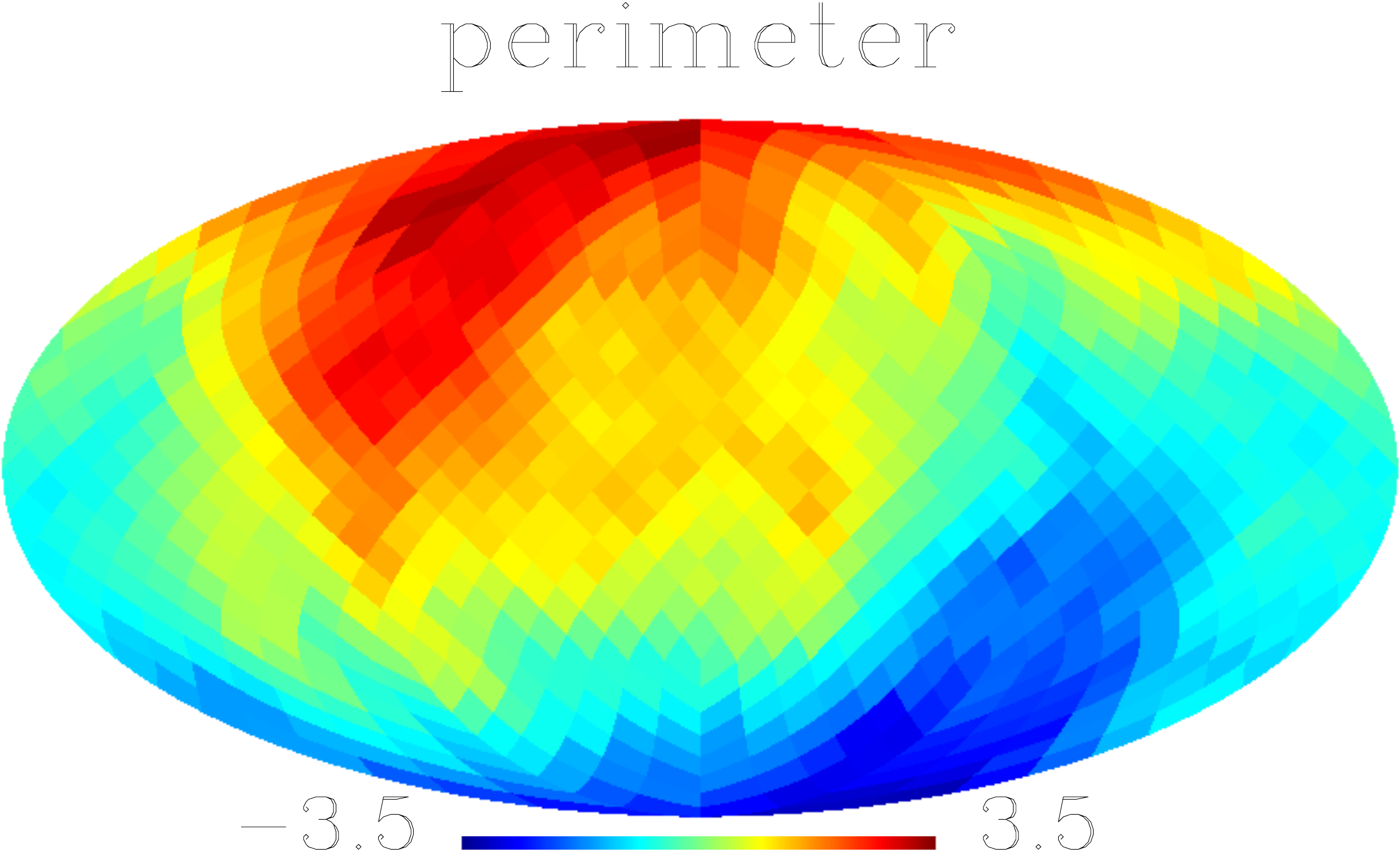}
\includegraphics[width=2.5cm, keepaspectratio=true,angle={90}]{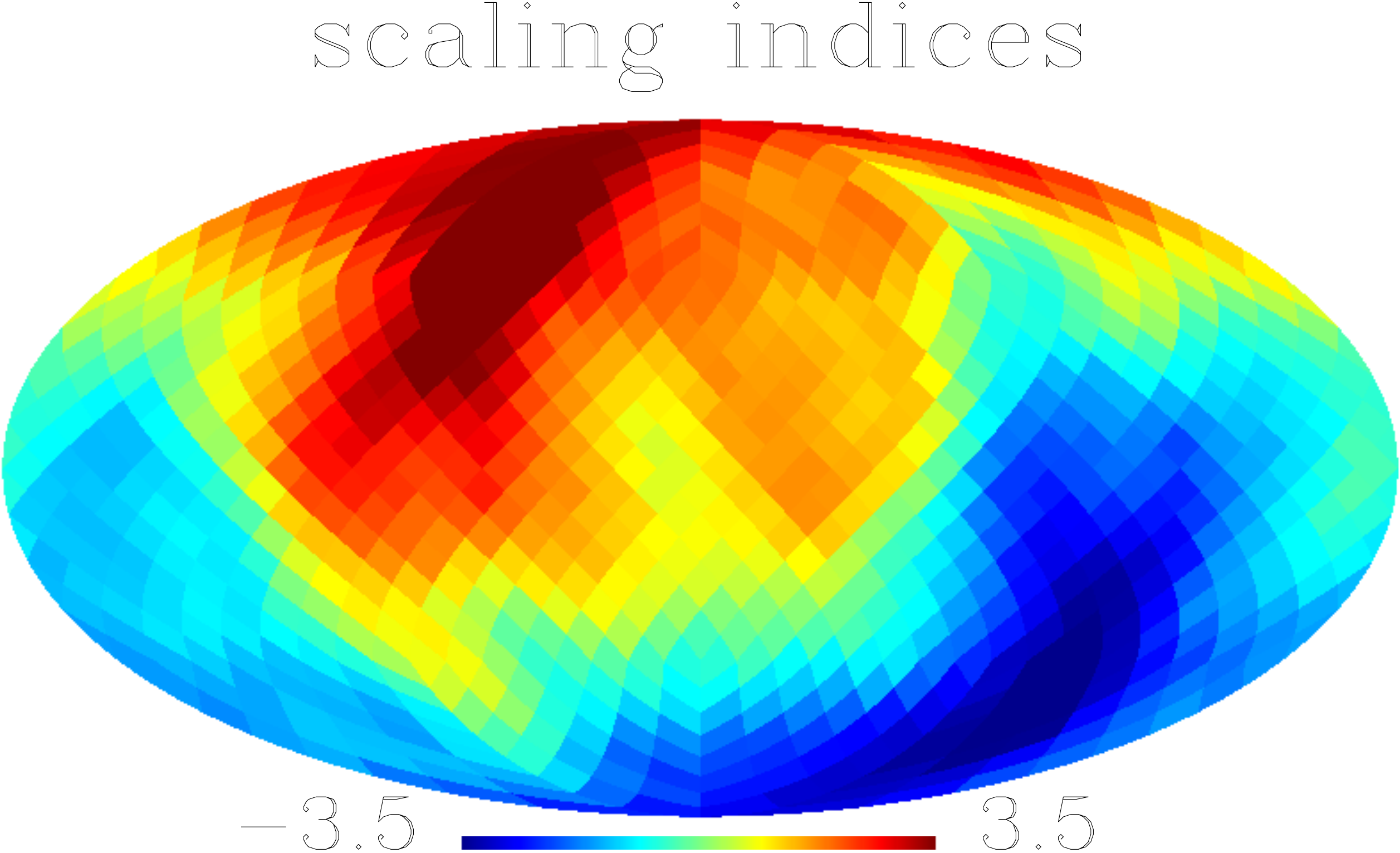}
\includegraphics[width=2.5cm, keepaspectratio=true,angle={90}]{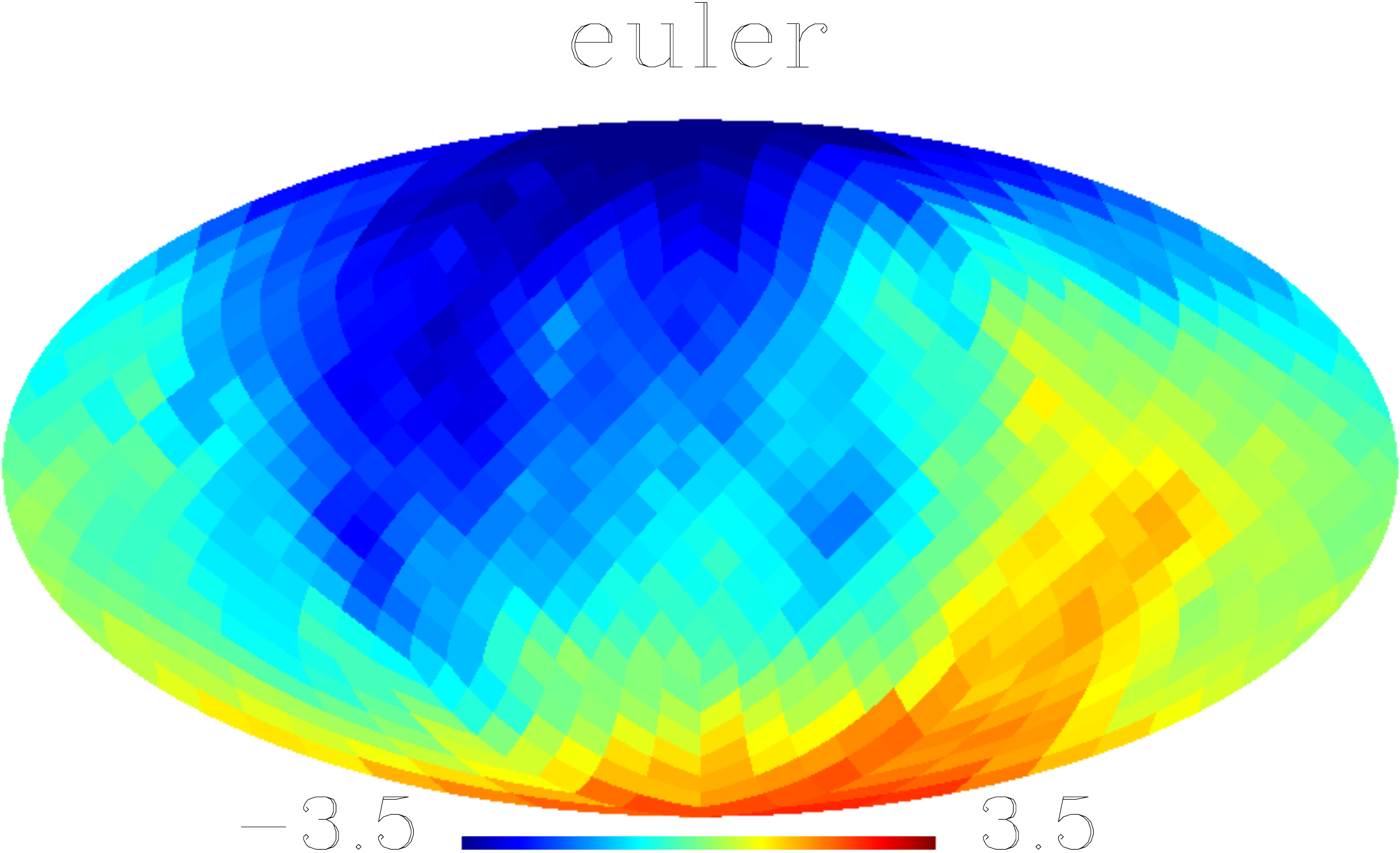}
\includegraphics[width=2.5cm, keepaspectratio=true,angle={90}]{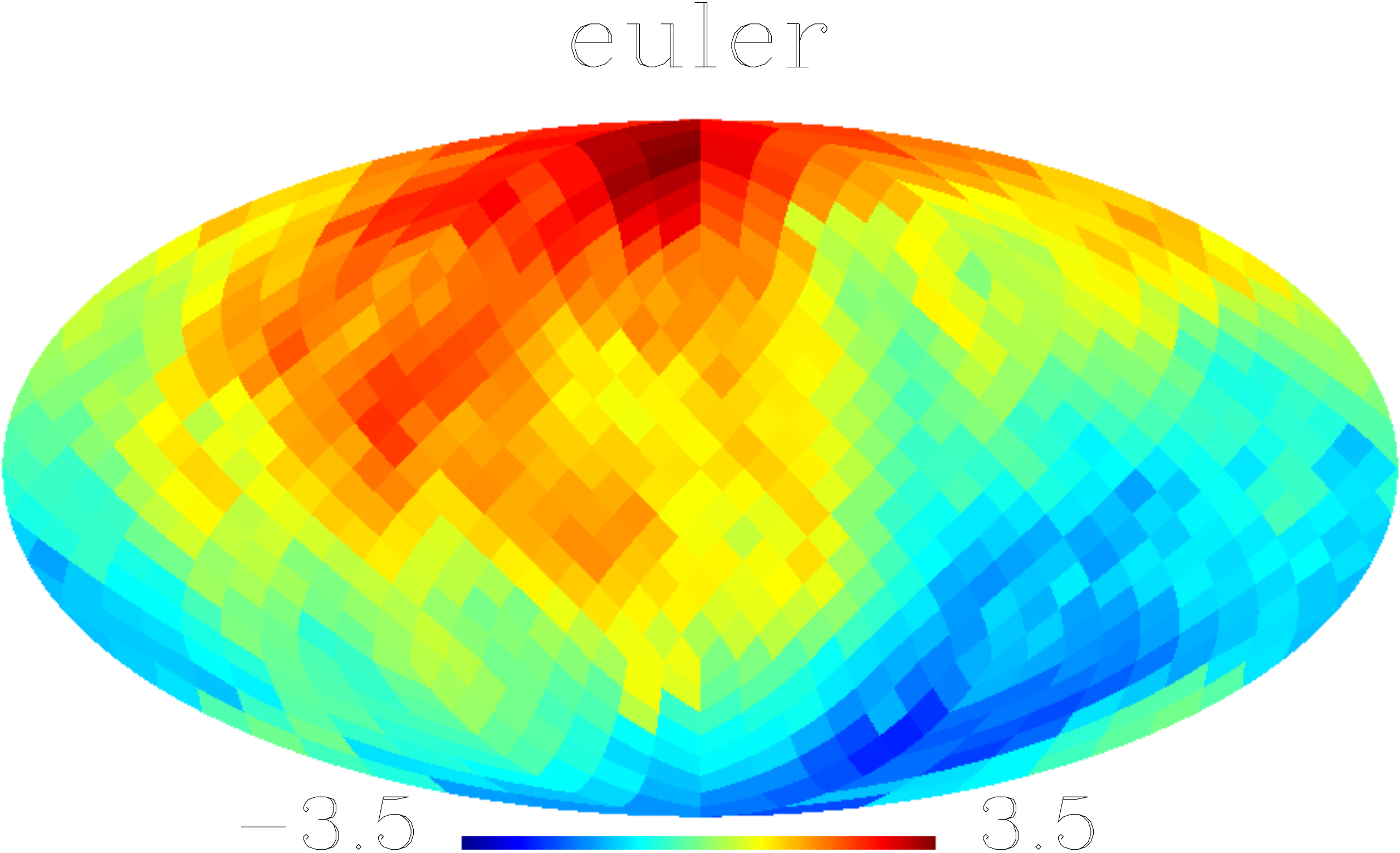}
\caption{Significance $S$ per hemisphere between surrogate of first and second order of NILC7 calculated for one threshold value $\nu_i$. For perimeter, $S$ is calculated at the threshold value where $M_{1,surro1}$ becomes maximal (upper left). For Euler, we chose the threshold values where $M_{2,surro1}$ becomes minimal (lower left) and maximal (lower right), respectively. For comparison, we show the deviations $S(\langle \alpha \rangle)$ for the mean of the scaling indices (upper right). The $\ell$-range for the method of the surrogates is $\Delta \ell = [2,20]$.}
\label{fig:signiMap_1threshold}
\end{figure}

The $\chi^2$ results discussed above do not depend on the algebraic sign anymore. In order to visualise the sign-dependence we plot the deviation $S$ per hemisphere between first and second order surrogates calculated with the Minkowski functionals for only one chosen threshold, shown in Figure \ref{fig:signiMap_1threshold}, and keep the algebraic signs of the deviations with this calculation. The chosen thresholds correspond to the minima or maxima of perimeter and Euler functional. For the SIM we choose to plot $S(Y)$ with $Y= \langle \alpha \rangle$ corresponding to Equation \ref{eqn:S(Y)} to keep the algebraic sign, as was done in previous works. For the Euler characteristic we see that the sign of the deviation depends on the threshold level $\nu$.  

\begin{figure}
\centering
\includegraphics[width=4.2cm, keepaspectratio=true,angle={90}]{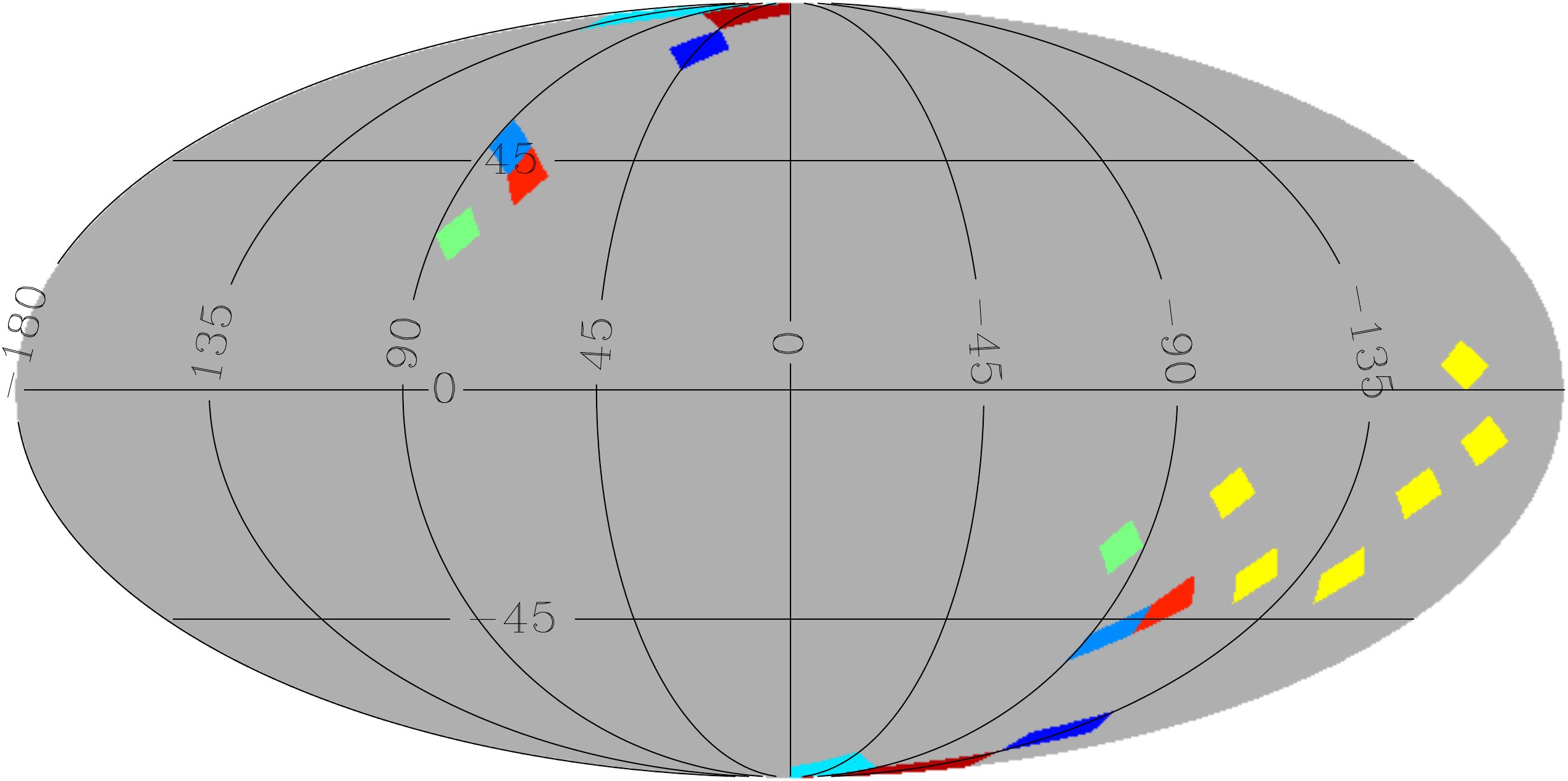}
\caption{A sky map in Galactic coordinates with the directions of maximal signal for NGs found in ILC7 (red) and NILC7 (blue) with Minkowski functionals (Area, Perimeter: blue, Euler: dark blue, Area, Perimeter, Euler: dark red) and SIM (light blue, light red). The yellow pixels show the dipole directions of the power distribution in 6 blocks of 100 multipoles each taken from \citet{2009ApJ...704.1448H}. The green pixels correspond to the ecliptic northern and southern pole, respectively.}
\label{fig:signiPunktePlot}
\end{figure}

An analysis of different $\ell$-ranges was also performed in \citet{2009ApJ...704.1448H}. In their work, the dipole directions of the power distributions of different multipole blocks of $100$ multipoles each were calculated. It was shown that for the $6$ multipole blocks in $\ell=[2,600]$ the dipole directions lie very close to each other. We compare our non-linear statistics for the method of surrogates with $\Delta \ell = [2,20]$ with these linear findings in Figure \ref{fig:signiPunktePlot}. We show our coordinates of the hemispherical pairs defined above and add the directions of the $6$ dipoles of the power distributions. Note that the linear and non-linear statistics can have different scale sensitivity. Although it is not clear to what extent the results can be reconciled in detail, interestingly, our large-scale investigations with Minkowski functionals and scaling indices as well as the results of  \citet{2009ApJ...704.1448H} show evidence of asymmetry. As mentioned above, the original power spectrum is exactly preserved for all our surrogate maps. The dipole direction of the multipole block $\ell = 2-100$ lies close to our large-scale SIM results for ILC7 and close to the Minkowski (Area, Perimeter) results for NILC7. Furthermore, these three pixels are close to the southern ecliptic pole. 

In \citet{2012arXiv1207.3235R} we introduced the method of surrogates for an incomplete sky. We find that even when the complete Galactic plane is removed, NGs and hemispherical asymmetries can still be detected in the CMB and conclude that the Galactic plane cannot be the dominant source for the found anomalies. Our results point in general towards a violation of statistical isotropy of the Universe and disfavor single-field slow roll inflation.

\begin{figure}
\centering
\includegraphics[width=2.4cm, keepaspectratio=true,angle={90}]{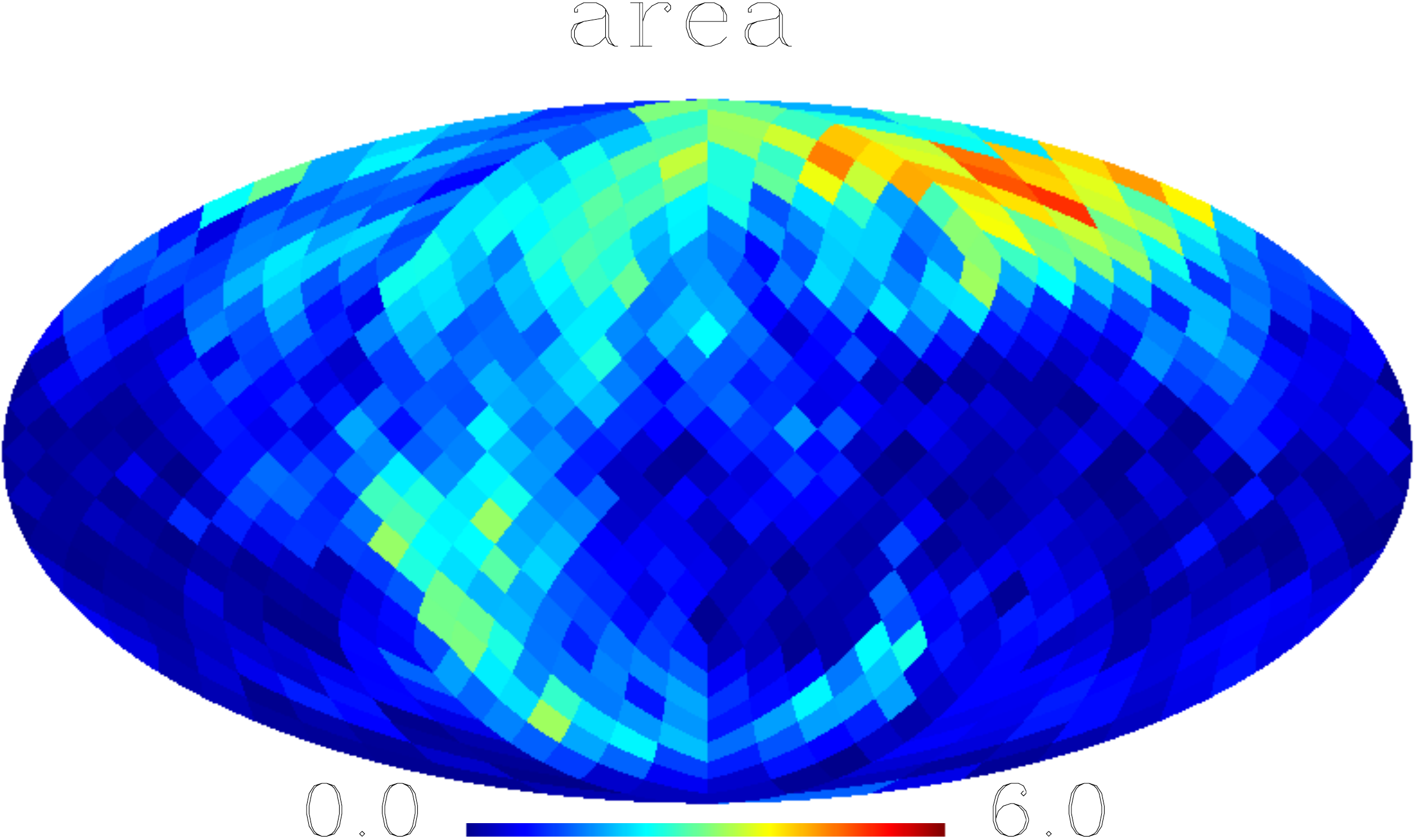}
\includegraphics[width=2.4cm, keepaspectratio=true,angle={90}]{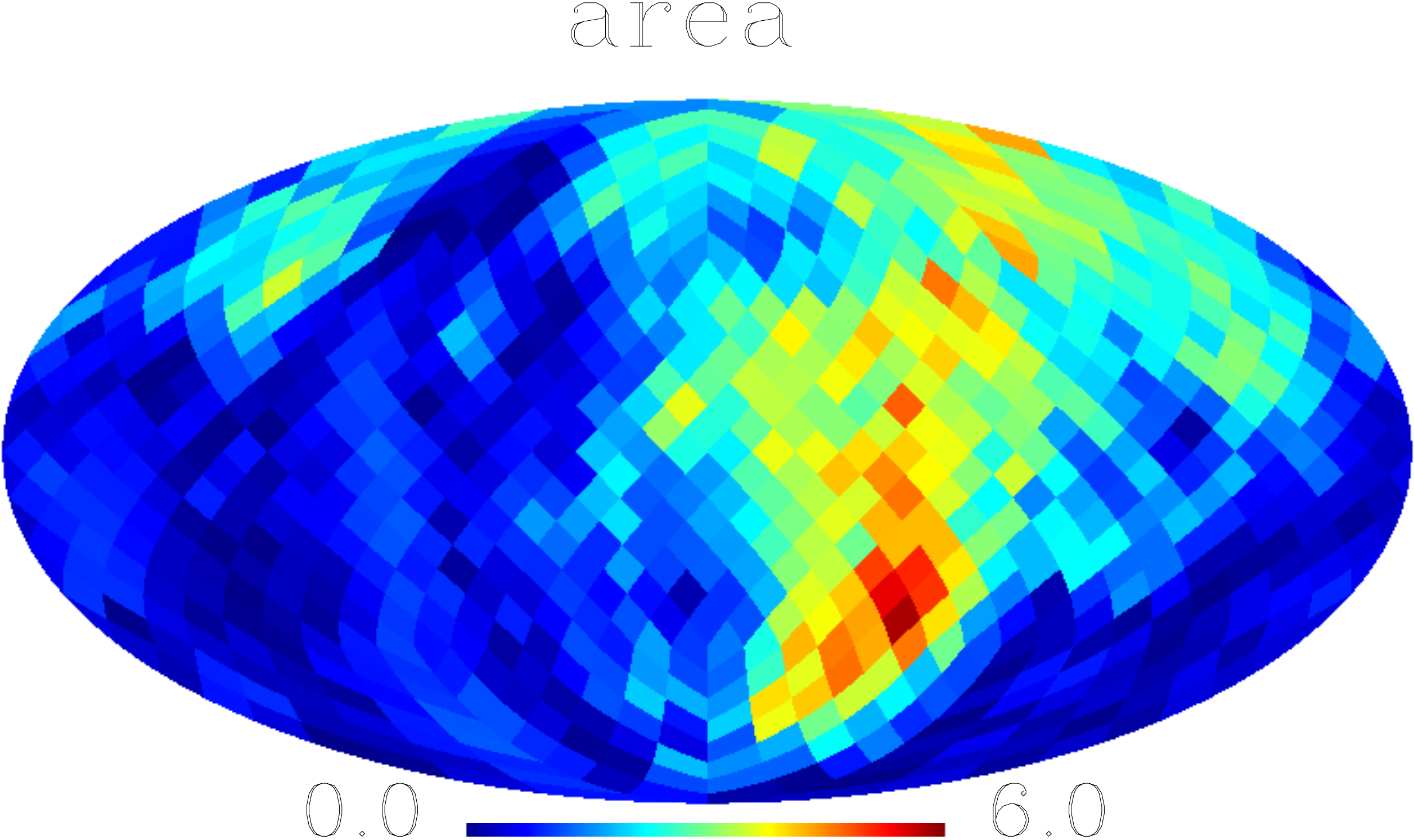}
\includegraphics[width=2.4cm, keepaspectratio=true,angle={90}]{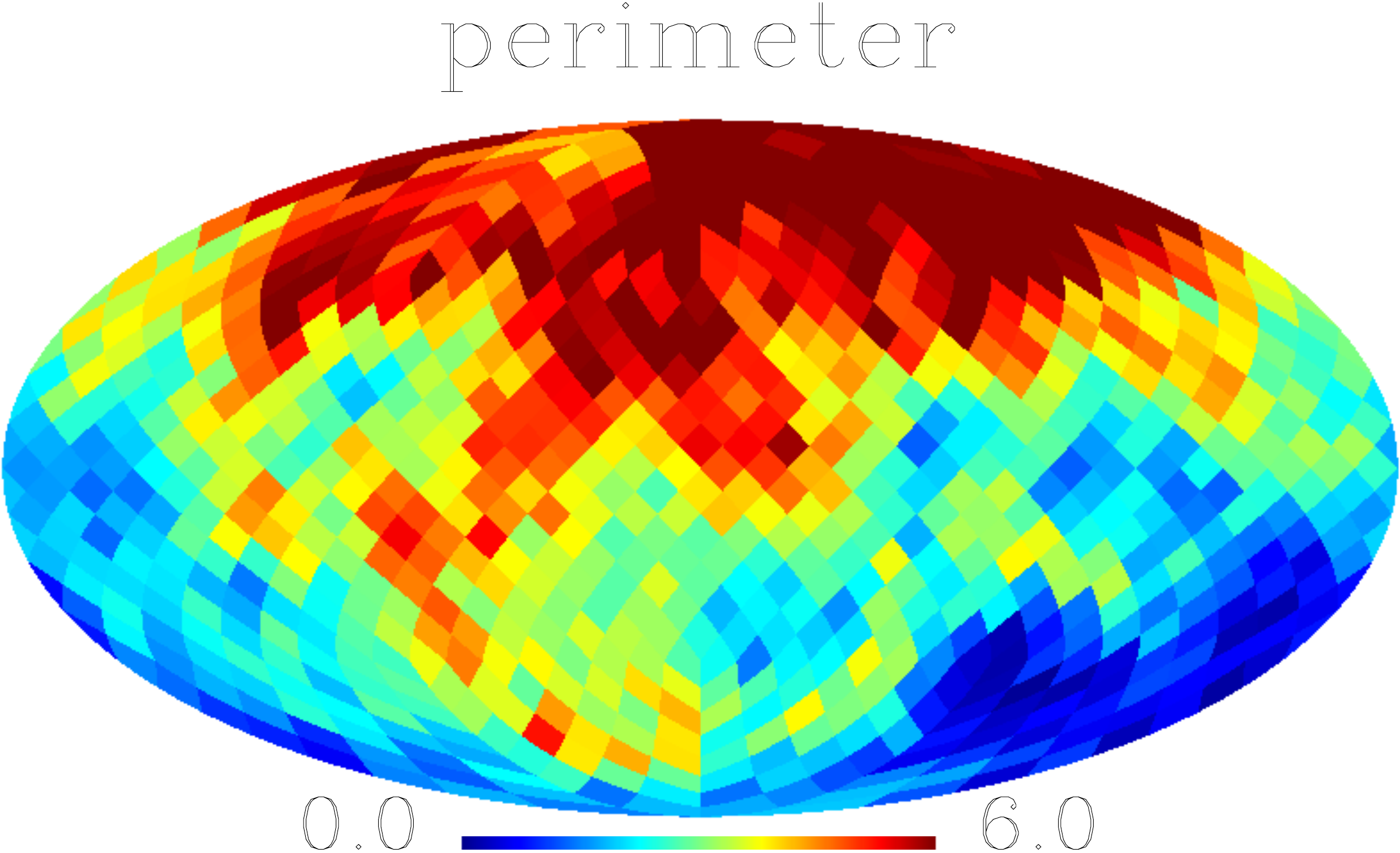}
\includegraphics[width=2.4cm, keepaspectratio=true,angle={90}]{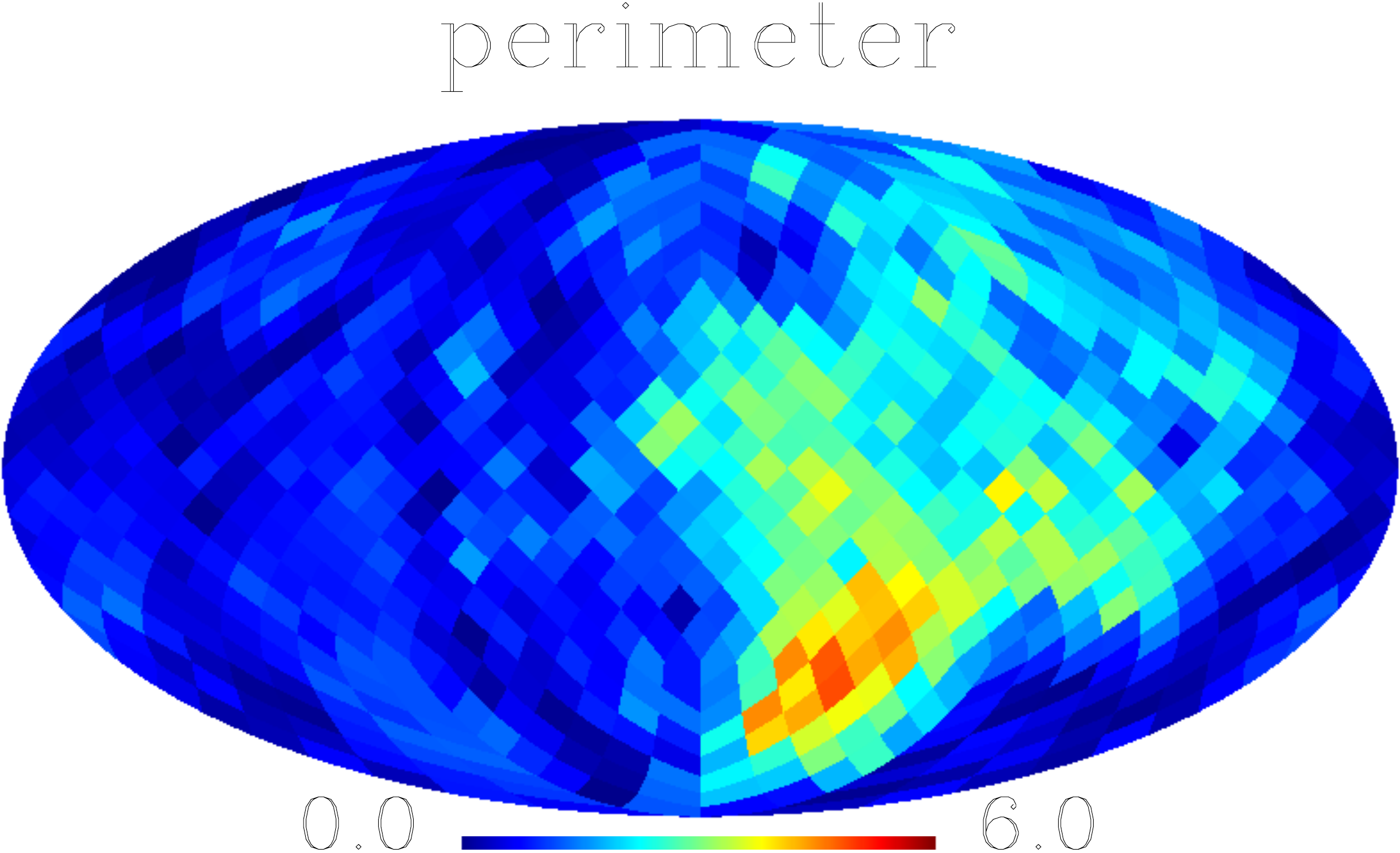}
\includegraphics[width=2.4cm, keepaspectratio=true,angle={90}]{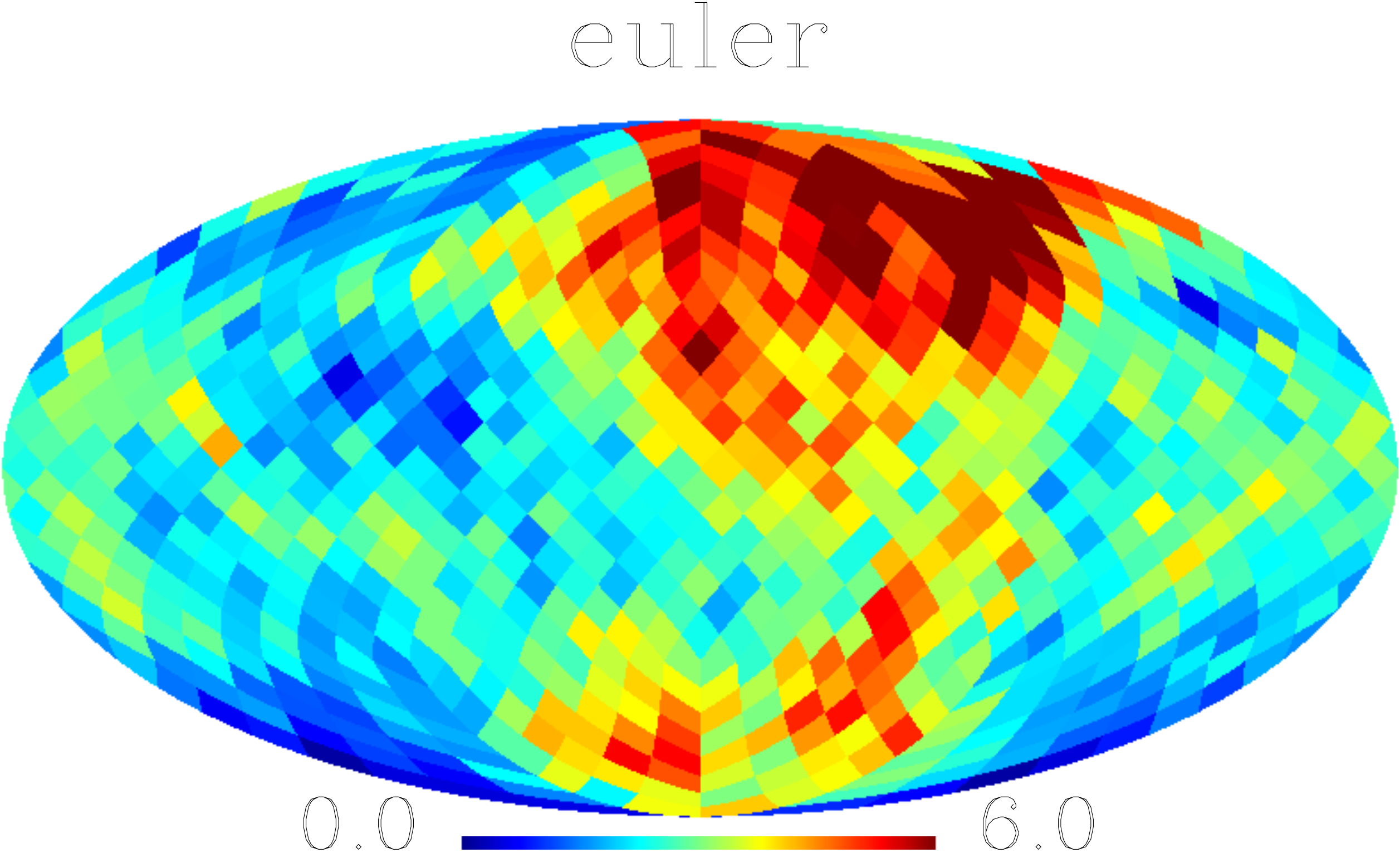}
\includegraphics[width=2.4cm, keepaspectratio=true,angle={90}]{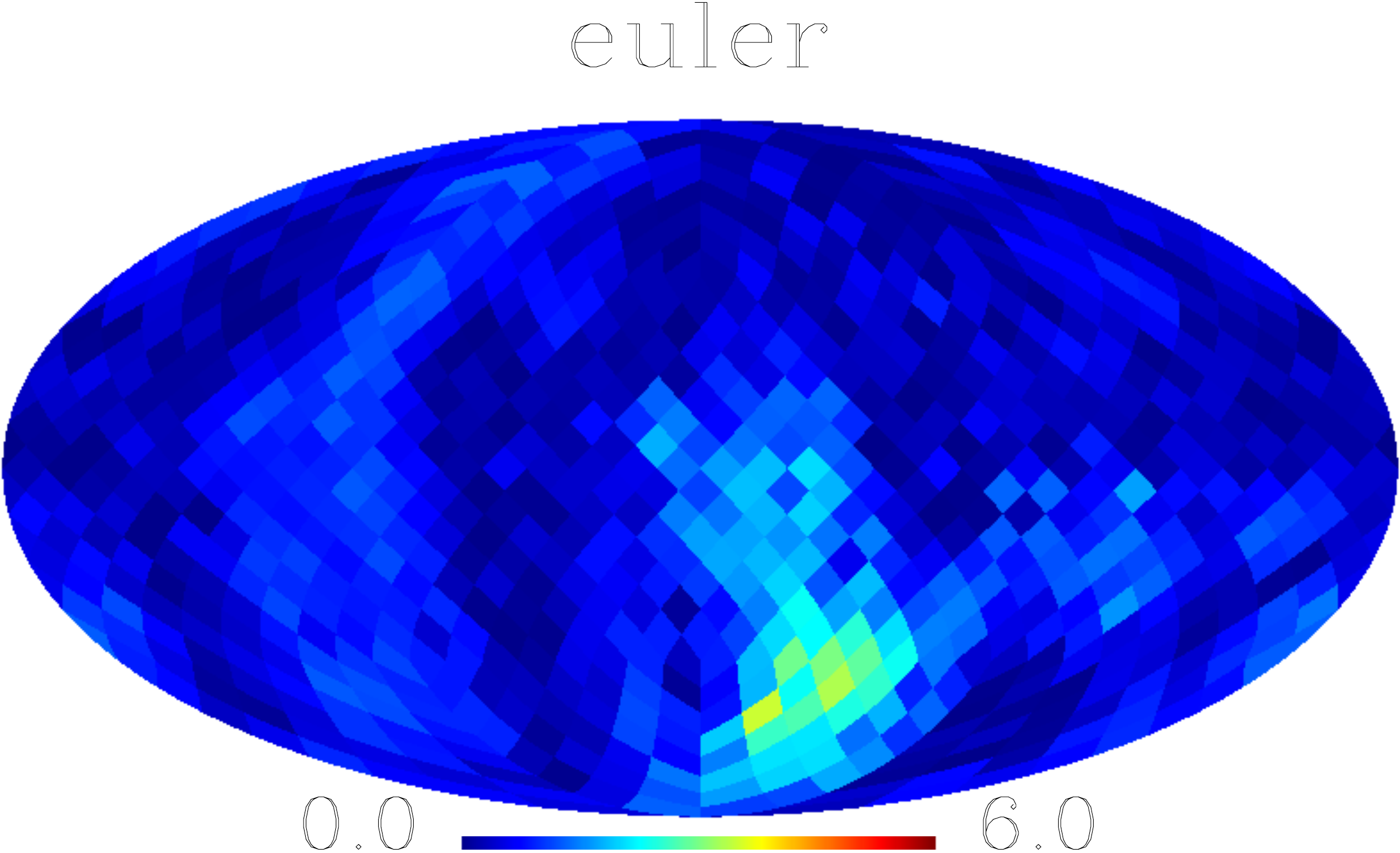}
\caption{Deviations  of the Minkowski Functionals $M_0$, $M_1$ and $M_2$ of the rotated hemispheres for the ILC7 (left column) and NILC7 (right column) map. The $\ell$-range for the method of the surrogates is $\Delta \ell = [120,300]$.}
\label{fig:signiMapNILC7120300}
\end{figure}

Supplemental to the results on largest scales, Figure \ref{fig:signiMapNILC7120300} shows the deviations $S(\chi^2)$ per hemisphere of the ILC7 and NILC7 map obtained by the Minkowski functionals for the surrogates $\Delta \ell$-range $[120,300]$, which is covering the first peak of the power spectrum. The three Minkowski functionals show indications for phase correlations for both maps. Yet the directions of these deviations on the sky totally differ from each other. Since the two compared maps differ in their resolution we extend the tests with a comparison of the NILC7 at a resolution decreased to one degree FWHM (not shown here). We generate the corresponding surrogate maps of first and second order and calculate the statistics for the Minkowski functionals and the scaling indices. The signatures for phase correlations detected by the Minkowksi functionals differ from the results of the fully resolved NILC7 map and they do not resemble the signatures of the ILC7 map. For the SIM there is more concordance between the findings of the differing resolutions of NILC7. We claim that the reason for disparities between the ILC and NILC maps on intermediate length-scales, $120\leq l \leq 300$, is not only the difference in beam resolutions but even more the difference in the foreground cleaning. As explained in \citet{2011MNRAS.415.2205R} foreground removal can induce phase correlations, especially on these intermediate $\ell$-ranges. The detected phase correlations in the $\Delta \ell$-range of $[120,300]$ depend on the foreground cleaning, the map resolutions and the response of the image analysis techniques. If there are additional deviations from Gaussianity with cosmological origin in the maps on these $\ell$-ranges we cannot distinguish them from systematics so far. Our findings clearly show that the found non-Gaussianities in WMAP data are scale-dependent and can have different origin.

\begin{figure}
\includegraphics[width=2.5cm, keepaspectratio=true,angle={90}]{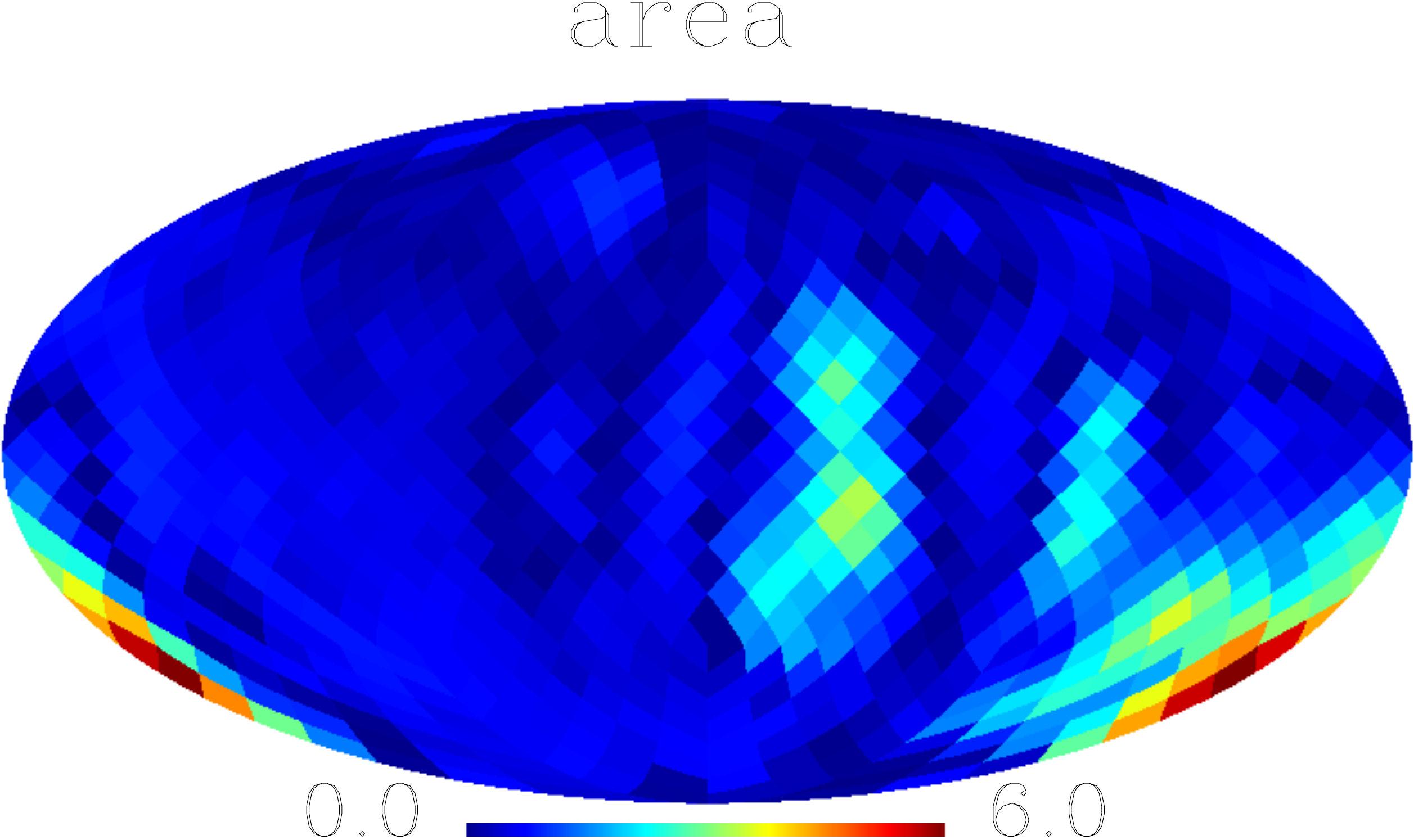}
\includegraphics[width=2.5cm, keepaspectratio=true,angle={90}]{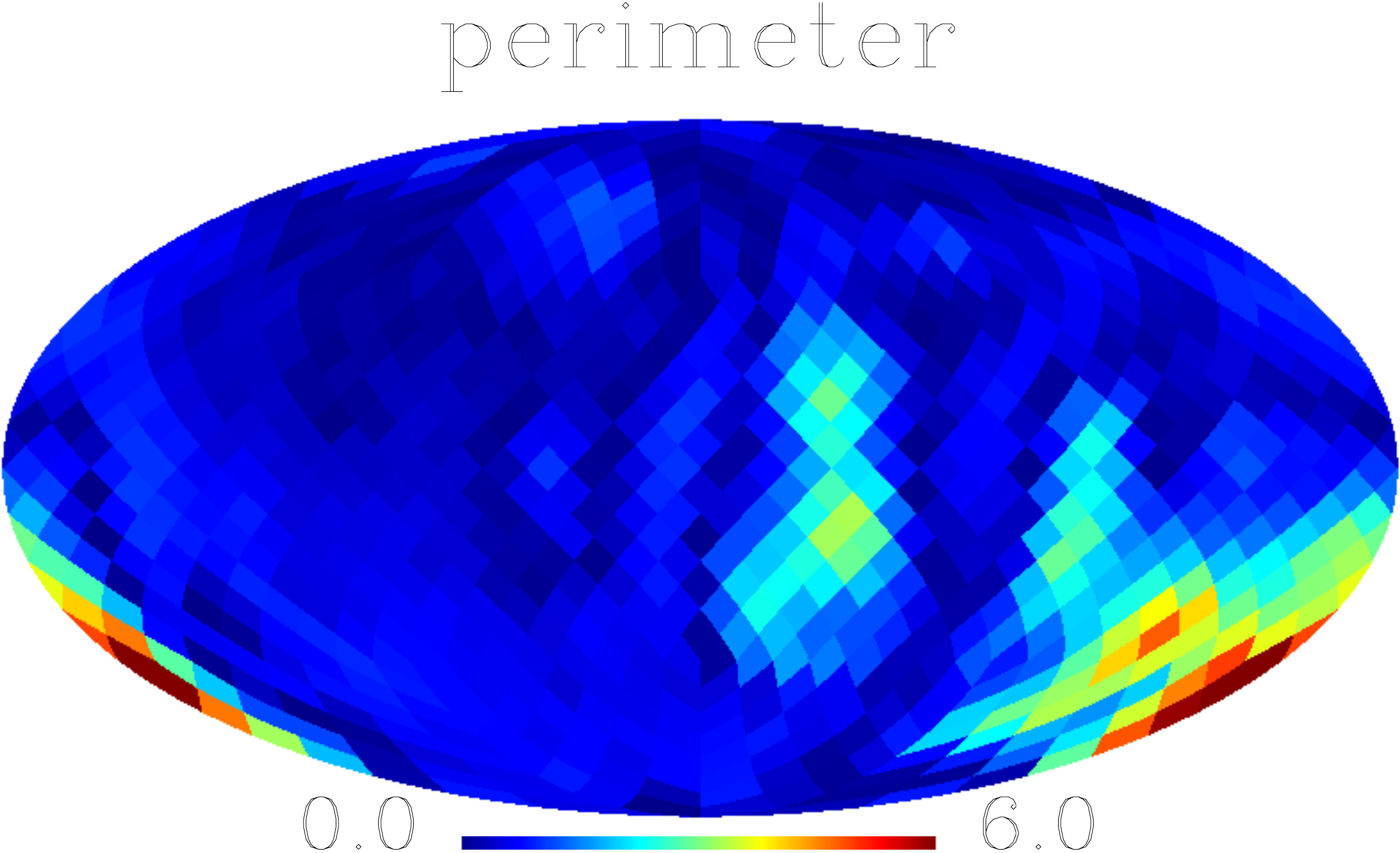}
\includegraphics[width=2.5cm, keepaspectratio=true,angle={90}]{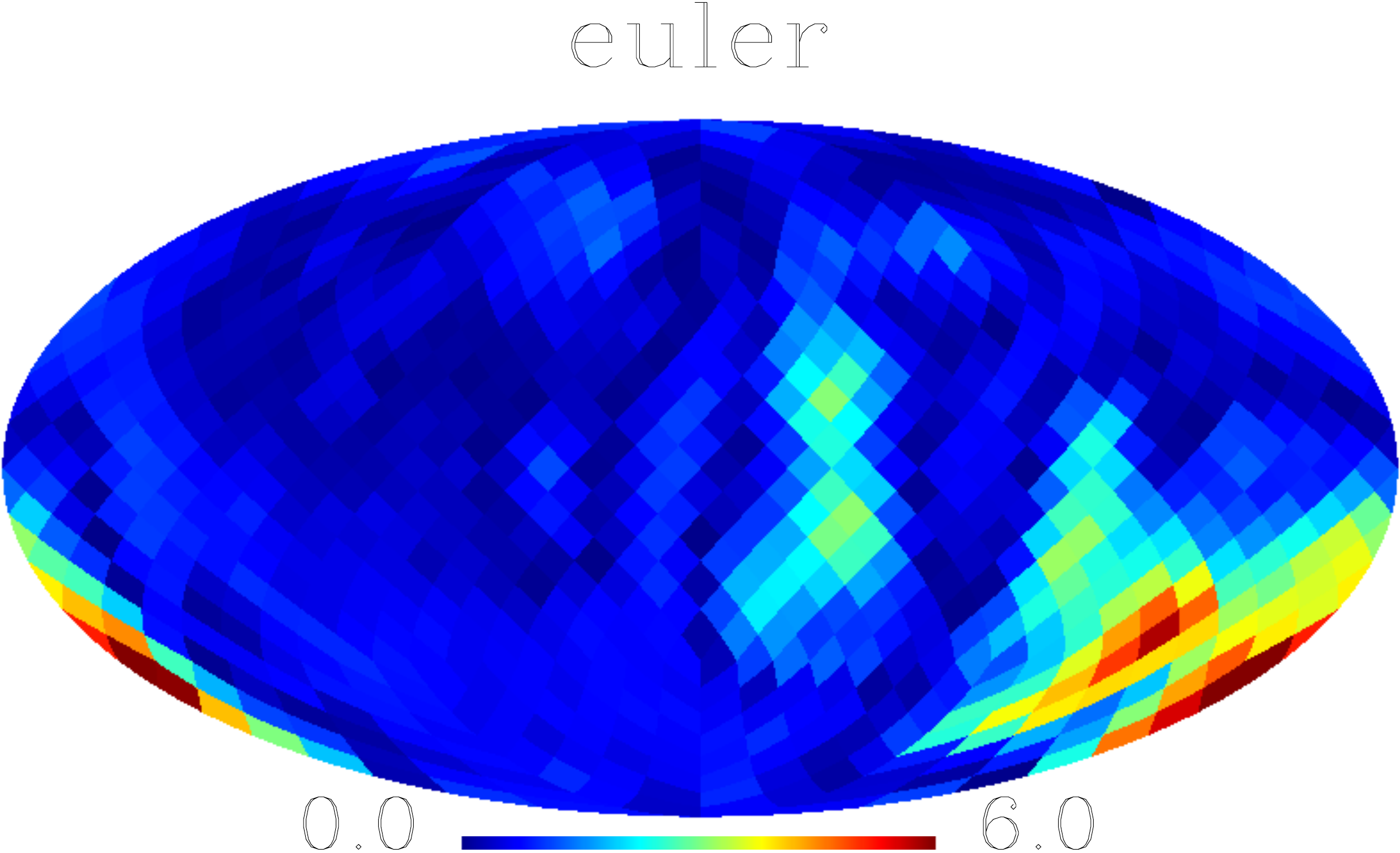}
\includegraphics[width=2.5cm, keepaspectratio=true,angle={90}]{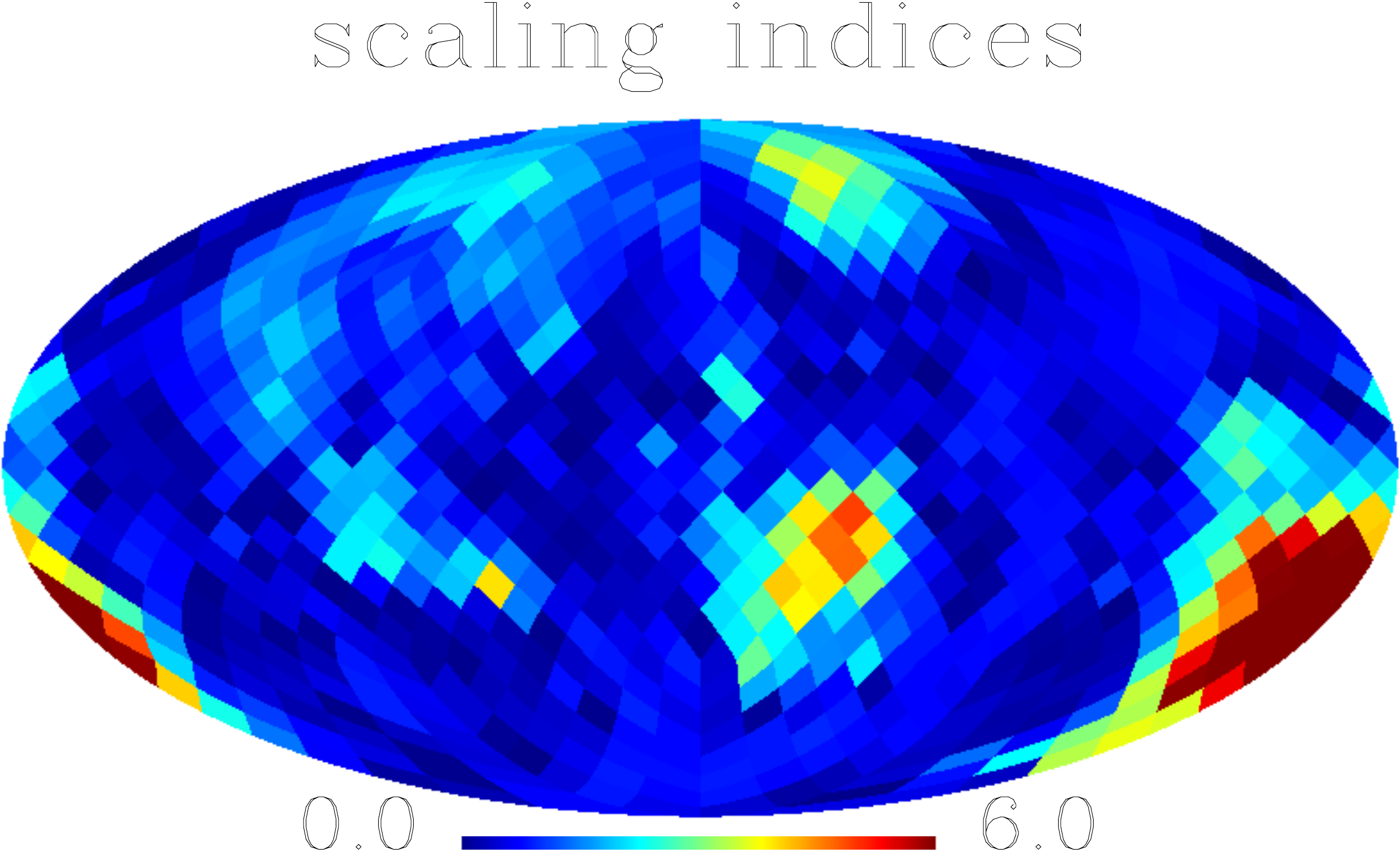}
\caption{Deviations $S(\chi^2)$ of the Minkowski Functionals $M_0$, $M_1$ and $M_2$ and the scaling indices (from left to right and top to bottom) of the rotated
hemispheres for the NILC7 map for the smaller apex angle $\pi/2$. The $\ell$-range for the method of the surrogate is $\Delta \ell = [2,20]$.}
\label{fig:signiMapPi4NILC7}
\end{figure}

So far, we have analysed the deviations from Gaussianity in areas of apex angle $\pi$ around the sky which leads to a solid angle of $2\pi$, so a full hemispherical analysis. We decrease the apex angle to $\pi/2$ and calculate the Minkowski functionals and scaling indices for areas with a size of only $30\%$ (solid angle $0.6\pi$) of the former hemispheres. The corresponding $S$-maps are shown in Figure \ref{fig:signiMapPi4NILC7} for the $\chi^2$ statistic of the three Minkowski functionals and the SIM. The $\pi/2$ study analyses smaller sky patches of the CMB. It can feature a better spatial localisation of the phase correlations. However, a smaller number of pixels might constrict the detection of a weaker signal. Comparing between the two classes of surrogates on smaller areas of the sky maps, we find no signal for non-Gaussianity in the northern ecliptic sky, whereas we detect individual spots in the southern sky which indicate phase correlations in these parts of the sky. The fact that we do not detect these spots in the northern sky makes ecliptic systematics from observations less likely. However, it cannot be ruled out that the detected spots are correlated with unknown foregrounds. Further study on the origin of the found spots and comparison with other detected anomalous spots in the CMB is needed.

\begin{figure*}
\includegraphics[width=5.3cm, keepaspectratio=true]{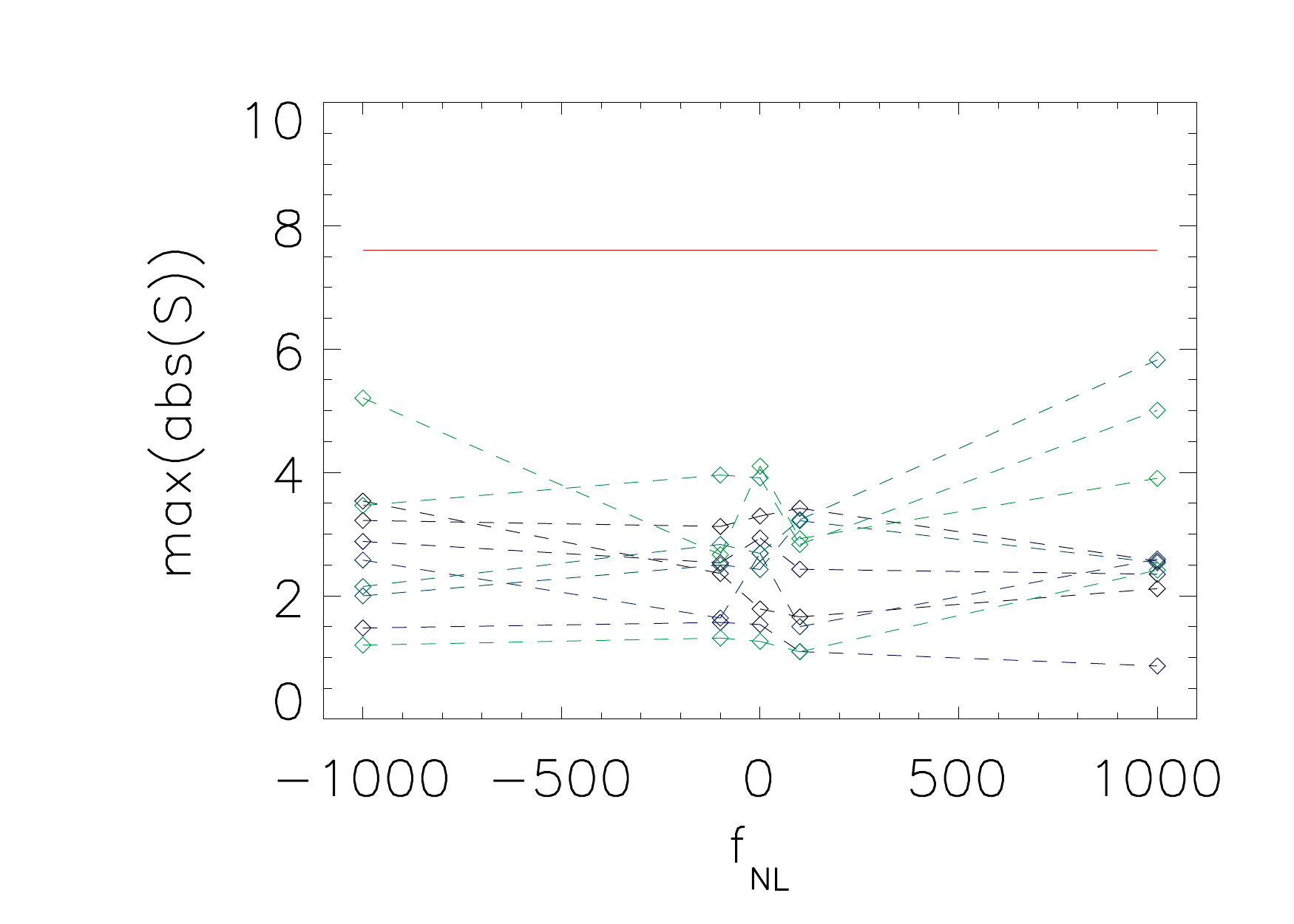}
\includegraphics[width=5.3cm, keepaspectratio=true]{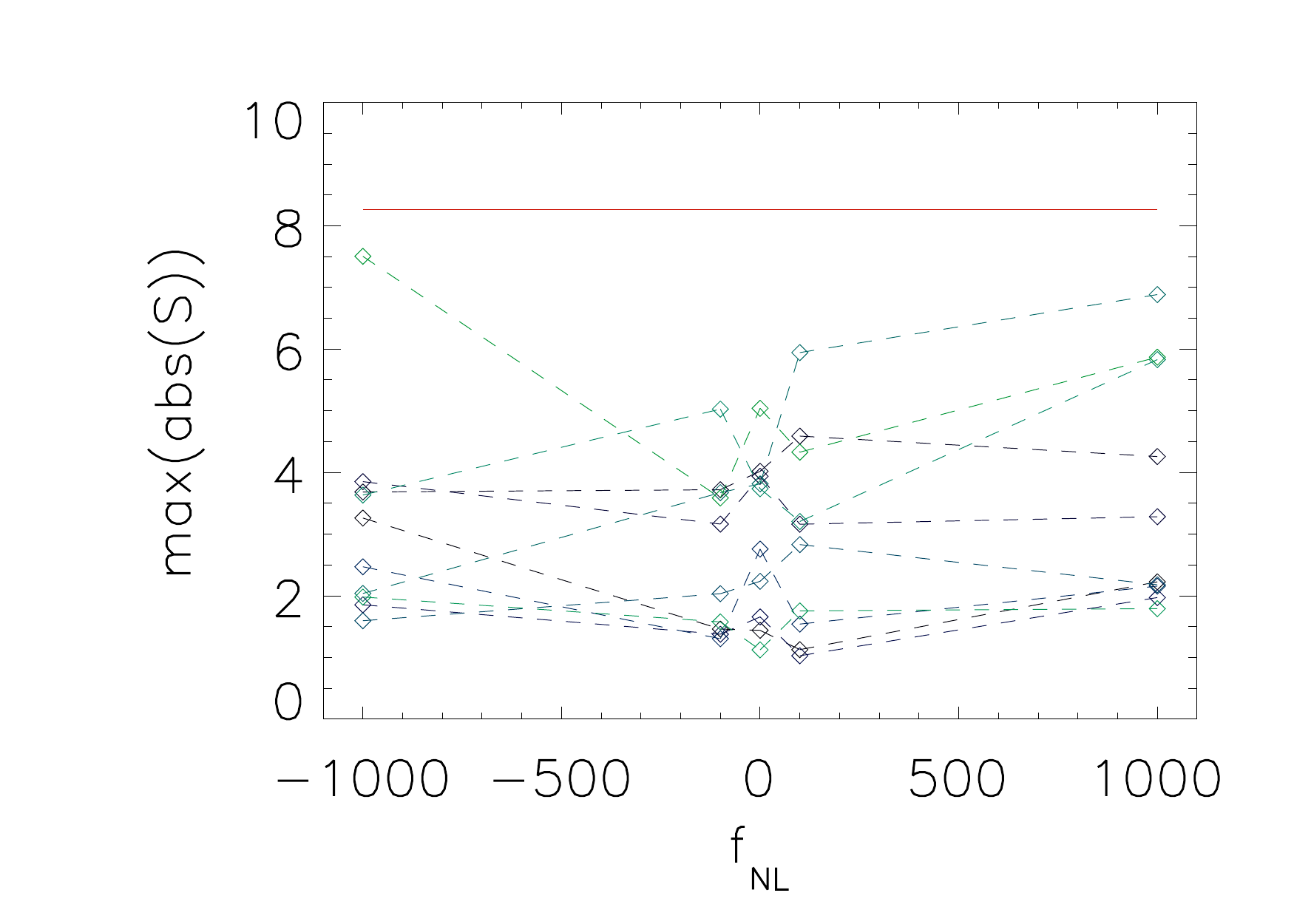}
\includegraphics[width=5.3cm, keepaspectratio=true]{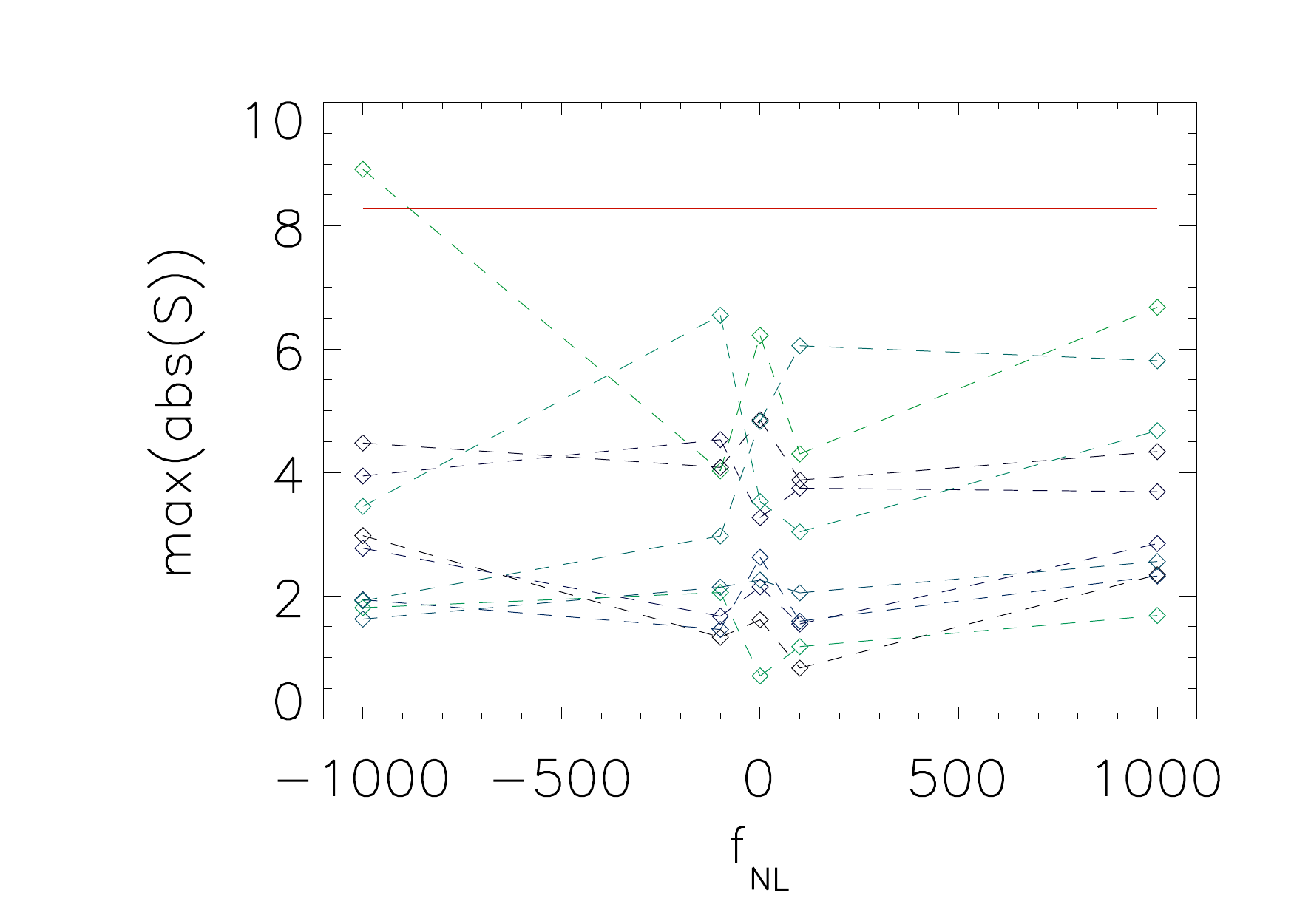}
\includegraphics[width=5.3cm, keepaspectratio=true]{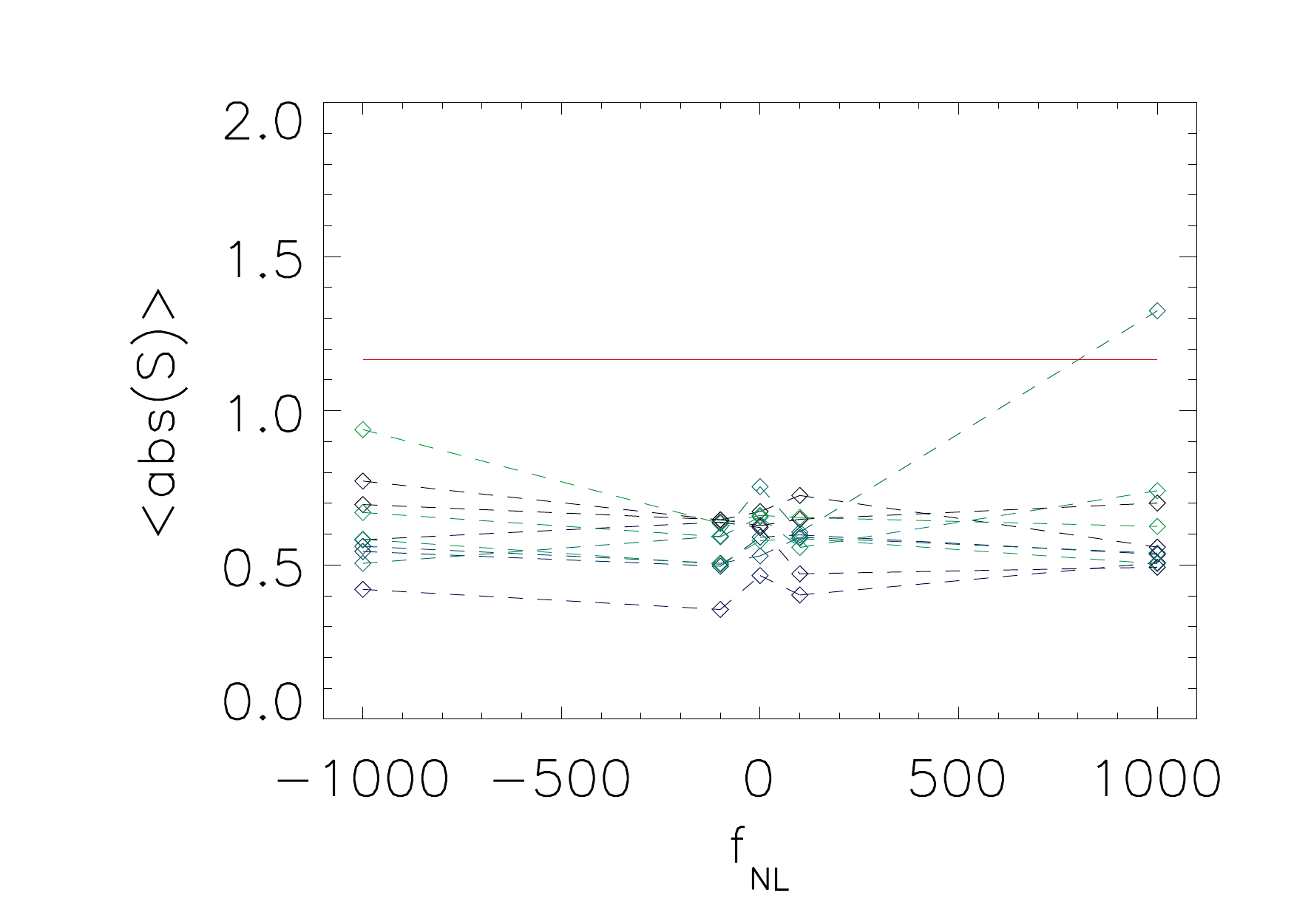}
\includegraphics[width=5.3cm, keepaspectratio=true]{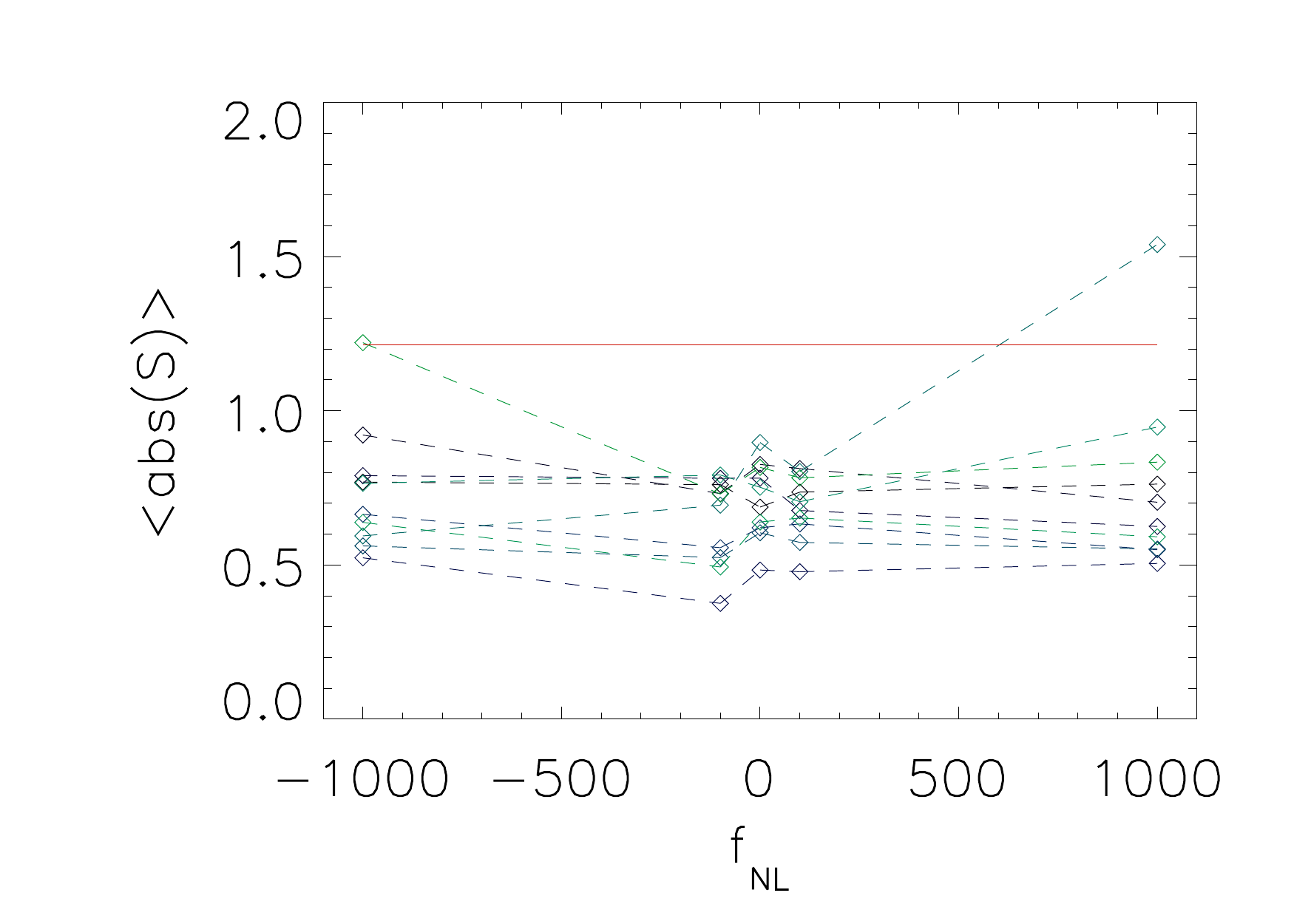}
\includegraphics[width=5.3cm, keepaspectratio=true]{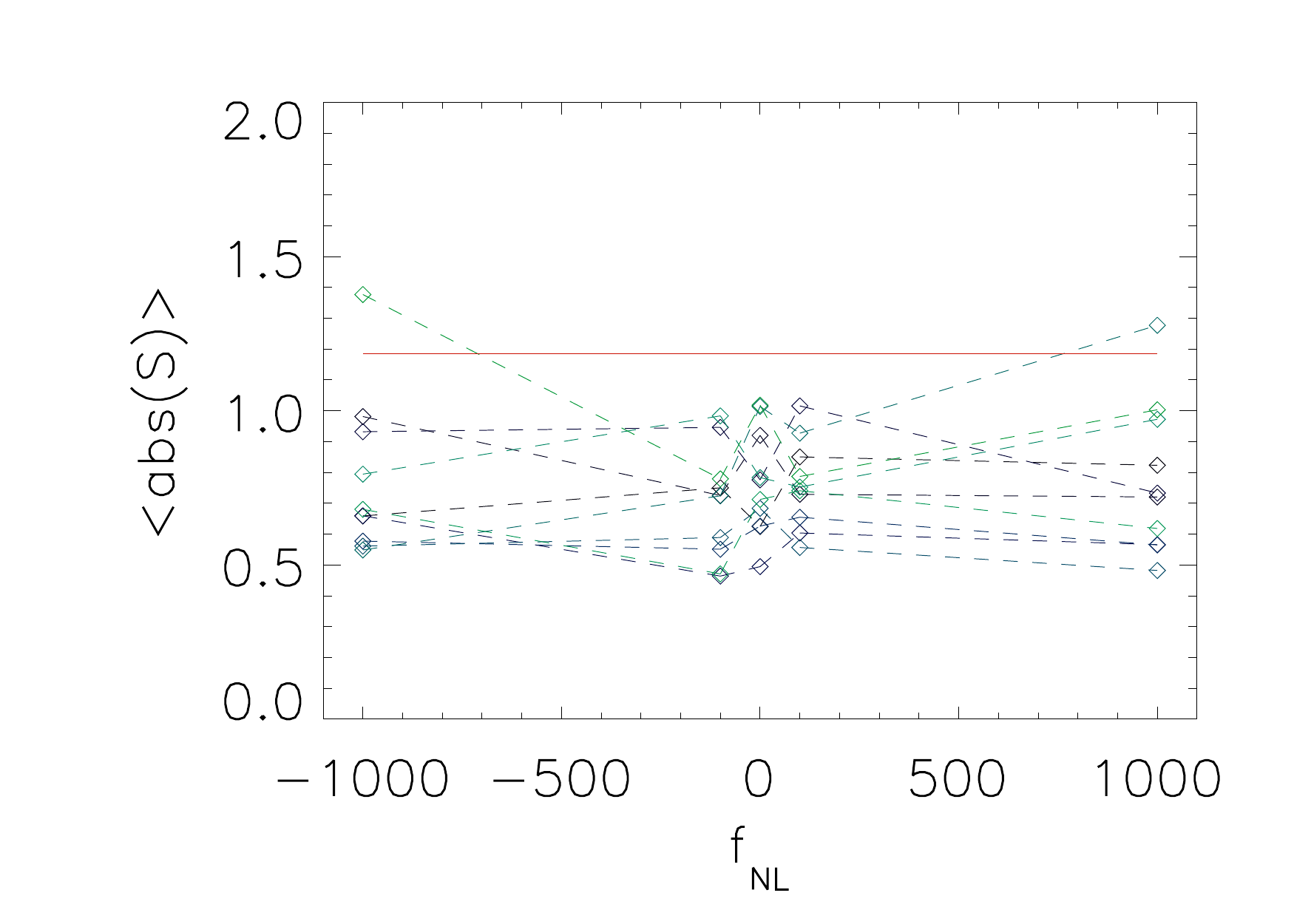}
\includegraphics[width=5.3cm, keepaspectratio=true]{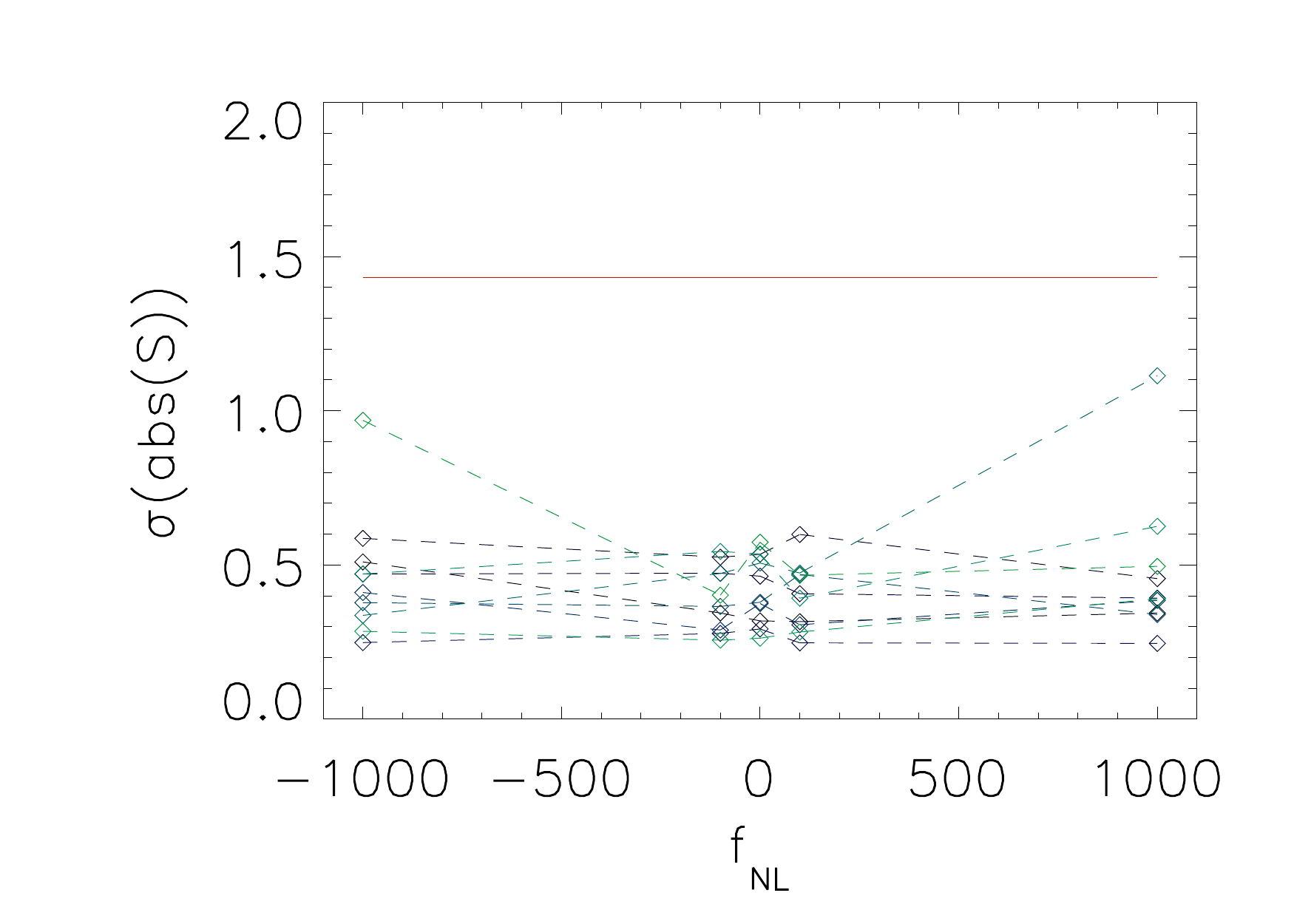}
\includegraphics[width=5.3cm, keepaspectratio=true]{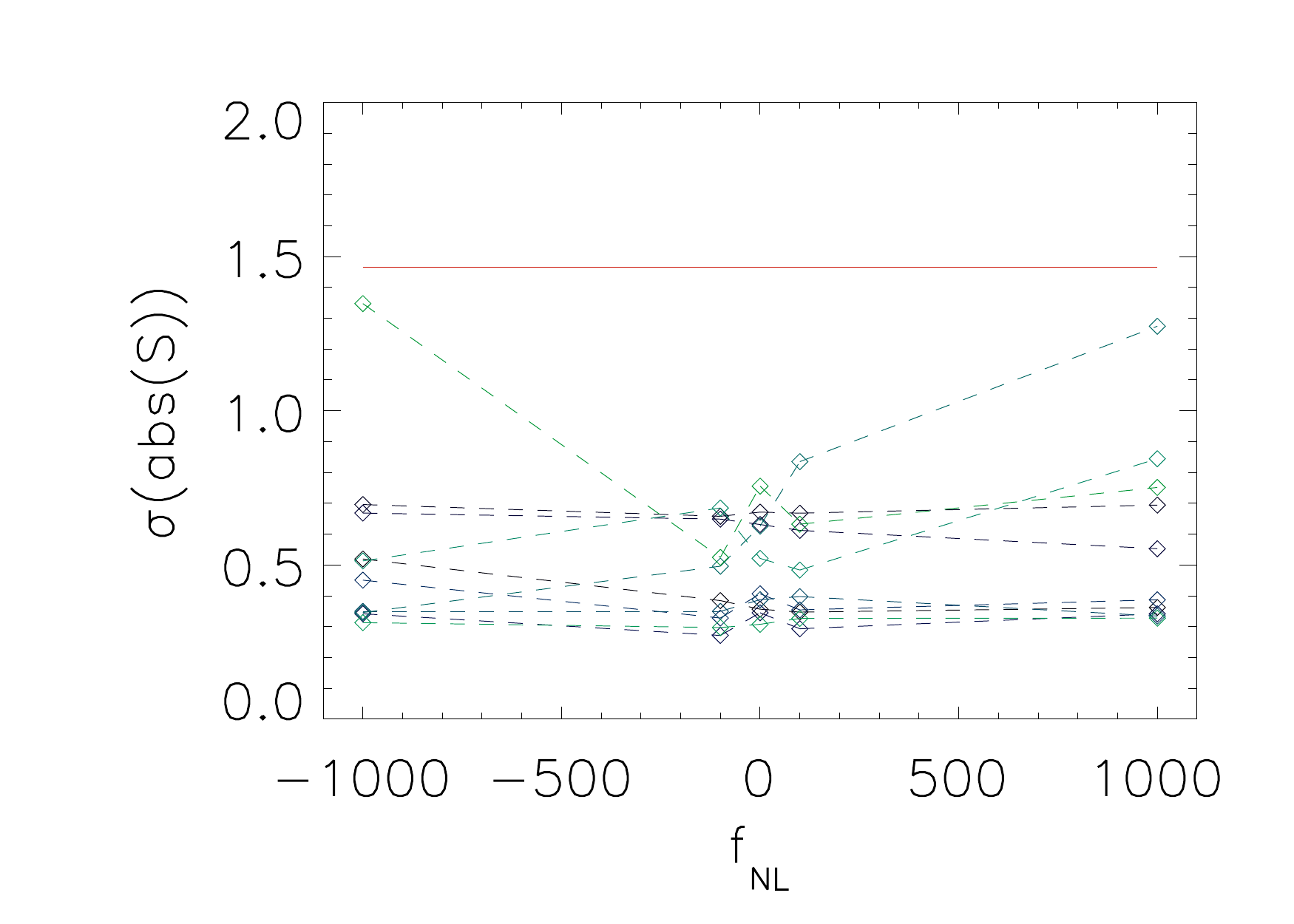}
\includegraphics[width=5.3cm, keepaspectratio=true]{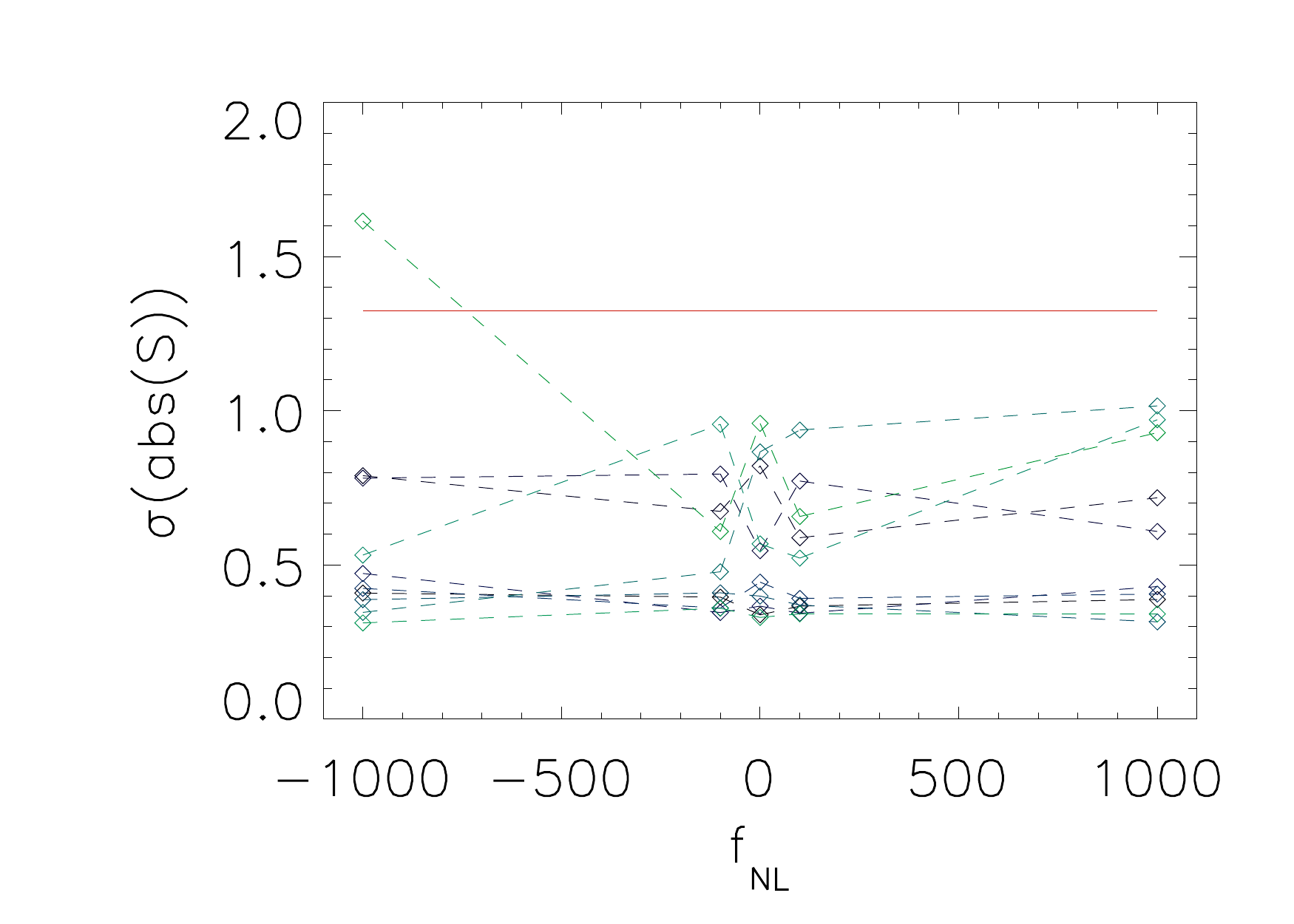}
\caption{Maximum, mean and standard deviation of the absolute values  of $S(\chi^2)_{M_\star}$ of ten $f_{\mathrm{NL}}^{\mathrm{local}}$ simulations compared to the NILC7 original map (red line) for area (left column), perimeter (middle) and Euler (right).  The HEALPix resolution of the maps is $N_{\mathrm{side}}=64$. The $\ell$-range for the method of the surrogates is $\Delta \ell = [2,20]$.}
\label{fig:FnlNILC7minkos}
\end{figure*}

\begin{figure}
\includegraphics[width=8cm, keepaspectratio=true]{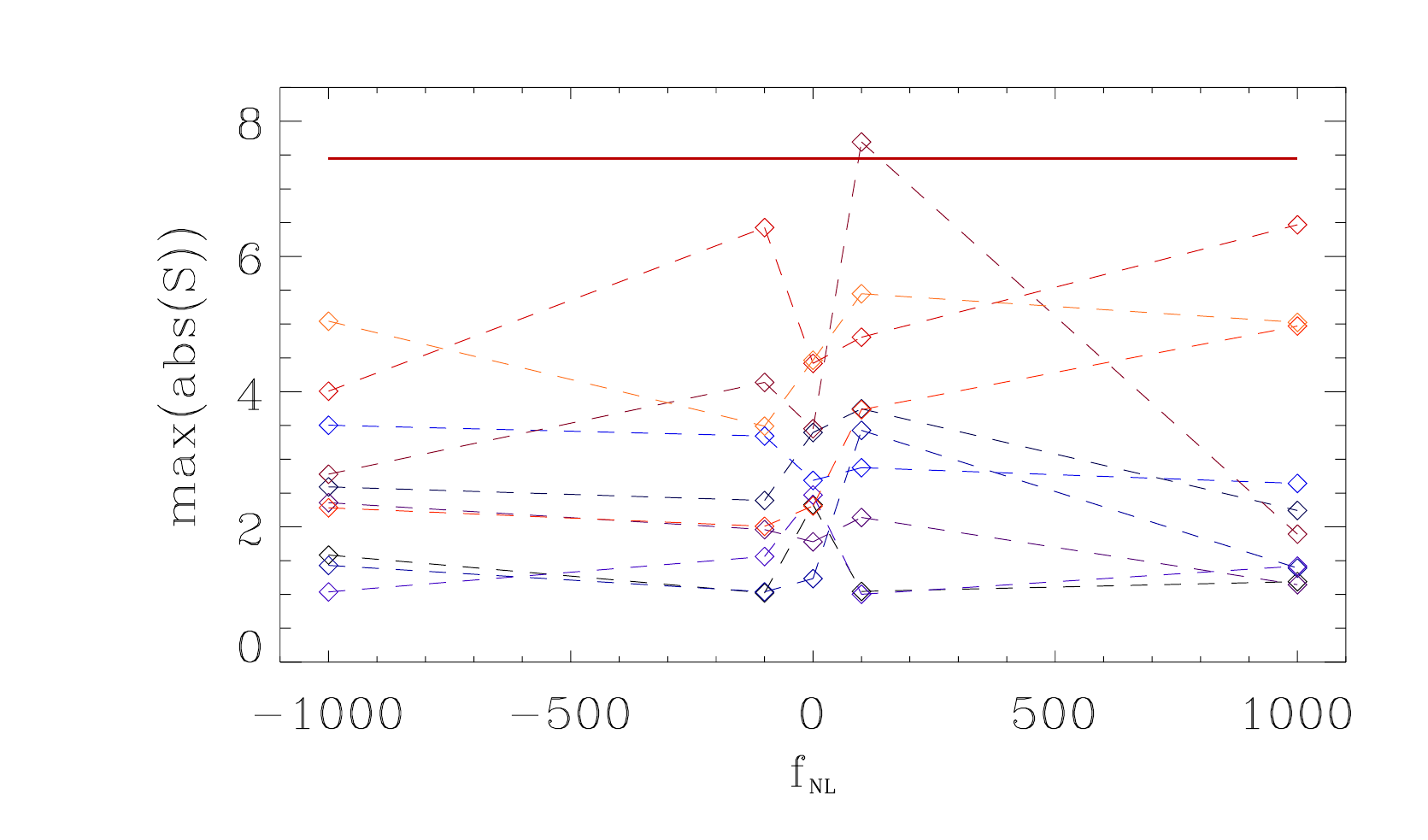}
\includegraphics[width=8cm, keepaspectratio=true]{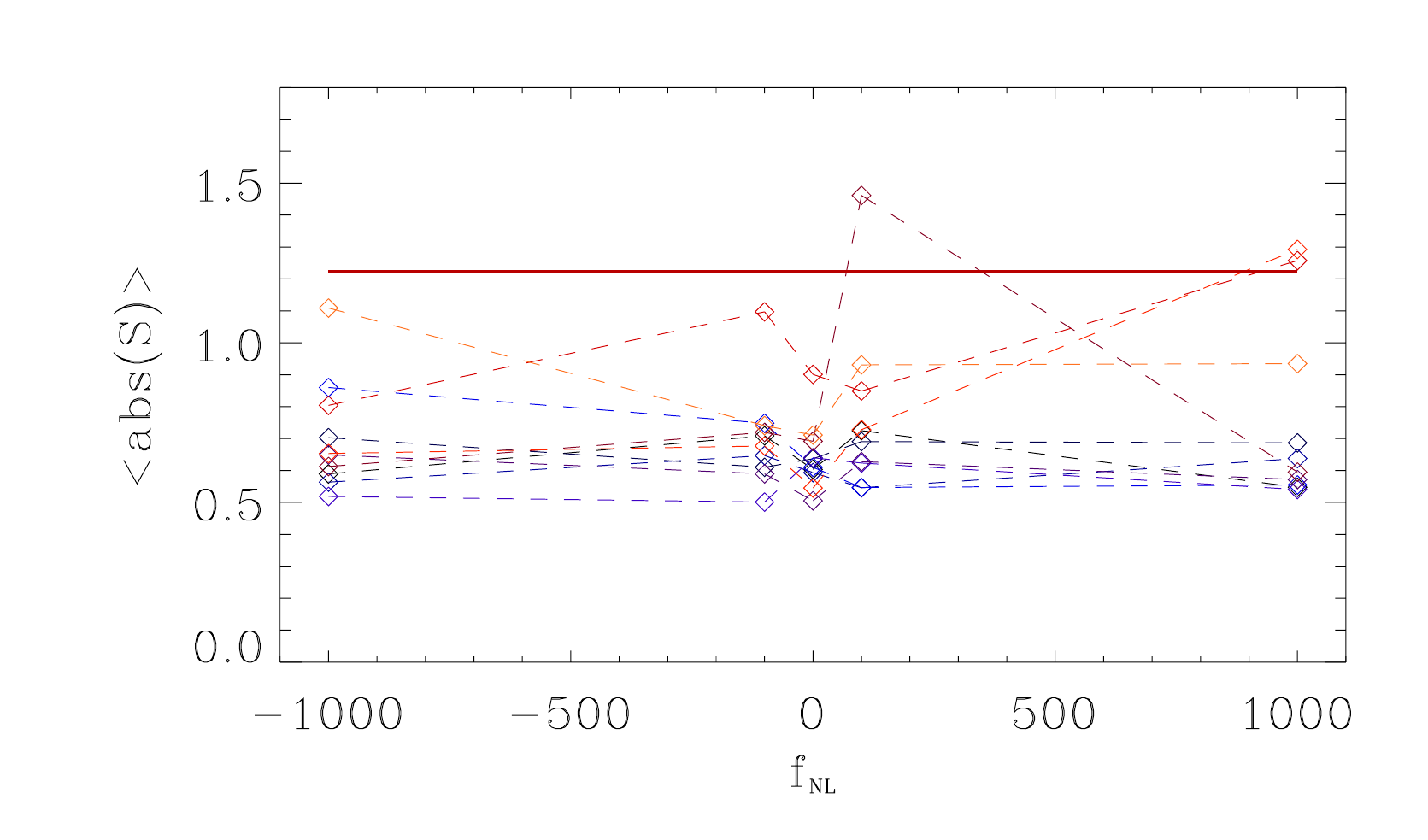}
\includegraphics[width=8cm, keepaspectratio=true]{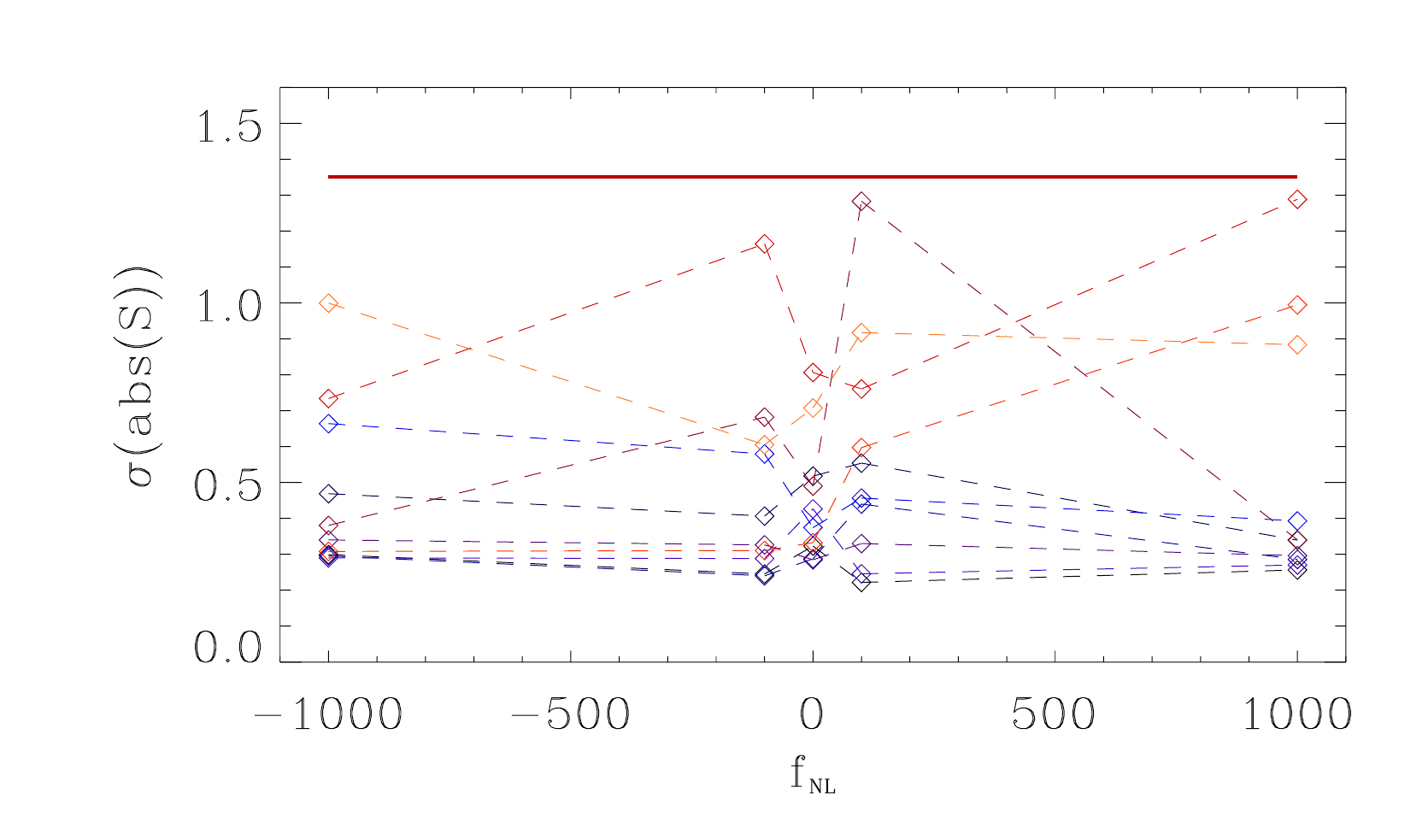}
\caption{Maximum, mean and and standard deviation of the absolute values of $S(\chi^2)_{\alpha}$ of ten $f_{\mathrm{NL}}^{\mathrm{local}}$ simulations compared to one NILC7 original map (red line). The HEALPix resolution of the maps is $N_{\mathrm{side}}=64$. The $\ell$-range for the method of the surrogates is $\Delta \ell = [2,20]$.}
\label{fig:FnlNILC7SIM}
\end{figure}

\subsection{$f_{\mathrm{NL}}$-dependent Simulations}
The analysis of the ILC and NILC7 surrogate maps with the Minkowski functionals as well as the scaling indices reveals HOCs in the original maps. In order to test whether these phase correlations can be reproduced by $f_{\mathrm{NL}}^{\mathrm{local}}$-models we analyse constrained realisations of the temperature maps with varying $f_{\mathrm{NL}}^{\mathrm{local}}$-parameter with values $f_{\mathrm{NL}}^{\mathrm{local}} = [0,\pm 100, \pm 1000]$ by means of the surrogate method for $\Delta \ell = [2,20]$. Figure \ref{fig:FnlNILC7minkos} shows the results for the corresponding $S(\chi^2)$ statistics for the three Minkowski functionals. We plot the maximum, mean and standard deviation of $S(\chi^2)$ for ten simulations and the NILC7 original map. The HEALPix resolution parameter of the maps is decreased to $N_{\mathrm{side}} =64$ which does not change the results for low-$\ell$ ranges. One can immediately see that the $S$-values of the  $f_{\mathrm{NL}}^{\mathrm{local}}$-simulations are nearly always smaller compared to the original data set. The few exceptions where the signal of the simulations lies close to the NILC7 data feature a value of $f_{\mathrm{NL}}^{\mathrm{local}}=\pm 1000$. These extreme values are already ruled out by recent analyses that resulted in current constraints for $|f_{\mathrm{NL}}^{\mathrm{local}}|$ well below $100$ (e.g. $f_{\mathrm{NL}}^{\mathrm{local}}= 32\pm21$ ($68\%$ CL), \cite{2011ApJS..192...18K}). For the scaling indices, we obtain similar results, which are shown in Figure \ref{fig:FnlNILC7SIM}. For the maximum plot, only one simulation lies above the results for the NILC7 map, which is at $f_{\mathrm{NL}}^{\mathrm{local}}=100$. For the mean, we obtain three points with a slightly higher value than the data, and none for the standard deviation. 

One has to conclude that the CMB simulations with $f_{\mathrm{NL}}^{\mathrm{local}}$ cannot reproduce the low-$\ell$ anomalies we found in the WMAP data. This means that the detected large-scale NGs and asymmetries in the data do not correspond to the type of NG which is described by $f_{\mathrm{NL}}^{\mathrm{local}}$ but stem from a different origin. The search for the source of the deviations from Gaussianity has to be continued in future analyses.

\section{Conclusions}
In addition to our previous work based on the use of surrogate maps we analysed latest WMAP experiment CMB maps with respect to asymmetries and scale-dependent non-Gaussianity. The surrogates are generated by a scale-dependent shuffling of Fourier phases while all other properties of the maps are preserved. In this work we focus on the Minkowski functionals calculated for the ILC7 and NILC7 maps as a scale-independent measure being sensitive on the HOCs of the maps. We compare these new results with the scale-dependent scaling index method calculated for the NILC7 map in this work and for the ILC7 map in previous works. We find that both measures detect highly significant signatures for phase-correlations and therefore deviations from Gaussianity, and furthermore ecliptic hemispherical asymmetries for the interval $\Delta \ell = [2,20]$ in both ILC and NILC maps. The reduction of the analysed sky region down to $30\%$ of the former hemispheres shows no signal for NG in the northern ecliptic sky. In the south we detect individual spots of NG. 

If the findings are indeed of intrinsic nature they would disagree with the predictions of isotropic cosmologies with single-field slow-roll inflation.

The two different image analysis techniques reveal very consistent results on the low-$\ell$ range for both maps. The signatures we find for $\Delta \ell = [120,300]$ show less  agreement between ILC and NILC and differ from the low-$\ell$ range results. They also depend on the image analysis technique and the resolution of the map. As discussed in our earlier works NGs on these scales can be induced by foreground cleaning and can be subject to secondary anisotropies. 

The constrained realisations of the CMB with varying $f_{\mathrm{NL}}^{\mathrm{local}}$ cannot parametrise the detected NGs and asymmetries on largest scales. $f_{\mathrm{NL}}^{\mathrm{local}}$ might still be a suitable parametrisation for smaller scales (larger-$\ell$ ranges). Also, other types of $f_{\mathrm{NL}}$, as the equilateral and orthogonal type, also the $g_{\mathrm{NL}}$ parameter could characterise the found NGs. A deeper study of different inflationary models, as for example Bianchi type models, that represent appropriate test candidates for the found anomalies, is required. 

Future investigations, e.g. of upcoming data of the Planck satellite, will shed more light on the open questions regarding instrumental constraints, observational systematics, map making influences and resolution problems.

\section*{Acknowledgements}
Many of the results in this paper have been derived using the HEALPix \citep{2005ApJ...622..759G} software and analysis package. The WMAP data are taken from the Legacy Archive for Microwave Background Data Analysis (LAMBDA). Support for LAMBDA is provided by the NASA Office of Space Science. HM thanks the Christiane N\"usslein-Volhard (CNV) foundation for financial support and acknowledges the support of the International Max Planck Research School. 

%%%%%%%%%%%%%%%%%%%%%%%%%%%%%%%%%%%%

% \bsp % ``This paper has been produced using the ...''
\bibliographystyle{mn2e}
\bibliography{/Users/hmeyer/Documents/Literatur}

\begin{thebibliography}{58}
\expandafter\ifx\csname natexlab\endcsname\relax\def\natexlab#1{#1}\fi

\bibitem[{{Acquaviva} {et~al}\mbox{.}(2003){Acquaviva}, {Bartolo}, {Matarrese},
  \& {Riotto}}]{2003NuPhB.667..119A}
{Acquaviva} V., {Bartolo} N., {Matarrese} S., {Riotto} A., 2003, Nuclear
  Physics B, 667, 119

\bibitem[{{Albrecht} \& {Steinhardt}(1982)}]{1982PhRvL..48.1220A}
{Albrecht} A., {Steinhardt} P.~J., 1982, Physical Review Letters, 48, 1220

\bibitem[{{Alishahiha}, {Silverstein} \& {Tong}(2004){Alishahiha},
  {Silverstein}, \& {Tong}}]{2004PhRvD..70l3505A}
{Alishahiha} M., {Silverstein} E., {Tong} D., 2004, \prd, 70, 123505

\bibitem[{{Arkani-Hamed} {et~al}\mbox{.}(2004){Arkani-Hamed}, {Creminelli},
  {Mukohyama}, \& {Zaldarriaga}}]{2004JCAP...04..001A}
{Arkani-Hamed} N., {Creminelli} P., {Mukohyama} S., {Zaldarriaga} M., 2004,
  \jcap, 4, 1

\bibitem[{Bardeen(1980)}]{PhysRevD.22.1882}
Bardeen J.~M., 1980, \prd, 22, 1882

\bibitem[{{Bartolo} {et~al}\mbox{.}(2004){Bartolo}, {Komatsu}, {Matarrese}, \&
  {Riotto}}]{2004PhR...402..103B}
{Bartolo} N., {Komatsu} E., {Matarrese} S., {Riotto} A., 2004, \physrep, 402,
  103

\bibitem[{{Bartolo}, {Matarrese} \& {Riotto}(2002){Bartolo}, {Matarrese}, \&
  {Riotto}}]{2002PhRvD..65j3505B}
{Bartolo} N., {Matarrese} S., {Riotto} A., 2002, \prd, 65, 103505

\bibitem[{{Basak} \& {Delabrouille}(2011)}]{2011MNRAS.tmp.1768B}
{Basak} S., {Delabrouille} J., 2011, \mnras, 1768

\bibitem[{{Bennett} {et~al}\mbox{.}(2003){Bennett}, {Halpern}, {Hinshaw},
  {Jarosik}, {Kogut}, {Limon}, {Meyer}, {Page}, {Spergel}, {Tucker}, {Wollack},
  {Wright}, {Barnes}, {Greason}, {Hill}, {Komatsu}, {Nolta}, {Odegard},
  {Peiris}, {Verde}, \& {Weiland}}]{2003ApJS..148....1B}
{Bennett} C.~L. {et~al.}, 2003, \apjs, 148, 1

\bibitem[{{Bernardeau} \& {Uzan}(2002)}]{2002PhRvD..66j3506B}
{Bernardeau} F., {Uzan} J.-P., 2002, \prd, 66, 103506

\bibitem[{{Buchbinder}, {Khoury} \& {Ovrut}(2008){Buchbinder}, {Khoury}, \&
  {Ovrut}}]{2008PhRvL.100q1302B}
{Buchbinder} E.~I., {Khoury} J., {Ovrut} B.~A., 2008, Physical Review Letters,
  100, 171302

\bibitem[{{Chen}(2010)}]{2010AdAst2010E..72C}
{Chen} X., 2010, Advances in Astronomy, 2010

\bibitem[{{Chiang}, {Naselsky} \& {Coles}(2007){Chiang}, {Naselsky}, \&
  {Coles}}]{2007ApJ...664....8C}
{Chiang} L.-Y., {Naselsky} P.~D., {Coles} P., 2007, \apj, 664, 8

\bibitem[{{Chiang} {et~al}\mbox{.}(2003){Chiang}, {Naselsky}, {Verkhodanov}, \&
  {Way}}]{2003ApJ...590L..65C}
{Chiang} L.-Y., {Naselsky} P.~D., {Verkhodanov} O.~V., {Way} M.~J., 2003,
  \apjl, 590, L65

\bibitem[{{Coles} {et~al}\mbox{.}(2004){Coles}, {Dineen}, {Earl}, \&
  {Wright}}]{2004MNRAS.350..989C}
{Coles} P., {Dineen} P., {Earl} J., {Wright} D., 2004, \mnras, 350, 989

\bibitem[{{de Oliveira-Costa} {et~al}\mbox{.}(2004){de Oliveira-Costa},
  {Tegmark}, {Zaldarriaga}, \& {Hamilton}}]{2004PhRvD..69f3516D}
{de Oliveira-Costa} A., {Tegmark} M., {Zaldarriaga} M., {Hamilton} A., 2004,
  \prd, 69, 063516

\bibitem[{{Delabrouille} {et~al}\mbox{.}(2009){Delabrouille}, {Cardoso}, {Le
  Jeune}, {Betoule}, {Fay}, \& {Guilloux}}]{2009A&A...493..835D}
{Delabrouille} J., {Cardoso} J.-F., {Le Jeune} M., {Betoule} M., {Fay} G.,
  {Guilloux} F., 2009, \aap, 493, 835

\bibitem[{{Elsner} \& {Wandelt}(2009)}]{2009ApJS..184..264E}
{Elsner} F., {Wandelt} B.~D., 2009, \apjs, 184, 264

\bibitem[{{Enqvist} \& {Sloth}(2002)}]{2002NuPhB.626..395E}
{Enqvist} K., {Sloth} M.~S., 2002, Nuclear Physics B, 626, 395

\bibitem[{{Eriksen} {et~al}\mbox{.}(2007){Eriksen}, {Banday}, {G{\'o}rski},
  {Hansen}, \& {Lilje}}]{2007ApJ...660L..81E}
{Eriksen} H.~K., {Banday} A.~J., {G{\'o}rski} K.~M., {Hansen} F.~K., {Lilje}
  P.~B., 2007, \apjl, 660, L81

\bibitem[{{Eriksen} {et~al}\mbox{.}(2005){Eriksen}, {Banday}, {G{\'o}rski}, \&
  {Lilje}}]{2005ApJ...622...58E}
{Eriksen} H.~K., {Banday} A.~J., {G{\'o}rski} K.~M., {Lilje} P.~B., 2005, \apj,
  622, 58

\bibitem[{{Eriksen} {et~al}\mbox{.}(2004){Eriksen}, {Hansen}, {Banday},
  {G{\'o}rski}, \& {Lilje}}]{2004ApJ...605...14E}
{Eriksen} H.~K., {Hansen} F.~K., {Banday} A.~J., {G{\'o}rski} K.~M., {Lilje}
  P.~B., 2004, \apj, 605, 14

\bibitem[{{Gold} {et~al}\mbox{.}(2011){Gold}, {Odegard}, {Weiland}, {Hill},
  {Kogut}, {Bennett}, {Hinshaw}, {Chen}, {Dunkley}, {Halpern}, {Jarosik},
  {Komatsu}, {Larson}, {Limon}, {Meyer}, {Nolta}, {Page}, {Smith}, {Spergel},
  {Tucker}, {Wollack}, \& {Wright}}]{2011ApJS..192...15G}
{Gold} B. {et~al.}, 2011, \apjs, 192, 15

\bibitem[{{G{\'o}rski} {et~al}\mbox{.}(2005){G{\'o}rski}, {Hivon}, {Banday},
  {Wandelt}, {Hansen}, {Reinecke}, \& {Bartelmann}}]{2005ApJ...622..759G}
{G{\'o}rski} K.~M., {Hivon} E., {Banday} A.~J., {Wandelt} B.~D., {Hansen}
  F.~K., {Reinecke} M., {Bartelmann} M., 2005, \apj, 622, 759

\bibitem[{Grassberger \& Procaccia(1983)}]{Grassberger1983189}
Grassberger P., Procaccia I., 1983, Physica D: Nonlinear Phenomena, 9, 189

\bibitem[{{Guth}(1981)}]{1981PhRvD..23..347G}
{Guth} A.~H., 1981, \prd, 23, 347

\bibitem[{Hadwiger(1957)}]{hadwiger:1957:vio}
Hadwiger H., 1957, Vorlesungen {\"u}ber Inhalt, Oberfl{\"a}che und
  Isoperimetrie. Springer-Verlag, Berlin

\bibitem[{{Hansen}, {Banday} \& {G{\'o}rski}(2004){Hansen}, {Banday}, \&
  {G{\'o}rski}}]{2004MNRAS.354..641H}
{Hansen} F.~K., {Banday} A.~J., {G{\'o}rski} K.~M., 2004, \mnras, 354, 641

\bibitem[{{Hansen} {et~al}\mbox{.}(2009){Hansen}, {Banday}, {G{\'o}rski},
  {Eriksen}, \& {Lilje}}]{2009ApJ...704.1448H}
{Hansen} F.~K., {Banday} A.~J., {G{\'o}rski} K.~M., {Eriksen} H.~K., {Lilje}
  P.~B., 2009, \apj, 704, 1448

\bibitem[{{Jarosik} {et~al}\mbox{.}(2011){Jarosik}, {Bennett}, {Dunkley},
  {Gold}, {Greason}, {Halpern}, {Hill}, {Hinshaw}, {Kogut}, {Komatsu},
  {Larson}, {Limon}, {Meyer}, {Nolta}, {Odegard}, {Page}, {Smith}, {Spergel},
  {Tucker}, {Weiland}, {Wollack}, \& {Wright}}]{2011ApJS..192...14J}
{Jarosik} N. {et~al.}, 2011, \apjs, 192, 14

\bibitem[{{Komatsu} {et~al}\mbox{.}(2009){Komatsu}, {Afshordi}, {Bartolo},
  {Baumann}, {Bond}, {Buchbinder}, {Byrnes}, {Chen}, {Chung}, {Cooray},
  {Creminelli}, {Dalal}, {Dore}, {Easther}, {Frolov}, {Khoury}, {Kinney},
  {Kofman}, {Koyama}, {Leblond}, {Lehners}, {Lidsey}, {Liguori}, {Lim},
  {Linde}, {Lyth}, {Maldacena}, {Matarrese}, {McAllister}, {McDonald},
  {Mukohyama}, {Ovrut}, {Peiris}, {Riotto}, {Rodrigues}, {Sasaki},
  {Scoccimarro}, {Seery}, {Sefusatti}, {Smith}, {Starobinsky}, {Steinhardt},
  {Takahashi}, {Tegmark}, {Tolley}, {Verde}, {Wandelt}, {Wands}, {Weinberg},
  {Wyman}, {Yadav}, \& {Zaldarriaga}}]{2009astro2010S.158K}
{Komatsu} E. {et~al.}, 2009, in Astronomy, Vol. 2010, astro2010: The Astronomy
  and Astrophysics Decadal Survey, pp. 158--+

\bibitem[{{Komatsu} {et~al}\mbox{.}(2011){Komatsu}, {Smith}, {Dunkley},
  {Bennett}, {Gold}, {Hinshaw}, {Jarosik}, {Larson}, {Nolta}, {Page},
  {Spergel}, {Halpern}, {Hill}, {Kogut}, {Limon}, {Meyer}, {Odegard}, {Tucker},
  {Weiland}, {Wollack}, \& {Wright}}]{2011ApJS..192...18K}
---, 2011, \apjs, 192, 18

\bibitem[{{Lehners} \& {Steinhardt}(2008)}]{2008PhRvD..78b3506L}
{Lehners} J.-L., {Steinhardt} P.~J., 2008, \prd, 78, 023506

\bibitem[{{Linde} \& {Mukhanov}(1997)}]{1997PhRvD..56..535L}
{Linde} A., {Mukhanov} V., 1997, \prd, 56, 535

\bibitem[{{Linde}(1982)}]{1982PhLB..108..389L}
{Linde} A.~D., 1982, Physics Letters B, 108, 389

\bibitem[{{Lyth}, {Ungarelli} \& {Wands}(2003){Lyth}, {Ungarelli}, \&
  {Wands}}]{2003PhRvD..67b3503L}
{Lyth} D.~H., {Ungarelli} C., {Wands} D., 2003, \prd, 67, 023503

\bibitem[{{Maldacena}(2003)}]{2003JHEP...05..013M}
{Maldacena} J., 2003, Journal of High Energy Physics, 5, 13

\bibitem[{{Mecke}, {Buchert} \& {Wagner}(1994){Mecke}, {Buchert}, \&
  {Wagner}}]{1994A&A...288..697M}
{Mecke} K.~R., {Buchert} T., {Wagner} H., 1994, \aap, 288, 697

\bibitem[{{Michielsen} \& {De Raedt}(2001)}]{michielsen01}
{Michielsen} K., {De Raedt} H., 2001, Physics Reports, 347, 461

\bibitem[{{Moroi} \& {Takahashi}(2001)}]{2001PhLB..522..215M}
{Moroi} T., {Takahashi} T., 2001, Physics Letters B, 522, 215

\bibitem[{{Naselsky} {et~al}\mbox{.}(2005){Naselsky}, {Chiang}, {Olesen}, \&
  {Novikov}}]{2005PhRvD..72f3512N}
{Naselsky} P., {Chiang} L.-Y., {Olesen} P., {Novikov} I., 2005, \prd, 72,
  063512

\bibitem[{{Park}(2004)}]{2004MNRAS.349..313P}
{Park} C.-G., 2004, \mnras, 349, 313

\bibitem[{{R{\"a}th} {et~al}\mbox{.}(2011){R{\"a}th}, {Banday}, {Rossmanith},
  {Modest}, {S{\"u}tterlin}, {G{\'o}rski}, {Delabrouille}, \&
  {Morfill}}]{2011MNRAS.415.2205R}
{R{\"a}th} C., {Banday} A.~J., {Rossmanith} G., {Modest} H., {S{\"u}tterlin}
  R., {G{\'o}rski} K.~M., {Delabrouille} J., {Morfill} G.~E., 2011, \mnras,
  415, 2205

\bibitem[{{R{\"a}th} {et~al}\mbox{.}(2002){R{\"a}th}, {Bunk}, {Huber},
  {Morfill}, {Retzlaff}, \& {Schuecker}}]{2002MNRAS.337..413R}
{R{\"a}th} C., {Bunk} W., {Huber} M.~B., {Morfill} G.~E., {Retzlaff} J.,
  {Schuecker} P., 2002, \mnras, 337, 413

\bibitem[{{R{\"a}th} {et~al}\mbox{.}(2009){R{\"a}th}, {Morfill}, {Rossmanith},
  {Banday}, \& {G{\'o}rski}}]{2009PhRvL.102m1301R}
{R{\"a}th} C., {Morfill} G.~E., {Rossmanith} G., {Banday} A.~J., {G{\'o}rski}
  K.~M., 2009, Physical Review Letters, 102, 131301

\bibitem[{{R{\"a}th} \& {Schuecker}(2003)}]{2003MNRAS.344..115R}
{R{\"a}th} C., {Schuecker} P., 2003, \mnras, 344, 115

\bibitem[{{R{\"a}th}, {Schuecker} \& {Banday}(2007){R{\"a}th}, {Schuecker}, \&
  {Banday}}]{2007MNRAS.380..466R}
{R{\"a}th} C., {Schuecker} P., {Banday} A.~J., 2007, \mnras, 380, 466

\bibitem[{{Rossmanith} {et~al}\mbox{.}(2011){Rossmanith}, {Modest}, {R{\"a}th},
  {Banday}, {G{\'o}rski}, \& {Morfill}}]{2011AdAst2011E..11R}
{Rossmanith} G., {Modest} H., {R{\"a}th} C., {Banday} A.~J., {G{\'o}rski}
  K.~M., {Morfill} G., 2011, Advances in Astronomy, 2011

\bibitem[{{Rossmanith} {et~al}\mbox{.}(2012){Rossmanith}, {Modest}, {R{\"a}th},
  {Banday}, {Gorski}, \& {Morfill}}]{2012arXiv1207.3235R}
{Rossmanith} G., {Modest} H., {R{\"a}th} C., {Banday} A.~J., {Gorski} K.~M.,
  {Morfill} G., 2012, ArXiv e-prints

\bibitem[{{Rossmanith} {et~al}\mbox{.}(2009){Rossmanith}, {R{\"a}th}, {Banday},
  \& {Morfill}}]{2009MNRAS.399.1921R}
{Rossmanith} G., {R{\"a}th} C., {Banday} A.~J., {Morfill} G., 2009, \mnras,
  399, 1921

\bibitem[{{Rubakov} \& {Vlasov}(2010)}]{2010arXiv1008.1704R}
{Rubakov} V., {Vlasov} A., 2010, ArXiv e-prints

\bibitem[{{Schmalzing} \& {Gorski}(1998)}]{1998MNRAS.297..355S}
{Schmalzing} J., {Gorski} K.~M., 1998, \mnras, 297, 355

\bibitem[{{Silverstein} \& {Tong}(2004)}]{2004PhRvD..70j3505S}
{Silverstein} E., {Tong} D., 2004, \prd, 70, 103505

\bibitem[{{Szapudi}(2009)}]{2009LNP...665..457S}
{Szapudi} I., 2009, in Lecture Notes in Physics, Berlin Springer Verlag, Vol.
  665, Data Analysis in Cosmology, {Mart{\'{\i}}nez} V.~J., {Saar} E.,
  {Mart{\'{\i}}nez-Gonz{\'a}lez} E., {Pons-Border{\'{\i}}a} M.-J., eds., pp.
  457--492

\bibitem[{Theiler {et~al}\mbox{.}(1992)Theiler, Eubank, Longtin, Galdrikian, \&
  Farmer}]{Theiler199277}
Theiler J., Eubank S., Longtin A., Galdrikian B., Farmer J.~D., 1992, Physica
  D: Nonlinear Phenomena, 58, 77

\bibitem[{{Vernizzi} \& {Wands}(2006)}]{2006JCAP...05..019V}
{Vernizzi} F., {Wands} D., 2006, \jcap, 5, 19

\bibitem[{{Vielva} {et~al}\mbox{.}(2004){Vielva},
  {Mart{\'{\i}}nez-Gonz{\'a}lez}, {Barreiro}, {Sanz}, \&
  {Cay{\'o}n}}]{2004ApJ...609...22V}
{Vielva} P., {Mart{\'{\i}}nez-Gonz{\'a}lez} E., {Barreiro} R.~B., {Sanz} J.~L.,
  {Cay{\'o}n} L., 2004, \apj, 609, 22

\bibitem[{{Winitzki} \& {Kosowsky}(1998)}]{1998NewA....3...75W}
{Winitzki} S., {Kosowsky} A., 1998, \na, 3, 75

\end{thebibliography}
\label{lastpage}

\end{document}